# Structure and dynamics of molecular networks:
# A novel paradigm of drug discovery
# A comprehensive review

Peter Csermely[a,*], Tamás Korcsmáros[a,b], Huba J.M. Kiss[a,c], Gábor London[d], Ruth Nussinov[e,f]

[a] Department of Medical Chemistry, Semmelweis University, P.O. Box 260, H-1444 Budapest 8, Hungary
[b] Department of Genetics, Eötvös University, Pázmány P. s. 1C, H-1117 Budapest, Hungary
[c] Department of Ophthalmology, Semmelweis University, Tömő str. 25-29, H-1083 Budapest, Hungary
[d] Department of Chemistry and Applied Biosciences, Swiss Federal Institute of Technology (ETH), Zurich, Switzerland
[e] Center for Cancer Research Nanobiology Program, SAIC-Frederick, Inc., National Cancer Institute, Frederick National laboratory for Cancer Research, Frederick, MD 21702, USA
[f] Sackler Institute of Molecular Medicine, Department of Human Genetics and Molecular Medicine, Sackler School of Medicine, Tel Aviv University, Tel Aviv 69978, Israel

**Abstract:** Despite considerable progress in genome- and proteome-based high-throughput screening methods and in rational drug design, the increase in approved drugs in the past decade did not match the increase of drug development costs. Network description and analysis not only give a systems-level understanding of drug action and disease complexity, but can also help to improve the efficiency of drug design. We give a comprehensive assessment of the analytical tools of network topology and dynamics. The state-of-the-art use of chemical similarity, protein structure, protein–protein interaction, signaling, genetic interaction and metabolic networks in the discovery of drug targets is summarized. We propose that network targeting follows two basic strategies. The "central hit strategy" selectively targets central nodes/edges of the flexible networks of infectious agents or cancer cells to kill them. The "network influence strategy" works against other diseases, where an efficient reconfiguration of rigid networks needs to be achieved by targeting the neighbors of central nodes/edges. It is shown how network techniques can help in the identification of single-target, edgetic, multi-target and allo-network drug target candidates. We review the recent boom in network methods helping hit identification, lead selection optimizing drug efficacy, as well as minimizing side-effects and drug toxicity. Successful network-based drug development strategies are shown through the examples of infections, cancer, metabolic diseases, neurodegenerative diseases and aging. Summarizing >1200 references we suggest an optimized protocol of network-aided drug development, and provide a list of systems-level hallmarks of drug quality. Finally, we highlight network-related drug development trends helping to achieve these hallmarks by a cohesive, global approach.

*Abbreviations: ADME, absorption, distribution, metabolism and excretion; ADMET, absorption, distribution, metabolism, excretion and toxicity; FDA, USA Food and Drug Administration; GWAS, genome-wide association study; mTOR, mammalian target of rapamycin; NME, new molecular entity; QSAR, quantitative structure–activity relationship; QSPR, quantitative structure–property relationship; PPAR, peroxisome proliferator-activated receptor; SNP, single-nucleotide polymorphism.*

* Corresponding author. Tel.: +36 1 459 1500; fax: +36 1 266 3802.
  E-mail address: csermely.peter@med.semmelweis-univ.hu (P. Csermely).

# 1. Introduction

'Business as usual' is no longer an option in drug industry (Begley & Ellis, 2012). There is a growing recognition that systems-level thinking is needed for the renewal of drug development efforts. However, interrelated data have grown to such an unforeseen complexity, which argues for novel concepts and strategies. The Introduction aims to convey to the Reader that the network description and analysis can be a suitable method to describe the complexity of human diseases and help the development of new drugs.

## 1.1. Drug design as an area requiring a complex approach

The population of Earth is growing and aging. Some of the major health challenges, such as many types of cancers and infectious diseases, diabetes and neurodegenerative diseases are in desperate need of innovative medicines. Despite of this challenge, fast and affordable drug development is a vision that contrasts sharply with the current state of drug discovery. It takes an average of 12 to 15 years and (depending on the therapeutic area) as much as 1 billion USD to bring a single drug into market. In the USA, pharmaceutical industry was the most R&D-intensive industry (defined as the ratio of R&D spending compared to total sales revenue) until 2003, when it was overtaken by communications equipment industry (Austin, 2006; Chong & Sullivan, 2007; Bunnage, 2011).

The increasingly high costs of drug development are partly associated

- with the high percentage of projects that fail in clinical trials,
- with the recent focus on chronic diseases requiring longer and more expensive clinical trials,
- with the increased safety concerns caused by catastrophic failures in the market and
- with more expensive research technologies.
- Moreover, direct costs are doubled, where the second half comes from the 'opportunity cost', i.e. the financial costs of tying up investment capital in multiyear drug development projects (Austin, 2006; Chong & Sullivan, 2007; Bunnage, 2011).

We have a few hundreds of targets of approved drugs from the >20,000 non-redundant proteins of the human proteome. Despite the considerably higher R&D investment after the millennium, the number of new molecular entities (NMEs) approved by the USA Food and Drug Administration (FDA) remained constant at an annual 20 to 30 compounds. The number of NMEs potentially offering a substantial advance over conventional therapies is an even more sobering number of 6 to 17 per year in the last decade (Fig. 1). However, it is worth to note that looking only at the number of new drugs without considering their therapeutic value omits an important factor in the analysis (Austin, 2006; Overington et al., 2006; Chong & Sullivan, 2007; Bunnage, 2011; Edwards et al., 2011; Scannell et al., 2012).

Part of the slow progress is related to the high risks of investments. The development of an NME-drug costs approximately four times more than that of a non-NME. Moreover, the 'curse of attrition' steadily remained the biggest issue of the pharmaceutical industry in the last decades (Fig. 2). Each NME launched to the market needs about 24 development candidates to enter the development pipeline. Attrition of phase II studies is the key challenge, where only 25% of the drug-candidates survive. The 25% survival includes new agents against known targets (the 'me-too' or 'me-better' drugs), and therefore may be a significant overestimate of the survival rate of drug-candidates directed towards new targets. The low survival rate is exacerbated further by the very high costs of a failing compound at this late development stage (Brown & Superti-Furga, 2003; Austin, 2006; Bunnage, 2011; Ledford, 2012). These high risks made the drug industry cautious, and sometimes perhaps over-cautious. As the pharmacologist and Nobel Laureate James Black said: "the most fruitful basis for the discovery of a new drug is to start with an old drug" (Chong & Sullivan, 2007). In fact, analysis of structure–activity relationship (SAR) pattern evolution, drug–target network topology and literature mining studies all showed the same behavior trend indicating that more than 80% of the new drugs tend to bind targets, which are connected to the network of previous drug targets (Cokol et al., 2005; Yildirim et al., 2007; Iyer et al., 2011a).

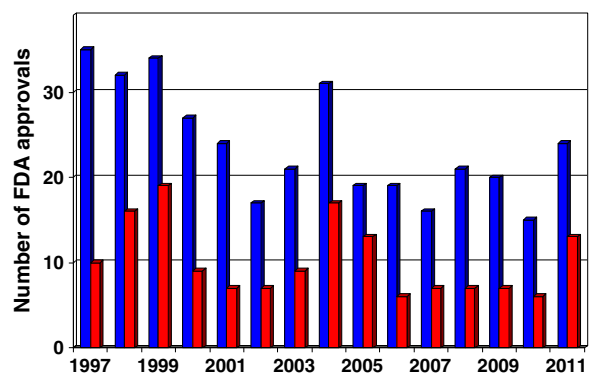

**Fig. 1.** Number of new molecular entities (NME, a drug containing an active ingredient that has not been previously approved by the US FDA) approved by the US Food and Drug Administration (FDA). Blue bars represent the total number of NMEs, whereas red bars represent "priority" NMEs that potentially offer a substantial advance over conventional therapies.
Source: http://www.fda.gov/Drugs/default.htm.

Improving the quality of target selection is widely considered as the single most important factor to improve the productivity of the pharmaceutical industry. From the 1970s target selection was increasingly separated from lead identification. Drug development process often fell to the 'druggability trap', where the attraction of working on a chemically approachable target encouraged development teams to push forward projects having a poor target quality. Additionally, chemical leads were often discovered to have unwanted side-effects and/or be toxic at later development phases (Brown & Superti-Furga, 2003; Hopkins, 2008; Bunnage, 2011).

The decline in the productivity of the pharmacological industry may stem partly from the underestimation of the complexity of cells, organisms and human disease (Lowe et al., 2010). We will illustrate the high level of this complexity by three examples.

- Under ideal conditions only 34% of single-gene deletions in yeast resulted in decrease in proliferation. However, when knockouts were screened against a diverse small-molecule library and a wide range of environmental conditions, 97% of the gene-deletions demonstrated a fitness defect (Hillenmeyer et al., 2008).
- Many of the most prevalent diseases, such as cancer, diabetes and coronary artery disease have a genetic background including a large number of genes (see Section 5 and Brown & Superti-Furga, 2003; Hopkins, 2008; Fliri et al., 2010). Following a treatment with a chemotherapeutic agent almost all of 1000 tagged proteins of cancer cells showed a dynamic response, when their temporal expression levels and localization were tracked (Cohen et al., 2008).
- As Loscalzo and Barabasi (2011) summarized in their excellent review, diseases are typically recognized and defined by their late-appearing manifestations in a partially dysfunctional organ-system. As a part of this, therapeutic strategies often do not focus on truly unique, targeted disease determinants, but (rightfully) address the patho-phenotypes of the already advanced disease stage. These advanced patho-phenotypes have a large number of symptoms, which are not primarily disease-specific (such as inflammation). This definition of disease may obscure subtle, but potentially important differences among patients with clinical presentations, and may also neglect pathobiological mechanisms extending the disease-defining organ system. Loscalzo and Barabasi (2011) argue that the complexity of

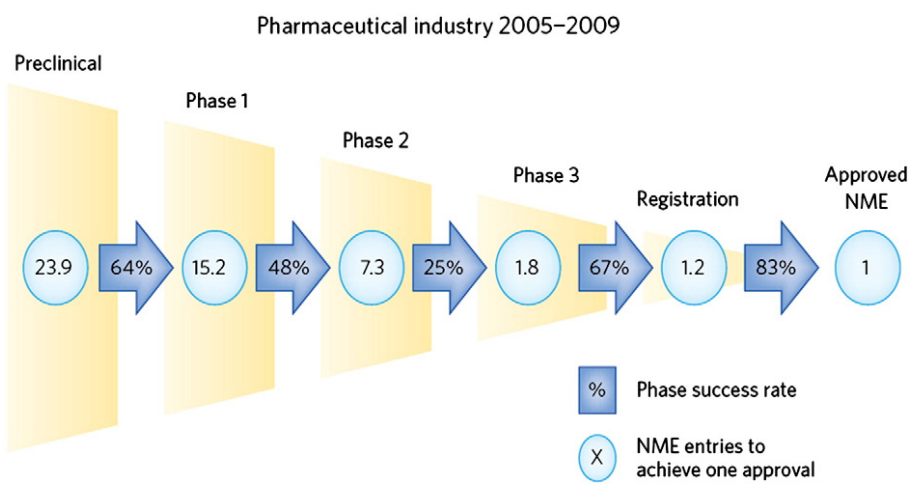

**Fig. 2.** Success rate of new molecular entities (NMEs) by R&D development phases. The figure shows the combined R&D survival by development phase for 14 large pharmaceutical companies. Note that attrition figures for early phases might be even higher, since an early problem might be first neglected making a failure only at a later phase (Brown & Superti-Furga, 2003).

Reprinted by permission from the Macmillan Publishers Ltd: Nature Chemical Biology, Bunnage, 2011, Copyright, 2011.

disease should be viewed as an emergent property of a pathobiological system, i.e. a property, which cannot be predicted by studying only the parts of the system, but emerges from the complex interrelationships of all system components. Kola and Bell (2011) arrive to the same conclusion urging the reform of the taxonomy of human disease.

These examples illustrate the extent of non-linearity and interdependence of cellular and organismal responses. To understand these observations and outcomes, we need novel approaches.

Over-reliance on inadequate animal or cellular models of disease has been considered to play a major part in the poor levels of Phase II drug candidate survival-rate. We illustrate the limitations and dangers of model-selection by three examples.

- 41% of the proteins expressed in rat lungs were absent from equivalent cultured cells (Lindsay, 2005).
- Animal strains are often in-bred, and are examined in a young age for diseases having an onset in elderly people (Lindsay, 2005).
- In psychological clinical studies 96% of the patients cover 12% of the world population (Henrich et al., 2010). A more equal coverage is also required by the geographic clustering of rare genetic variants affecting drug efficacy (Nelson et al., 2012).

There is a growing recognition that systems-level thinking may help to overcome many of the current troubles of drug development (Brown & Superti-Furga, 2003; Csermely et al., 2005; Lindsay, 2005; Korcsmáros et al., 2007; Henney & Superti-Furga, 2008; Hopkins, 2008; Westerhoff et al., 2008; Bunnage, 2011; Chua & Roth, 2011; Farkas et al., 2011; Penrod et al., 2011; Begley & Ellis, 2012). As a sign of this, leading systems biologists aim to construct a computer replica of the whole human body, called as the 'silicon human' by 2038 (Kolodkin et al., 2012).

In fact, systems-level thinking characterized drug development until the 1970s, when mechanistic drug-targets were unknown. Until the late 1970s even the concept of the receptor was not based on sequence and structural data, but on the chemical similarities of ligands exerting similar pharmacological actions (Brown & Superti-Furga, 2003; Keiser et al., 2010). It was only after the early 1980s, that the focus shifted from physiological observations to the molecular level (Pujol et al., 2010).

The renewal of systems-based thinking in drug discovery was helped by the following three factors. 1.) The development of robust high-throughput platforms to gather large amounts of comparable molecular data. 2.) The assembly and availability of curated databases integrating the knowledge of the field. 3.) The emergence of interdisciplinary research to understand these data (Arrell & Terzic, 2010). Most of the current largest pharmaceutical firms are products of horizontal mergers between two or more large drug companies which have been taking place since 1989. Though larger companies have the advantage to fund and sustain a broader range of larger research programs, the development of large firms and research enterprises was often considered to decrease flexible responses to novel development opportunities (Austin, 2006; Gros, 2012). An increased efficiency needs coordinated networking of large drug development firms, biotechnological companies and research institutions (Hasan et al., 2012; Heemskerk et al., 2012). Moreover, systems-level thinking needs a new behavior code of sharing data and approaches. This new alliance is characterized by the following behavior.

- In systems-level drug development, quality and not quantity of data is a key issue. A reliable data pipeline must be assembled using appropriate standards and quality control-metrics keeping in mind the needs of systems biology. This is all the more important, since it may also overcome the unreliability problems which surfaced recently, when Amgen tried to reproduce data from 53 published preclinical studies of potential anticancer drugs, and it failed in all but 6 cases (11% reproducibility rate), or Bayer Health Care could reproduce only 25% of previously published preclinical studies (Henney & Superti-Furga, 2008; Prinz et al., 2011; Begley & Ellis, 2012; Landis et al., 2012).
- Sharing of systems-level results led to a fast development of predictive toxicology, which is a key step of a more efficient progress (Henney & Superti-Furga, 2008).

Datasets are growing to dimensions, where the three billion nucleotides that comprise the human genome (International Human Genome Sequencing Consortium, 2004; ENCODE Project Consortium, 2012) became millionths of the ~1 petabyte data we had in 2008 (Schadt et al., 2009), which have grown well over 1 exabyte (billion times billion bytes) by 2012. These magnitudes require appropriate computational tools to understand them. Through this review we hope to convince the Reader that network description and analysis offer novel tools, which can help us to understand the complexity of human disease and enable the integration of knowledge towards a more efficient combat strategy for healthier life.

### 1.2. Molecular networks as efficient tools in the description of cellular and organism behavior

Complexity can be described through the rather simple saying that 'in a complex system the whole is more than the sum of its parts: cutting a horse to two will not result in two small horses' (Kolodkin et al., 2012; San Miguel et al., 2012). Newman (2011) summarized a number of excellent sources to study complexity. A recent summary listed the following hallmarks of complex systems and their behavior: many heterogeneous interacting parts; multiple scales; combinatorial explosion of possible states; complicated transition laws; unexpected or unpredicted emergent properties; sensitivity to initial conditions; path-dependent dynamics; networked hierarchical connectivity; interaction of autonomous agents; self-organization, collective shifts; non-equilibrium dynamics; adaptivity to changing environments; co-evolving subsystems; ill-defined boundaries and multilevel dynamics (San Miguel et al., 2012). Though this list is certainly still incomplete, and not all of its parts are characterizing the complex systems of drug discovery, the list shows the tremendous difficulties we face when trying to understand complex structures and their behavior. The same report (San Miguel et al., 2012) listed the following major challenges of complex system studies:

- data gathering by large-scale experiments, data sharing and data assembly using mutually agreed curation rules, management of huge, distributed, dynamic and heterogeneous databases;
- moving from data to dynamical models going beyond correlations to cause–effect relationships, understanding the relationship between simple and comprehensive models with appropriate choices of variables, ensemble modeling and data assimilation, modeling the 'systems of systems of systems' with many levels between micro and macro; and
- formulating new approaches to prediction, forecasting, and risk, especially in systems that can reflect on and change their behavior in response to predictions and in systems, whose apparently predictable behavior is disrupted by apparently unpredictable rare or extreme events.

Due to the complexity of the cells, organisms and diseases, extreme reductionism often fails in drug design. However, the other extreme, taking into account all possible variables of all possible components, is neither feasible, nor doable. Fortunately we do not have to challenge the impossible when thinking on complexity in drug design for two major reasons. On the one hand, the structure of complex systems is not only complicated, but also modular, and has a number of degenerate segments. This enables us to identify the most important system segments as we will show in Section 2. On the other hand, complex systems often determine a state space, which is also modular, and has a surprisingly low number of major attractors. In fact, this is what makes the discrimination of phenotypes possible at all. In other words: complexity has a side of simplicity. As fortunate 'side-effects' of the attractor-segmented, modular state space, many of the emergent properties of complex systems tolerate a number of errors in the individual data determining them. The above features of drug design-related complex systems make those descriptions successful, which are 'complex' themselves, meaning that they are neither too simplistic, nor go too much into details (Bar-Yam et al., 2009; Csermely, 2009; Huang et al., 2009; Mar & Quackenbush, 2009; Kolodkin et al., 2012). In agreement with these considerations, mathematical systems theory states that "the scale and complexity of the solution should match the scale and complexity of the problem" (Bar-Yam, 2004).

Network-approach is a description, which provides a good compromise between extreme reductionism and the 'knowledge of everything'. We are by far not alone sharing this view. Diseases have been perceived as network perturbations (Huang et al., 2009; Del Sol et al., 2010). In recent years network analysis became an increasingly acclaimed method in drug design (Hopkins, 2008; Ma'ayan, 2008; Pawson & Linding, 2008; Berger & Iyengar, 2009; Schadt et al., 2009; Baggs et al., 2010; Fliri et al., 2010; Lowe et al., 2010; Pujol et al., 2010). In agreement with the expert-opinions, network-applications show a steady increase of drug design-related publications (Fig. 3). We summarize the major network types (detailed in Section 3), network analysis types (detailed in Section 2), drug design areas helped by network studies (detailed in

Section 4) and the four key areas of drug design described in detail as the examples in Section 5 in Fig. 4.

We will detail the definition and types of networks in Section 2.1. The applicability of network analysis in drug design is determined by the following major factors: 1.) proper definition of network nodes, edges and edge weights; 2.) data quality and carefully defined, uniformly applied data inclusion criteria; 3.) data refinement by genetic variability, aging, environmental effects and compounding pathologies such as bacterial or viral infections (Arrell & Terzic, 2010; Kolodkin et al., 2012). However, we will not cover details of data acquisition, since this topic fits better into the broader area of systems biology, which is not the subject of the current review.

Networks are often viewed via their mathematical representations, i.e. graphs. However, this often proves to be an over-simplification in drug design for two major reasons. 1.) Network nodes of cellular systems are not exact 'points', as in graph theory, but macromolecules, having a network structure themselves, as we will show in Section 3.2.2) Network nodes have a lot of attributes in the rich biological context of the cell. 3.) Network dynamics is crucial in order to understand the complexity of diseases and the action of drugs (Pujol et al., 2010). Therefore, it is often useful to include edge directions, signs (activation or inhibition), conditionality (an edge is active only, if one of its nodes has another edge) and a number of dynamically changing quantitative measures in network descriptions. However, it is important to warn here that we should not include too many details in network descriptions, since we may shift our description from optimal towards the 'knowledge of everything'. Including more and more details in network science may lead to the trap of 'over-complication', where the beauty and elegance of the approach are lost. This may lead to the decline of the use of network description and analysis (similarly to the over-use of the explanatory power and decline of chaos theory, fractals, and many other approaches before).

The optimal simplicity of networks is also important, since networks give us a visual image. We summarize a rather long list of network visualization techniques in Table 1 showing the rich variety of approaches to solve this important task. A detailed comparison of some methods was described in several reviews (Suderman & Hallett, 2007; Pavlopoulos et al., 2008; Gehlenborg et al., 2010; Fung et al., 2012). A good visualization method provides a pragmatic trade-off between highlighting the biological concept and comprehensibility. Trying several methods is often advisable, since sampling scale and/or bias may lead to subjective interpretations of the network images obtained.

Correct visualization of networks is not only important for making a pleasing image. The right hemisphere of our brain works with images, and has the unique strength of pattern recognition. This complements the logical thinking of the left hemisphere. Regretfully, our logical thinking can deal with 5 to 6 independent pieces of information at the same time as an average. However, the complexity of human disease requires an information-handling capacity, which is by magnitudes higher than that of logical thinking. Pattern recognition by the right hemisphere copes with this complexity. This is why we also need to see networks, and may not only measure them. Besides the 'optimal simplicity', visualization is another advantage of networks over data-mining and other very useful, but highly detailed approaches (Csermely, 2009). To illustrate the network description and analysis in drug design, we compare the classic view and the network view of drug action in Fig. 5.

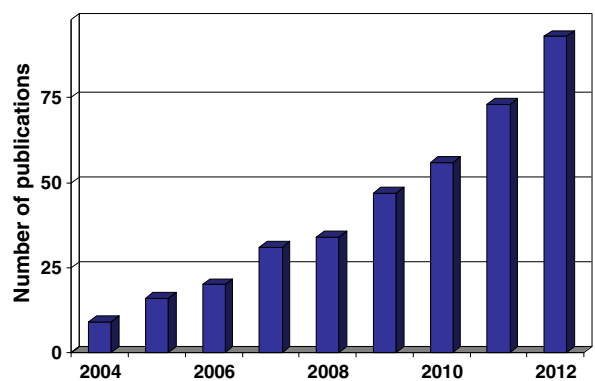

**Fig. 3.** Network-application in drug-design related publications. Data are from PubMed using the query of "network AND drug" for title and abstract words. The number of publications in 2012 is an extrapolation.

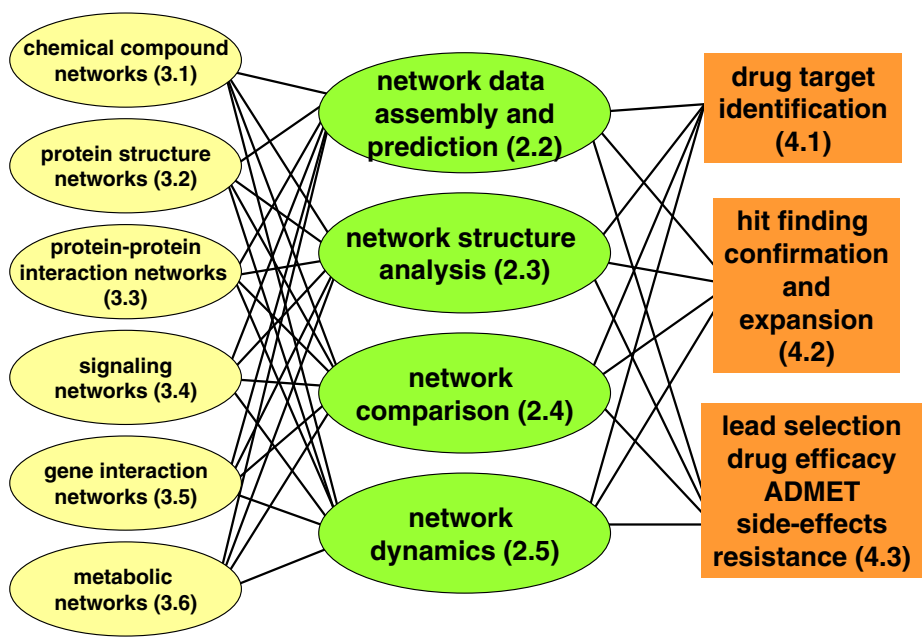

**Fig. 4.** Uses of network description and analysis in drug design. Numbers in parentheses refer to section numbers of this review.

As we have described in the previous paragraphs, network description and analysis offer a wide range of possibilities to understand the complexity of human disease and to develop novel drugs. As an example of the richness of networks, the 'semantic web' covers practically every conceptual entity appearing in the world-wide-web (Chen et al., 2009a). In the current review we cannot cover all. Therefore, with the exception of the network of human diseases described in Section 1.3, we will restrict ourselves to molecular networks ranging from the networks of chemical compounds and of protein structures to the various networks of the macromolecules constituting the cells. We will not cover the following areas, where we list a few reviews and papers of special interest:

- networked particles in drug delivery (Rosen & Elman, 2009; Luppi et al., 2010; Bysell et al., 2011);
- network of plants as resources of herbal remedies and traditional medicines (Saslis-Lagoudakis et al., 2012);
- cytoskeletal networks or membrane organelle networks (Michaelis et al., 2005; Escribá et al., 2008; Gombos et al., 2011);
- inter-neuronal, inter-lymphocyte and other intercellular networks including extracellular matrix, cytokine, endocrine or paracrine networks (Jerne, 1974, 1984; Cohen, 1992; Werner, 2003, 2005; Small, 2007; Jung et al., 2011; Acharyya et al., 2012; Margineanu, 2012);
- the ecological networks of the microorganisms living in human gut, oral cavity, skin, etc. and their interconnected networks with human cells (Ben-Jacob et al., 2012; Clemente et al., 2012; Mueller et al., 2012);

- social networks and their potential effects on spreading of epidemics, as well as disease-related habits such as drug abuse, smoking, over-eating, etc. (Christakis & Fowler, 2011);
- network-related modeling methods, such as: neural network models, differential equation networks, network-related Markov chain methods, Boolean networks, fuzzy logic-based network models, Bayesian networks and network-based data mining models (Huang, 2001; Ideker & Lauffenburger, 2003; Winkler, 2004; Fayos & Fayos, 2007; Fernandez et al., 2011).

At the end of the Introduction we will illustrate network thinking by showing the richness and usefulness of network representations of human diseases.

*1.3. The networks of human diseases*

Several diseases, such as cancer, or complex physiological processes, such as aging, were described as a network phenomenon quite a while ago (Kirkwood & Kowald, 1997; Hornberg et al., 2006; Sőti & Csermely, 2007). In this section we will not detail disease-related molecular networks (such as interactomes, or signaling networks changing in disease), since this will be the subject of Section 3. We will describe the large variety of options to build up the networks of human diseases, where diseases are nodes of the network, and will show how network-assembled bio-data can be used to predict novel disease biomarkers including novel disease-related genes.

*1.3.1. Network representations of diseases and their therapies*

In the network description, sets of interlined data need first to be structured by defining 'nodes'. This might already be rather difficult, as we will show in detail in Section 2.1. However, the definition of edges, i.e. connections between the nodes, may be an especially demanding task. Networks of human diseases provide a very good example, since a large number of data categories are related to the concept of disease enabling the construction of a large variety of networks (Goh et al., 2007; Rzhetsky et al., 2007; Feldman et al., 2008; Spiró et al., 2008; Hidalgo et al., 2009; Barabási et al., 2011; Zhang et al., 2011a; Janjic & Przulj, 2012).

Some of the major disease-related categories are shown in Fig. 6. Human disease can be conceptualized as a phenotype, i.e. an emergent property of the human body as a complex system (Kolodkin et al.,

**Table 1**
Network visualization resources.

| Name | Website | References |
|------|---------|-----------|
| Arena3D | http://arena3d.org | Secrier et al., 2012 |
| ArrayXPath | http://www.snubi.org/software/ArrayXPath | Chung et al., 2005 |
| AVIS | http://actin.pharm.mssm.edu/AVIS2 | Berger et al., 2007 |
| BioLayout Express 3D | http://www.biolayout.org | Freeman et al., 2007; Theocharidis et al., 2009 |
| BiologicalNetworks | http://biologicalnetworks.net | Kozhenkov & Baitaluk, 2012 |
| BioTapestry | http://www.biotapestry.org | Longabaugh, 2012 |
| BisoGenet | http://bio.cigb.edu.cu/bisogenet-cytoscape | Martin et al., 2010 |
| CellDesigner | http://www.celldesigner.org | Kitano et al., 2005 |
| Cell Illustrator | http://www.cellillustrator.com | Nagasaki et al., 2011 |
| CFinder | http://www.cfinder.org | Adamcsek et al., 2006 |
| Cytoscape | http://www.cytoscape.org | Smoot et al., 2011 |
| GenePro | http://wodaklab.org/genepro | Vlasblom et al., 2006 |
| GeneWays | http://anya.igsb.anl.gov/Geneways/GeneWays.html | Rzhetsky et al., 2004 |
| GEOMIi | http://sydney.edu.au/engineering/it/~visual/valacon/geomi/ | Ahmed et al., 2006 |
| Gephi | http://gephi.org | Bastian et al., 2009 |
| Graphviz | http://www.graphviz.org | Gansner & North, 2000 |
| Gridlayout | http://kurata21.bio.kyutech.ac.jp/grid/grid_layout.htm | Li & Kurata, 2005 |
| Guess | http://graphexploration.cond.org/index.html | Adar, 2006 |
| Hive Plots | http://www.hiveplot.com | Krzywinski et al., 2012 |
| Hybridlayout | http://www.cadlive.jp/hybridlayout/hybridlayout.html | Inoue et al., 2012 |
| Hyperdraw | http://www.bioconductor.org/packages/release/bioc/html/hyperdraw.html | Murrell, 2012 |
| IM Browser | http://proteome.wayne.edu/PIMdb.html | Pacifico et al., 2006 |
| IPath | http://pathways.embl.de | Yamada et al., 2011 |
| JNets | http://www.manchester.ac.uk/bioinformatics/jnets | Macpherson et al., 2009 |
| KGML-ED | http://kgml-ed.ipk-gatersleben.de | Klukas & Schreiber, 2007 |
| LEDA | http://www.algorithmic-solutions.com/leda/about/index.htm | Mehlhorn & Näher, 1999 |
| MAVisto | http://mavisto.ipk-gatersleben.de | Schwobbermeyer & Wunschiers, 2012 |
| Medusa | http://coot.embl.de/medusa | Hooper & Bork, 2005 |
| ModuLand | www.linkgroup.hu/modules.php | Szalay-Bekő et al., 2012 |
| Multilevel Layout | http://code.google.com/p/multilevellayout | Tuikkala et al., 2012 |
| NAVIGaTOR | http://ophid.utoronto.ca/navigator | Brown et al., 2009 |
| NetMiner | http://www.netminer.com/index.php | |
| Network Workbench | http://nwb.cns.iu.edu | NWB Team, 2006 |
| Ondex | http://www.ondex.org | Kohler et al., 2006 |
| Osprey | http://biodata.mshri.on.ca/osprey/servlet/Index | Breitkreutz et al., 2003 |
| Pajek | http://pajek.imfm.si/doku.php | Batagelj & Mrvar, 1998 |
| PathDraw | http://rospath.ewha.ac.kr/toolbox/PathwayViewerFrm.jsp | Paek et al., 2004 |
| Pathway Tools | http://bioinformatics.ai.sri.com/ptools | Karp et al., 2010 |
| PATIKA | http://www.patika.org | Dogrusoz et al., 2006 |
| PaVESy | http://pavesy.mpimp-golm.mpg.de/PaVESy.htm | Ludemann et al., 2004 |
| PhyloGrapher | http://www.atgc.org/PhyloGrapher | |
| PIMWalker | http://pimr.hybrigenics.com | Meil et al., 2005 |
| PIVOT | http://acgt.cs.tau.ac.il/pivot | Orlev et al., 2004 |
| PolarMapper | http://kdbio.inesc-id.pt/software/polarmapper | Goncalves et al., 2009 |
| ProteinNetVis | http://graphics.cs.brown.edu/research/sciviz/proteins/home.htm | Jianu et al., 2010 |
| ProteoLens | http://bio.informatics.iupui.edu/proteolens | Huan et al., 2008 |
| RedeR | http://bioconductor.org/packages/release/bioc/html/RedeR.html | Castro et al., 2012 |
| RING | http://protein.bio.unipd.it/ring | Martin et al., 2011 |
| SoNIA | http://www.stanford.edu/group/sonia | Bender-deMoll & McFarland, 2006 |
| Transcriptome-Browser | http://tagc.univ-mrs.fr/tbrowser | Lepoivre et al., 2012 |
| UCSF structureViz | http://www.cgl.ucsf.edu/cytoscape/structureViz | Morris et al., 2007 |
| VANTED | http://vanted.ipk-gatersleben.de | Rohn et al., 2012 |
| VisANT | http://visant.bu.edu | Hu et al., 2009 |
| VitaPad | http://sourceforge.net/projects/vitapad | Holford et al., 2005 |
| WebInterViewer | http://interviewer.inha.ac.kr | Han et al., 2004b |
| yFiles | http://www.yworks.com/en/index.html | Becker & Rojas, 2001 |
| yWays | http://www.yworks.com/en/products_yfiles_extensionpackages_ep2.htm | |

Summaries of Suderman and Hallett (2007), Pavlopoulos et al. (2008), Gehlenborg et al. (2010) and Fung et al. (2012) compare some of the options above.

2012). Some of the categories, such as symptoms, are related to this phenotype. Many other categories, such as

- disease-related genes (abbreviated as 'disease genes'),
- functions of disease genes (marked as gene ontology);
- the transcriptome (i.e. expression levels of all mRNAs + the cistrome, i.e. DNA-binding transcription factors + the epigenome, i.e. the actual chromatin status of the cell including DNA and histone modifications, as well as their 3D structure)
- the interactome, the signaling network and the metabolome,

are all related to the underlying genotype, i.e. the constituents of the human body related to the etiology of the disease. A third group of categories, such as therapies, drugs and other factors marked as

"environment", represents the effects of the environment (Fig. 6). Connections (uniformly defined, data-encoded relationships) between any two of these categories define a so-called bipartite network, where two different types of nodes are related to each other. Moreover, more than two categories may also form a network, which is called as a multi-partite network (Goh et al., 2007; Yildirim et al., 2007; Nacher & Schwartz, 2008; Spiró et al., 2008; Li et al., 2009a; Bell et al., 2011; Wang et al., 2011a).

We have three options for the visualization of bipartite networks. We will illustrate this in the example of the network of human diseases and human genes shown to be associated with a particular disease in Fig. 7 (Goh et al., 2007). We may include both types of nodes and all their connections to the visual image as shown in the center of

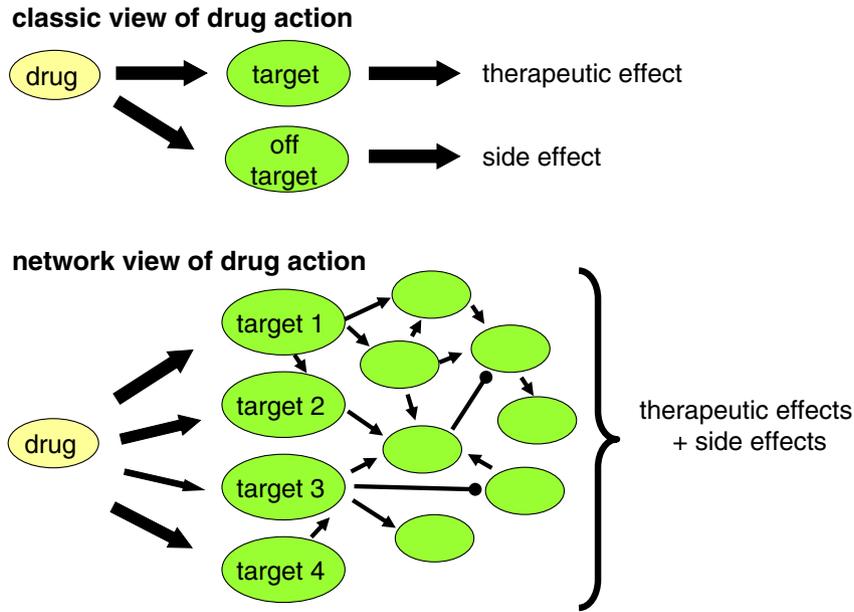

**Fig. 5.** Classic and network views of drug action. Made after the basic idea of Berger and Iyengar (2009).

Fig. 7. However, the selection of only a single node type results in a simpler network representation, which is easier to understand. We have two projections of the full, bipartite network as shown on the two sides of Fig. 7. In the first type of projection we connect two human diseases, if there is a human gene, which is participating in the etiology of both diseases (left side of Fig. 7). Edge weight may be derived here from the number of genes connecting the two diseases. Alternatively, we may construct a network of human genes, which are connected, if there is at least one human disease, where they both belong (right side of Fig. 7; Goh et al., 2007). Similar projections can be made with any category-pairs, or multiple category-sets of Fig. 6.

### 1.3.2. The human disease network

The landmark study of Goh et al. (2007) provided the first network map of the genetic relationship of 516 human diseases. This approach used the "shared gene formalism" recognizing that diseases sharing a gene or genes likely have a common genetic basis. Later, this concept was extended with the "shared metabolic pathway formalism" recognizing that enzymatic defects affecting the flux of "reaction A" in a metabolic pathway will lead to disease-conditions that are known to be associated with the metabolites situated downstream of "reaction A" in the same metabolic pathway. The shared metabolic pathway formalism proved to be better predictor of metabolic diseases than the shared gene formalism. Another approach is based on the "disease comorbidity formalism" connecting diseases, which have a co-occurrence in patients exceeding a predefined threshold. Subsequently, many other studies incorporated a number of other data including gene-expression levels, protein–protein interactions, signaling components, such as microRNAs, tissue-specificity, and a number of environmental effects including drug treatment and other therapies to construct disease similarity networks (Barabási et al., 2011; Goh & Choi, 2012; Janjic & Przulj, 2012). We summarize the disease-network types using two, three or more different datasets in Table 2. We will summarize drug target networks in Section 4.1.3.

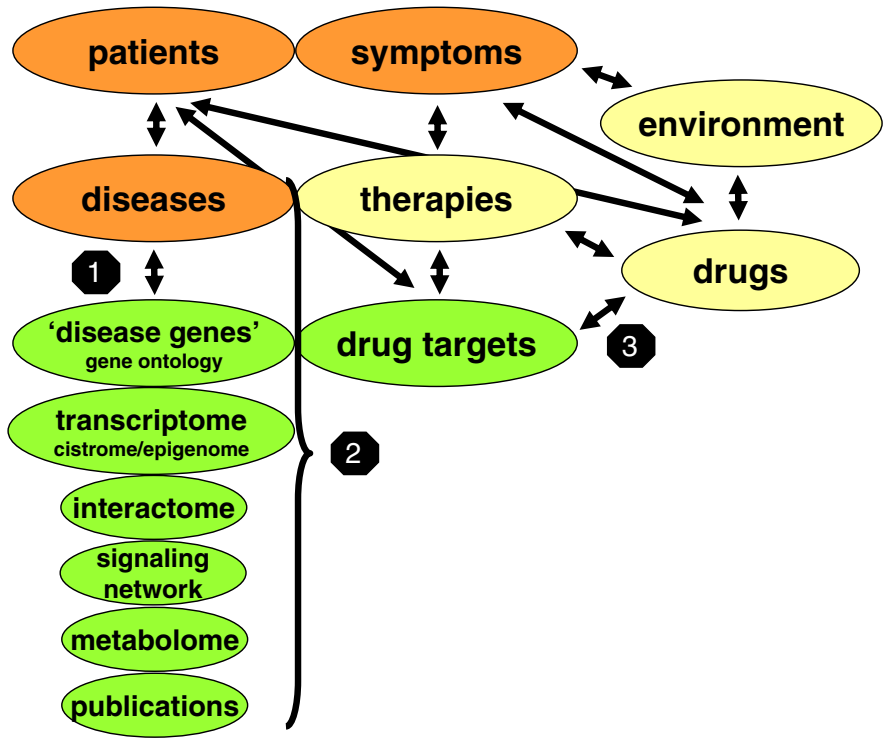

**Fig. 6.** Options for network representations of disease-related data. The figure summarizes some of the options to assess disease-related data using network description and analysis. Each ellipse represents a type of data. Arrows stand for possible network representations. 1: Human disease networks discussed in this section and in Table 2. 2: Additional network-related data helping the identification of disease-related human genes (acting like possible drug targets) detailed in Table 3. 3: Drug target networks discussed in Section 4.1.3.

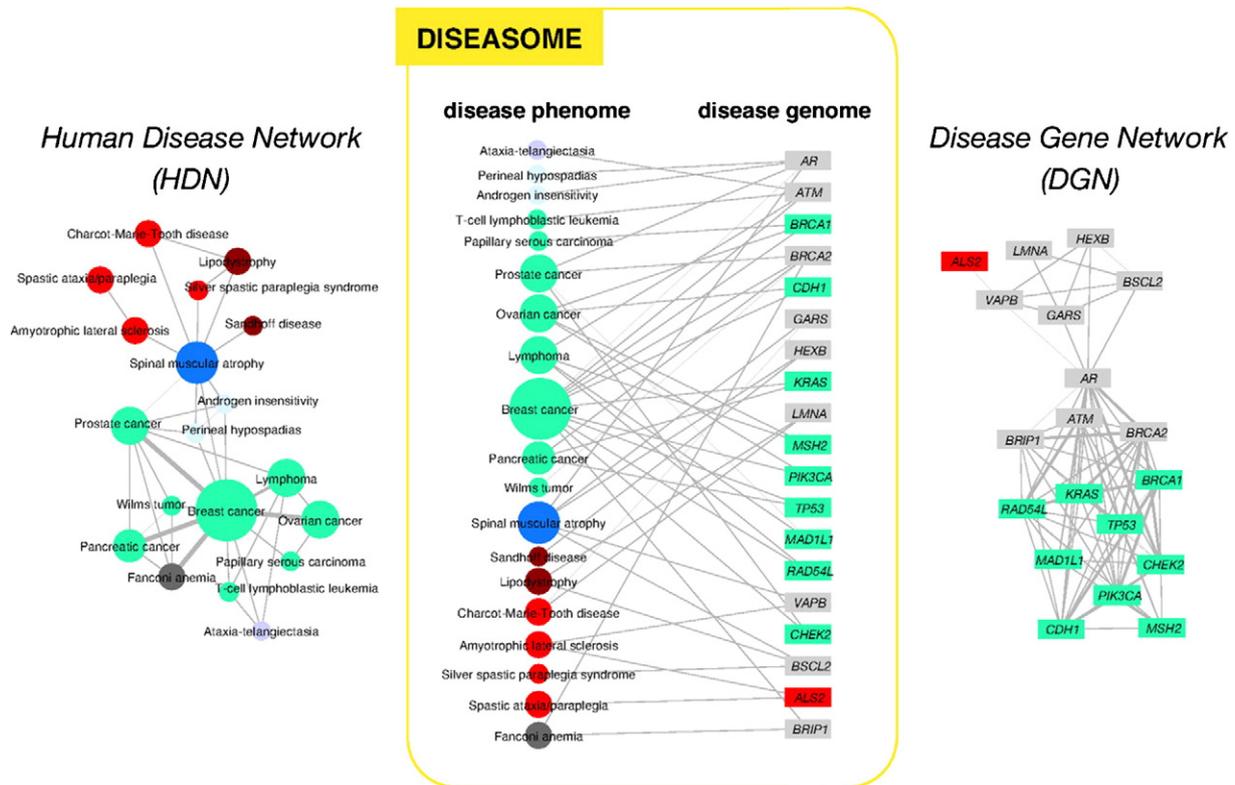

**Fig. 7.** Two projections of the human disease network. On the middle of the figure a segment of the bipartite network of human diseases and related human genes is shown. On the projection on the left side two diseases are connected, if they have at least one common gene. On the projection on the right side two genes are connected, if they have at least one common disease.
Reproduced with permission from Goh et al. (2007); Copyright, 2007, National Academy of Sciences, U.S.A.

Various data-associations listed in Table 2 enrich each other, as it has been shown in the example of the orphan diseases, Tay–Sachs disease and Sandhoff syndrome, which did not share any known disease genes in 2011, but were connected in a literature co-occurrence based network. The connection of the two diseases was in agreement with the shared metabolic pathway of their mutated genes. Zhang et al. (2011a) listed several other examples for such mutual enrichment of various datasets. Comparing Table 2 with Fig. 6 reveals several combinations of data, which have not been used to form human disease networks yet. We expect further advance in this rapidly growing field.

As take home messages from the studies listed in Table 2, we summarize the following observations.

- The intuitive assumption that "hubs (defined here as nodes with many more neighbors than average in the human interactome) play a major role in adult diseases" often fails due to the embryonic lethality of these key genes. In agreement with this, orphan diseases (which are often life-threatening or chronically debilitating, and affect less than 6.5 patients per 10,000 inhabitants) tend to be hubs, and are often associated with essential genes. Similarly, diseases having somatic mutations, such as cancer, have a central position in the human interactome. Germ-line mutations leading to more common diseases tend to be located in the functional periphery (but not in the utmost periphery) of the human interactome (Goh et al., 2007; Feldman et al., 2008; Barabási et al., 2011; Zhang et al., 2011a).

- Disease-related genes tend to be tissue specific, with the notable exception of most cancer-related genes, which are not overexpressed in the tissues from which the tumors emanate (Goh et al., 2007; Jiang et al., 2008; Lage et al., 2008; Barabási et al., 2011).

- Disease-related genes have a smaller than average clustering coefficient avoiding densely connected local structures (Feldman et al., 2008). Low clustering coefficient was successfully applied as a discriminatory feature in the prediction of disease-related genes (Sharma et al., 2010a).

- Disease-related genes tend to form overlapping disease modules in protein–protein interaction networks showing even a 10-fold increase of physical interactions relative to random expectation (Gandhi et al., 2006; Goh et al., 2007; Oti & Brunner, 2007; Feldman et al., 2008; Jiang et al., 2008; Stegmaier et al., 2010; Bauer-Mehren et al., 2011; Loscalzo & Barabasi, 2011; Xia et al., 2011). Overlaps of disease modules are also characteristic to comorbidity networks (Rzhetsky et al., 2007; Hidalgo et al., 2009).

- Genes bridging disease modules in the human interactome may provide important points of interventions (Nguyen & Jordan, 2010; Nguyen et al., 2011). Genes involved in the aging process often occupy such bridging positions (Wang et al., 2009).

- Diseases that share disease-associated cellular components (genes, proteins, metabolites, microRNAs, etc.) show phenotypic similarity and comorbidity (Lee et al., 2008a; Barabási et al., 2011).

- The above findings are recovered, if we go one level deeper in the network hierarchy than the human interactome, to the level of protein domains and their interactions (Sharma et al., 2010a; Song &

**Table 2**
Human disease-related networks and network datasets.

| Type of related data (types of network nodes)[a] | Name and additional description, website | References[b] |
|---|---|---|
| • Disease<br>• Disease related genes<br>• Disease<br>• Disease-related genes<br>• Interactome<br>• Publication | Human disease network (Cytoscape plug-in DisGeNET: http://ibi.imim.es/DisGeNET/DisGeNETweb.html)<br>Gene-based, interactome-enriched and scientific publication based human disease networks | Goh et al., 2007; Feldman et al., 2008; Bauer-Mehren et al., 2010; Stegmaier et al., 2010 Zhang et al., 2011a |
| • Disease<br>• Interactome module<br>• mRNA changes | Disease-responsive interactome module-based human disease network (disease correlations based on disease-induced changes in mRNA expression of interactome modules) | Suthram et al., 2010 |
| • Disease<br>• mRNA changes at the transcriptome level<br>• Drugs | A Bayesian network-based disease-responsive transcriptome analysis to construct a human disease network | Huang et al., 2010a |
| • Disease<br>• Disease-related genes<br>• Interactome<br>• Protein/DNA interaction<br>• Tissue<br>• Drug | iCTNet: a Cytoscape plug-in to construct an integrative network of diseases, associated genes, drugs and tissues (http://www.cs.queensu.ca/ictnet) | Wang et al., 2011b |
| • Disease<br>• Disease-related genes<br>• Interactome<br>• Gene Ontology terms | Biomine: an integrated bio-entity network with more than a million entities and 8 million edges (http://biomine.cs.helsinki.fi) | Eronen & Toivonen, 2012 |
| • Disease<br>• Expression patterns<br>• MicroRNA targets<br>• Network modules of interactome, transcriptome | PAGED: an integrated bio-entity network with more than a million entities from 20 organisms (http://bio.informatics.iupui.edu/PAGED) | Huang et al., 2012b |
| • Disease<br>• Disease-related genes<br>• Interactome<br>• Protein gene regulation pathways<br>• Gene Ontology terms<br>• Small molecule (drug)<br>• Species | An integrated bio-entity network | Bell et al., 2011 |
| • Disease<br>• Adjacent members of metabolic pathways | Metabolic pathway-corrected human disease network | Lee et al., 2008a |
| • Disease<br>• MicroRNA | MicroRNA/disease association-based disease network obtained from publication data | Lu et al., 2008 |
| • Patient<br>• Disease | Disease comorbidity network | Rzhetsky et al., 2007; Hidalgo et al., 2009 |
| • Disease<br>• Environmental factor<br>• Disease-related genes | Etiome: a database + clustering analysis of environmental + genetic (= etiological) factors of human diseases | Liu et al., 2009 |

[a] Here we included only those networks and datasets, which contained human diseases. Drug target networks and network datasets will be summarized in Section 4.1.3.
[b] References containing direct network analysis are marked with italic. All other references are referring to datasets, which are potential sources of future network representations.

Lee, 2012). Diseases occurring more frequently are associated with longer proteins (Lopez-Bigas & Ouzounis, 2004; Lopez-Bigas et al., 2005). Disease-associated proteins tend to have 'younger' folds, developed later in evolution, and have a smaller 'family' of similar folds. These protein folds are less designable (i.e. a smaller number of possible representations by different amino acid sequences) weakening the robustness against mutations, and the fitness of the hosting organism during evolution (Wong & Frishman, 2006).

• Going one level higher in the network hierarchy than the human interactome, to the level of comorbidity networks, patients tend to develop diseases in the vicinity of diseases they already had (Rzhetsky et al., 2007; Hidalgo et al., 2009; Barabási et al., 2011).

• Disease-hubs of comorbidity networks show a higher mortality than less well connected diseases, and are often successors of more peripheral diseases. The progression of diseases is different for patients of different genders and ethnicities (Lee et al., 2008a; Hidalgo et al., 2009; Barabási et al., 2011).

Human disease networks are expected to reveal more on the interrelationships of diseases using both additional data-associations and novel network analysis tools, listed in Section 2. These advances will not only enrich our integrated view on human diseases, but will also lead to the following potential uses of human disease networks:

• better classification of diseases (e.g. for putatively useful drugs and therapies) and predictions for understudied or unknown diseases;
• disease diagnosis and identification of disease biomarkers as described in detail in Section 1.3.3;
• identification of drug target candidates (including multi-target drugs, drug repositioning, etc.) as described in detail in Section 4.1;
• help in hit finding and expansion as described in detail in Section 4.2;
• enrich background data for lead optimization (including ADME, side-effects and toxicity, etc.) as described in detail in Section 4.3.

An increasing number of publications describe various molecular networks characterizing the cellular state in a certain type of disease. We have not included their direct description in this section, since here we only review the networks of the diseases as network nodes. In Section 5 we will summarize the drug-design related applications of these molecular networks in case of four disease families: infections, cancer, diabetes and neurodenegerative diseases. In the next section

**Table 3**

Network-based predictions of disease-related genes as biomarkers.

| • Type of prediction methods[a]<br>• Type of data used | Name and additional description, website | References |
|---|---|---|
| • Similarity-based<br>• Protein structure descriptor-related QSAR | New disease-related proteins are predicted by their structural similarity to known disease-related proteins | Vilar et al., 2009 |
| • Interaction-based<br>• (Predicted) interactome | New disease-related genes are predicted by their interactome neighborhood | Krauthammer et al., 2004; Chen et al., 2006a; Oti et al., 2006; Xu & Li, 2006 |
| • Iterative summary of interactome and disease neighborhood<br>• Disease similarity network, interactome | Measures the neighborhood association in both the interactome and disease similarity networks and iteratively calculates the similarity of the node to diseases | Guo et al., 2011 |
| • Semantic similarity score<br>• Semantic similarity networks of diseases and related genes | Calculates a semantic similarity score between gene ontology terms as well as human genes associated with them | Jiang et al., 2011 |
| • Summarized network neighborhood of several candidate genes<br>• Disease, gene-descriptions, disease related genes, interactome, mRNA co-expressions, pathways | Constructs an integrative network and predicts candidate genes by their network closeness to known disease-related genes; Prioritizer: http://129.125.135.180/prioritizer | Franke et al., 2006 |
| • Shortest path length<br>• Disease, gene-descriptions, disease related genes, interactome, mRNA co-expressions | Uses a maximum expectation gene cover algorithm finding small gene sets to predict associated new disease-related genes | Karni et al., 2009 |
| • User-defined path distance from known disease-related genes<br>• Up to 10 integrated interactomes | New disease-related genes are predicted by their interactome closeness to known disease-related proteins; Genes2Networks: http://actin.pharm.mssm.edu/genes2networks | Berger et al., 2007 |
| • Interaction-based<br>• Disease-related mutations, domain–domain resolved interactome | New disease-related genes are predicted by their association to previously known disease-related genes at protein–protein domains affected by the disease-associated mutations of the known disease related gene | Sharma et al., 2010a; Song & Lee, 2012 |
| • Interaction-based<br>• Disease-related mutations, 3D structurally resolved interactome | New disease-related genes are predicted by their association to previously known disease-related genes at 3D modeled protein–protein interfaces affected by the disease-associated mutations of the known disease related gene | Wang et al., 2012b |
| • Clustering<br>• Disease-related genes, interactome | New disease-related genes are predicted by their common protein–protein interaction network module with previous disease-related genes | Navlakha & Kingsford, 2010; |
| • Closeness<br>• Disease-related genes, disease network, interactome | Closeness of unrelated proteins is calculated in the interactome from protein products of disease-related genes, and compared with phenotype similarity profile: large closeness marks a potential new disease-related gene; CIPHER: http://rulai.cshl.edu/tools/cipher | Wu et al., 2008 |
| • Random walk<br>• Disease-related genes, disease network, interactome | Random walks in the interactome are started from protein products of disease-related genes: frequent visits of a previously unrelated protein mark a potential new disease-related gene; Cytoscape plug-in GPEC: http://sourceforge.net/p/gpec | Kohler et al., 2008; Chen et al., 2009b; Le & Kwon, 2012 |
| • Iterative network propagation<br>• Disease-related genes, disease network, interactome | Iterative steps of information flow from disease-related and between interacting proteins: after convergence a large flow of a previously unrelated protein marks potential new disease-related gene; Cytoscape plug-in PRINCIPLE/PRINCE: http://www.cs.tau.ac.il/~bnet/software/PrincePlugin | Vanunu et al., 2010; Gottlieb et al., 2011b |
| • Random walk with re-starts in both networks<br>• Disease-related genes, disease network, interactome | Random walk in both the interactome and the disease networks: number of frequent visits marks candidate genes | Li & Patra, 2010 |
| • NetworkBlast algorithm to align interactome and disease networks<br>• Disease-related genes, disease network, interactome | After alignment of the interactome and disease networks finds high scoring subnetworks (bi-modules); candidate genes have the highest scoring bi-modules | Wu et al., 2009a |
| • Information flow with statistical correction<br>• Disease-related genes, interactome | Statistically corrects random walk-based prediction with the degree distribution of the network; DADA: http://compbio.case.edu/dada | Erten et al., 2011a |
| • Topological network similarity<br>• Disease-related genes | Calculates neighborhood similarity in the interactome and prioritizes candidate genes; VAVIEN: http://diseasegenes.org | Erten et al., 2011b |
| • Neighborhood similarity<br>• Disease-related genes, interactome, expression patterns | Calculates expression weighted neighborhood similarity (using Katz centrality or other methods) in the interactome | Zhao et al., 2011b; Wu et al., 2012 |
| • Semantic-based centrality<br>• Disease-related genes, interactome, pathways | Calculates data-type weighted centrality in the integrated network and uses it as a rank of candidate genes | Gudivada et al., 2008 |
| • Direct neighbor-based Bayesian predictor<br>• Disease-related genes, disease network, interactome, pathways | Constructs candidate protein complexes in a virtual pull-down experiment, and scores candidates by measuring the similarity between the phenotype in the complex and disease phenotype | Lage et al., 2007 |
| • Genetic linkage analysis of gene network clusters<br>• Disease-related genes, text mining-based associations (binding, phosphorylation, methylation, etc.) | Calculates genetic linkage analysis of connected clusters in a text mining-derived direct interaction network | Iossifov et al., 2008 |
| • Random forest learning<br>• Disease, disease related genes, disease networks, single-nucleotide polymorphisms (SNPs) | Predict deleterious SNPs and disease genes using the random forest learning method, uses interactomes and deleterious SNPs to predict disease-related genes by random forest learning | Care et al., 2009 |
| • Random walk, iterative network propagation (PRINCE/PRINCIPLE)<br>• Disease, disease related genes, interactome, protein/DNA interaction, tissue, drug | A Cytoscape plug-in to construct an integrative network of diseases, associated genes, drugs and tissues; iCTNet: http://www.cs.queensu.ca/ictnet | Wang et al., 2011b |
| • Machine learning<br>• Disease, disease related genes, gene annotations, interactome, expression levels, sequences | Integrative methods using similarities of neighbors or shortest paths in multiple data sources including interactomes; Endeavour: http://esat.kuleuven.be/endeavour; Phenopred: http://www.phenopred.org | Radivojac et al., 2008; Tranchevent et al., 2008; Linghu et al., 2009; Costa et al., 2010 |

**Table 3** (continued)

| • Type of prediction methods[a]<br>• Type of data used | Name and additional description, website | References |
|---|---|---|
| • Rank coherence with target disease and unrelated disease networks<br>• Disease, disease related genes, gene annotations, interactome, expression levels, genome-wide association studies | Calculates rank coherences between the integrated network characteristic in the target disease and unrelated diseases; rcNet: http://phegenex.cs.umn.edu/Nano | Hwang et al., 2011 |

[a] The table summarizes methods using networks as data representations. We excluded those methods, like neural network or Bayesian network-based methods, which decipher associations between various, not network-assembled data. Several methods are included in the excellent reviews of Wang et al. (2011a) and Doncheva et al. (2012a).

we will illustrate the help of network analysis in the diagnosis and therapy of human diseases by the network-based identification of disease biomarkers.

### 1.3.3. Network-based identification of disease biomarkers

Network-based identification of disease related genes was suggested by relatively early studies (Krauthammer et al., 2004; Chen et al., 2006a; Franke et al., 2006; Gandhi et al., 2006; Oti et al., 2006; Xu & Li, 2006). In the last few years several network-based methods have been developed helping the identification of genes related to a particular disease as reviewed by the excellent summaries of Wang et al. (2011a) and Doncheva et al. (2012a). Table 3 summarizes methods for prediction of disease-related genes using networks as data representations. We excluded those network-related methods, like those neural network-based or Bayesian network-based methods, which decipher associations between various, not network-assembled data. Network prediction methods, which can also be used for prediction of disease-associated genes will be discussed in Section 2.2.2.

Most of the methods listed in Table 3 identify novel disease-related genes as disease biomarkers. Several network-based methods outperform former, sequence-based methods in the identification of novel, disease-related genes. Methods including non-local information of network topology usually perform better than methods based on local network properties. As a general trend the more information the method includes, the better prediction it may achieve. However, with the multiplication of datasets, biases and circularity may also be introduced, which will lead to an overestimation of the performance. Moreover, it is difficult to dissect the performance-contribution of the datasets and the prediction method itself. Additionally, each type of dataset may require a different method for optimal analysis. Therefore, the separate analysis of each data source was suggested with a subsequent combination of the ranking lists using rank aggregation algorithms. This procedure also facilitates backtracking the origin of the most relevant information. Functional GO-term annotations usually bring crucially important information to the analysis. The inclusion of interactome edge-based disease perturbations may improve the performance of these methods even further in the future (Kohler et al., 2008; Navlakha & Kingsford, 2010; Sharma et al., 2010a; Vanunu et al., 2010; Jiang et al., 2011; Wang et al., 2011a; Cho et al., 2012; Doncheva et al., 2012a). Importantly, several of the methods in Table 3 are not only able to diagnose known diseases, but may also identify important features of understudied or unknown diseases (Huang et al., 2010a; Wang et al., 2011a).

'Disease-related gene-hunting' became a very powerful area of medical studies. However, Erler and Linding (2010) warned that network models, and not their individual nodes, should be used as biomarkers, since thresholds and changes of individual nodes (such as the protein phosphorylation at a certain site) may be related to entirely different outcomes in different network contexts of different patients. We will summarize the concepts treating networks (and their segments) as drug targets in Section 4.1.7.

Very similar methods to those listed in Table 3 may be applied to network-based identification of disease-related signaling network, such as phosphorylation or microRNA profiles, or metabolome profiles. As part of these approaches, metabolic network analysis was applied to identify metabolites, which may serve as biomarkers of a certain disease (Fan et al., 2012). Shlomi et al. (2009) identified 233 metabolites, whose concentration was elevated or reduced as a result of 176 human inborn dysfunctional enzymes affecting of metabolism. Their network-based method can provide a 10-fold increase in biomarker detection performance. Mass spectrometry phosphoproteome analysis combined with signaling networks and bioinformatics sources like NetworKIN and NetPhorest may provide biomarker profiles of several diseases such as cancer or cardiovascular disease (Linding et al., 2007; Yu et al., 2007a; Jin et al., 2008; Miller et al., 2008; Ummanni et al., 2011; Savino et al., 2012).

## 2. An inventory of network analysis tools helping drug design

Even the best network analytical methods will fail, if applied to a network constructed with a crude definition. Therefore, we start this section listing the major points of network definition including network-related questions of data collection, such as sampling, prediction and reverse engineering. The latter two methods are important network-related tools to find novel drug target candidates. We will continue and conclude this section by listing an inventory of the major concepts used in the analysis of network topology, comparison and dynamics evaluating their potential use in drug design. The section will give just the essence of the methods, and will provide the interested Reader a number of original references for further information.

### 2.1. Definition(s) and types of networks

To define a network we have to define its nodes and edges (Barabási & Oltvai, 2004; Boccaletti et al., 2006; Zhu et al., 2007; Csermely, 2009; Lovász, 2012). Network nodes are the entities building up the complex system represented by the network. Nodes are often called as vertices, or network elements. Classical, graph-type network descriptions do not consider the original character of nodes. (A node of such a graph will be "ID-234", which is characterized by its contact structure only.) Thus node definition requires a clear sense of those node properties, which discriminate network nodes from other entities, and make them 'equal'. Recently, node-weights were successfully applied to characterize the node structure of a network in a simple form (Wiedermann et al., 2013). In the case of molecular networks, where nodes are amino acids, proteins or other macromolecules such discrimination is rather easy. However, subtle problems may still remain. For example, should we include extracellular proteins as well? If not, what happens, if an extracellular protein is just about to be secreted? What if it is engulfed by the cell and internalized? Node definition may become especially difficult in the case of complex data structures, like those we mentioned in Section 1.3. Accurate node definitions are time consuming, but lead to benefits at subsequent stages.

Network edges are often called interactions, connections, or links. In the molecular networks discussed in this review edges represent physical or functional interactions of two network nodes (Zlatic et al., 2009). However, in hypergraph representations meta-edges often connect more than two nodes. Edge definition often inherently contains a threshold determined by the detection limit and by the time-window of detection. Two nodes may become connected, if the sensitivity and/or duration of detection are increased. A number of recent publications explored the effect of time-window changes on the structure of social networks (Krings et al., 2012; Perra et al., 2012). Several concepts of network dynamics detailed in Section 2.5 are inherently related to time-window of detection. As an example, the distinction of the popular date hubs (Han et al., 2004a), i.e. hubs changing their partners over time, clearly depends on the time-window of observation.

Weights of network edges may give an answer to the "where-to-set-the-detection-threshold" dilemma offering a continuous scale of interactions. Edge weights represent the intensity (strength, probability, affinity) of the interaction. Edges may also be directed, where a sequence of action and/or a difference in node influence are included in the edge definition. Lovász (2012) gives an excellent summary of the basic dilemmas of network definition problems.

However, we have many more options than defining network nodes, edges, weights and directions. Recent network descriptions started to explore the options to include edge reciprocity (Squartini et al., 2012), or to preserve multiple node attributes (Kim & Leskovec, 2011). Moreover, in reality networks are seldom directed in an unequivocal way. (When CEOs and VPs are talking to each other, it is not always the case that CEOs influence VPs, and VPs do not influence CEOs.) However, to date, a continuous scale of edge direction has not been introduced to molecular networks. Edges may also be colored, where different types of interactions are discriminated. A special subset of colored networks is signed networks, where edges are either positive (standing for activation) or negative (representing inhibition). Edges may also be conditional, i.e. being only active, if one of their nodes accommodated another edge previously. There are a number of potential uses of these network representations e.g. in signaling, or in genetic interaction networks.

As a closing remark, the definition of edges often hides one of two fundamentally different concepts. Network connections may either restrict the connected nodes (this is the case, where connections represent physical contacts), or may enrich connected nodes (this is the case, where connections represent channels of transport or information transmission). These constraint-type or transmission-type network properties may appear in the same network, where they may be simplified to activation or inhibition like those in signal transduction networks. Though there were initial explorations of the differences of constraint-type and transmission-type network properties (Guimera et al., 2007a), an extended application of this concept is missing.

## 2.2. Network data, sampling, prediction and reverse engineering

Lovász (2012) gives an excellent summary of the network sampling problem. In most biological systems data coverage has technical limitations, and experimental errors are rather prevalent. As part of these uncertainties and errors, not all of the possible interactions are detected, and a large number of false-positives may also appear (Zhu et al., 2007; De Las Rivas & Fontanillo, 2010; Sardiu & Washburn, 2011). However, it is often a question of judgment, whether the investigator believes that only 'high-fidelity' interactions are valid, and discards all other data as potential artifacts, or uses the whole spectrum of data considering low-confidence interactions as low affinity and/or low probability interactions (Csermely, 2004, 2009). The highest quality interactions are reliable, but may not be representative of the whole network (Hakes et al., 2008). The unavailability of complete datasets can be circumvented by a number of methods which 1.) help the correct sampling of networks; 2.) enable the prediction of nodes/edges and 3.) infer network structure from the behavior of the complex system by reverse engineering. We will discuss these methods in this section.

### 2.2.1. Problems of network incompleteness, network sampling

Since complex networks are not homogeneous, their segments may display different properties than the whole network (Han et al., 2005; Stumpf et al., 2005; Tanaka et al., 2005; Stumpf & Wiuf, 2010; Annibale & Coolen, 2011; Son et al., 2012). Therefore, the use of a representative sample of the network is a key issue. In the last few years several methods became available to assess whether the available part of an unknown complete network is a representative sample. These methods also allow the extrapolation of the partially available network data to the total dataset (Wiuf et al., 2006; Stumpf et al., 2008). Radicchi et al. (2011) introduced a GloSS filtering technique preserving both the weight distribution and network topology. Recently a comparison of several (re)-sampling methods was given (Mirshahvalad et al., 2013; Wang, 2012). Guimera and Sales-Pardo (2009) provided a method to detect missing interactions (false negatives) and spurious interactions (false positives). Riera-Fernandez et al. (2012) gave numerical quality scores to network edges based on the Markov–Shannon entropy model. However, data purging methods should be applied with caution, since unexpected edges of 'creative nodes' may also be identified as 'spurious' edges, and may be removed (Csermely, 2008; Lü & Zhou, 2011). Network sampling methods were recently reviewed by Ahmed et al. (2012).

### 2.2.2. Prediction of missing edges and nodes, network predictability

Prediction of missing edges and nodes is not only important to assess network reliability, but can also be used for predictions of e.g. heretofore undetected interactions of disease-related proteins, or extension of drug target networks helping drug design (Spiró et al., 2008). In Section 1.3.3 and Tables 2 and 3 we already listed several methods for the efficient prediction of new edges and nodes from complex human disease-related datasets. Prediction is not only a discovery tool, but it also helps to avoid the unpredictable, which is considered as dangerous. However, as we will see at the end of this section, in complex systems the least predictable constituents are the most exciting ones.

Lü and Zhou (2011) provided an excellent review of edge prediction. Referring to this paper for details here we will summarize only the major points of this field.

- Edges can be predicted by the properties of their nodes, e.g. protein sequences, or domain structures (Smith & Sternberg, 2002; Li & Lai, 2007; Shen et al., 2007; Hue et al., 2010).
- The similarity of the edge neighborhood in the network is widely used in edge prediction. Edge neighborhood may be restricted to the common neighbors of the connected nodes, may include all first neighbors, all first and second neighbors, cliques, the nodes' network modules, or the whole network. Consequently, similarity indices may be local (like the Adamic–Adar index, common neighbors index, hub promoted index, hub suppressed index, Jaccard index, Leicht–Holme–Newman index, preferential attachment index, resource allocation index, Salton index, or the Sørensen index) mesoscopic (like the local path index or the local random walk index), or global (like the average commute time index, cosine-based index, Katz index, Leicht–Holme–Newman index, matrix forest index, random walk with restart index, or the SimRank index). Edge neighborhood may be compared by using the network degree, preferential attachment methods, fitness values, community structure, network hierarchy, a stochastic bloc model, a probabilistic model, or by using hypergraphs (Albert & Albert, 2004; Liben-Nowell & Kleinberg, 2007; Yan et al., 2007a; Guimera & Sales-Pardo, 2009; Lu et al., 2009; Zhou et al., 2009; Chen et al., 2012a; Eronen & Toivonen, 2012; Hu et al., 2012; Musmeci et al., in press; Yan & Gregory, 2012; Liu et al., 2013). It is important to note that methods may perform differently, if the missing edge is in a dense network core or in a sparsely connected network periphery (Zhu et al., 2012a). The optimal method also depends on the average length of shortest paths in the network. Edge prediction methods often require a large increase in computational time to achieve a higher accuracy (Lü & Zhou, 2011).

• Edge prediction can be performed by comparing the network to an appropriately selected model network, to a similar real world network, or to an ensemble of networks (Liben-Nowell & Kleinberg, 2007; Clauset et al., 2008; Nepusz et al., 2008; Xu et al., 2011a; Gutfraind et al., 2012).

• Edges can also be predicted by the analysis of sequential snapshots of network topology (also called as network dynamics, or network evolution, see Section 2.5; Hidalgo & Rodriguez-Sickert, 2008; Lü & Zhou, 2011). In network time-series older events might have less influence on the formation of a new edge than newer ones. Additionally, all network evolution models can be used as edge-predictors. However, one has to keep in mind that network evolution models always include guesses about the factors influencing the generation of a novel edge (Lü & Zhou, 2011).

Edge prediction of drug–target networks allows the discovery of new drug target candidates and the repositioning of existing drugs (van Laarhoven et al., 2011). Prediction methods may combine several data-sources, like mRNA expression patterns, genotypic data, DNA–protein and protein–protein interactions (Zhu et al., 2008; Pandey et al., 2010). Dataset combination may help the precision of edge prediction. However, prediction of directed, weighted, signed, or colored edges of these combined datasets is still a largely unsolved task (Lü & Zhou, 2011).

Node prediction is even more difficult, than edge prediction (Getoor & Diehl, 2005; Liben-Nowell & Kleinberg, 2007). Predicted nodes may occupy structural holes, i.e. bridging positions between multiple network modules (Burt, 1995; Csermely, 2008), or may be identified by methods, like chance-discovery. Chance-discovery uses an iterative annealing process, and extends the dense clusters observed at lower annealing 'temperatures' (Maeno & Ohsawa, 2008). In fact, the well developed methodology of the identification of disease-related genes that we detailed in Section 1.3 can be regarded as a node prediction problem, and may give exciting clues for node prediction in networks other than those of disease-related data.

The predictability of network edges is not only a function of data coverage and network structure, but also depends on network dynamics. The mistaken identification of unexpected edges as spurious edges (Lü & Zhou, 2011), and the better predictability of edges in dense cores than those in network periphery (Zhu et al., 2012a) are both related to the inherent unpredictability caused by network dynamics. As an example, the edge-structure of date hubs, where hubs change their neighbors (Han et al., 2004a), is certainly less predictable than that of party hubs, i.e. hubs preserving a rather constant neighborhood. Date hubs mostly reside in inter-modular positions (Han et al., 2004a; Komurov & White, 2007; Kovács et al., 2010). Predictability is also related to network rigidity and flexibility (Gáspár & Csermely, 2012): an edge or node in a more flexible network position is less predictable than others situated in a rigid network environment.

Bridging positions are often more flexible and less predictable than intra-modular edges. If a node is connecting multiple, distant modules with approximately the same, low intensity, and continuously changing its position, like the recently described 'creative nodes' do (Csermely, 2008), its predictability will be exceptionally low. A shift towards lower predictability (higher network flexibility) is often accompanied by an increased adaptation capability at the system level. Moreover, a complex system lacking flexibility is unable to change, to adapt and to learn (Gyurkó et al., in press). Thus it is not surprising that highly unpredictable, 'creative' nodes characterize all complex systems. Importantly, these highly unpredictable nodes help in delaying critical transitions of the systems, i.e. postponing market crash, ecological disaster or death (Csermely, 2008; Scheffer et al., 2009; Farkas et al., 2011; Sornette & Osorio, 2011; Dai et al., 2012). In fact, the most unpredictable nodes are the most exciting nodes of the system having a hidden influence on the fate of the whole system at critical situations. The prediction of their unpredictable behavior remains a major challenge of network science.

### 2.2.3. Prediction of the whole
#### network, reverse engineering, network-inference

There are situations, when the network is so incomplete that we do not know anything on the network structure. However, we often have a detailed knowledge of the behavior of the complex system encoded by the network. The elucidation of the underlying network from the emergent system behavior is called reverse engineering or network-inference.

In a typical example of reverse engineering we know the genome-wide mRNA expression pattern and its changes after various perturbations (including drug action, malignant transformation, development of other diseases, etc.), but we have no idea of the gene–gene interaction network, which is causing the changes in mRNA expression pattern. As a rough estimate, a network of 10,000 genes can be predicted with reasonable precision using less than a hundred genome-wide mRNA datasets. Network prediction can be greatly helped using previous knowledge, e.g. on the modules of the predicted network. The correct identification of the relatedness of mRNA expression sets (position in time series, tissue-specificity, etc.) may often be a more important determinant of the final precision of network prediction than the precise measurement of the mRNA expression levels. Models of network dynamics, probabilistic graph models and machine learning techniques are often incorporated in reverse engineering methods. Some of these approaches, like Bayesian methods, require a rather intensive computational time. Therefore, computationally less expensive methods such as the copula method, or the simultaneous expression model with Lasso regression were also introduced. The topology of the predicted network often determines the type of the best method. This is one reason, why combination of various methods (or the use of iterative approaches) may outperform individual methodologies (Liang et al., 1998a; Akutsu et al., 1999; Ideker et al., 2000; Kholodenko et al., 2002; Yeung et al., 2002; Segal et al., 2003; Tegnér et al., 2003; Friedman, 2004; Tegnér & Bjorkegren, 2007; Cosgrove et al., 2008; Kim et al., 2008; Ahmed & Xing, 2009; Stokic et al., 2009; Marbach et al., 2010, 2012; Yip et al., 2010; Pham et al., 2011; Schaffter et al., 2011; Altay, 2012; Crombach et al., 2012; Kotera et al., 2012). Jurman et al. (2012a) designed a network sampling stability-based tool to assess network reconstruction performance.

Reverse engineering techniques were successfully applied to reconstruct drug-affected pathways (Gardner et al., 2003; di Bernardo et al., 2005; Chua & Roth, 2011; Gosline et al., 2012). Besides the identification of gene regulatory networks from the transcriptome, reverse engineering methods may also be used to identify signaling networks from the phosphorome or signaling network (Kholodenko et al., 2002; Sachs et al., 2005; Zamir & Bastiaens, 2008; Eduati et al., 2010; Prill et al., 2011), metabolic networks from the metabolome (Nemenman et al., 2007), or drug action mechanisms and drug target candidates from various datasets (Gardner et al., 2003; di Bernardo et al., 2005; Lehár et al., 2007; Lo et al., 2012; Madhamshettiwar et al., 2012).

Though the number of reverse-engineering methods has been doubled every two years, 1.) the inclusion of non-linear system dynamics, of multiple data sources and of multiple methods; 2.) distinguishing between direct and indirect regulations; 3.) a better discrimination between causal relationships and coincidence; as well as 4.) network prediction in case of multiple regulatory inputs per node remain major challenges of the field (Tegnér & Bjorkegren, 2007; Marbach et al., 2010).

### 2.3. Key segments of network structure

In this section we will give a brief summary of the major concepts and analytical methods of network structure starting from local network topology and proceeding towards more and more global network structures. Selection of key network positions as drug target options has a major dilemma. On the one hand, the network position has to be important enough to influence the diseased body; on the other hand, the selected network position must not be so important that its attack would lead to toxicity. The successful solution of this dilemma requires a detailed knowledge on the structure and dynamics of complex networks.

### 2.3.1. Local topology: hubs, motifs and graphlets

A minority of nodes in a large variety of real world networks is a hub, i.e. a node having a much higher number of neighbors than average. Real world networks often have a scale-free degree distribution providing a non-negligible probability for the occurrence of hubs, as it was first generalized to real world networks by the seminal paper of Barabási and Albert (1999). If hubs are selectively attacked, the information transfer deteriorates rapidly in most real world networks. This property made hubs attractive drug targets (Albert et al., 2000). However, some of the hubs are essential proteins, and their attack may result in increased toxicity. This narrowed the use of major hubs as drug targets mostly to antibiotics, to other anti-infectious drugs and to anticancer therapies. In agreement with these, on average, targets of FDA-approved drugs tend to have more connections than peripheral nodes, but fewer connections than hubs (Yildirim et al., 2007). Cancer-related proteins have many more interaction partners than non-cancer proteins making the targeting of cancer-specific hubs a reasonable strategy in anti-cancer therapies (Jonsson & Bates, 2006). Besides the direct count of interactome neighbors algorithms have been developed to identify hubs using Gene Ontology terms (Hsing et al., 2008). Going one level deeper in the network hierarchy, amino acids serving as hubs of protein structure networks play a key role in intra-protein information transmission (Pandini et al., 2012), and may provide excellent target points of drug interactions.

The emerging picture of using hubs as drug targets can be summarized by two opposite effects. On the one hand, hubs are so well connected that their attack may lead to cascading effects compromising the function of a major segment of the network; on the other, nodes with limited number of connections are at the 'ends' of the network, and their modulation may have only limited effects (Penrod et al., 2011). There are several important remarks refining this conclusion.

- Not all hubs are equal. Weighted and directed networks are extremely important in discriminating between hubs. A hub having 20 neighbors connected with an equal edge-weight is different from a hub having the same number of 20 neighbors having a highly uneven edge-structure of a single, dominant edge and 19 low intensity edges. A sink-hub with 20 incoming edges is not at all the same than a source-hub with the same number 20 outgoing edges. Soluble proteins possess more contacts on average than membrane proteins (Yu et al., 2004a) warning that the hub-defining threshold of neighbors cannot be set uniformly.
- Hub-connectors, i.e. edges or nodes connecting major hubs also offer very interesting drug targeting options (Korcsmáros et al., 2007; Farkas et al., 2011).
- Not all peripheral nodes are unimportant. There are peripheral nodes called 'choke points', which uniquely produce or consume an important metabolite. The inhibition of 'choke points' often leads to a lethal effect (Yeh et al., 2004; Singh et al., 2007).
- Importantly, interdependent networks, i.e. at least two interconnected networks, were shown to be much more vulnerable to attacks than single network structures (Buldyrev et al., 2010). We have several interdependent networks in our cells, such as the networks of signaling proteins and transcription factors, or the interactome of membrane proteins and the network of the interacting nuclear, plasma, mitochondrial and endoplasmic reticulum membranes. The excessive vulnerability of interdependent networks should make us even more cautious in the selection of drug target nodes. The options of edgetic drugs, multi-target drugs and allo-network drugs, we will describe in Section 4.1.6 (Nussinov et al., 2011), may circumvent the worries and problems related to the single and direct targeting of network nodes with drugs.

Network motifs are circuits of 3 to 6 nodes in directed networks that are highly overrepresented as compared to randomized networks (Milo et al., 2002; Kashtan et al., 2004). Graphlets are similar to motifs but are defined as undirected networks (Przulj et al., 2006). Motifs proved to be efficient in predicting protein function, protein–protein interactions and development of drug screening techniques (Bu et al., 2003; Albert & Albert, 2004; Luni et al., 2010; Cloutier & Wang, 2011). Rito et al. (2010) made an extensive search for graphlets in protein–protein interaction networks and concluded that interactomes may be at the threshold of the appearance of larger motifs requiring 4 or 5 nodes. Such a topology would make interactomes both efficient having not too many edges and robust harboring alternative pathways.

### 2.3.2. Broader network topology: modules,
### bridges, bottlenecks, hierarchy, core, periphery, choke points

Network modules (or in other words: network communities) are the primary examples of mesoscopic network structures, which are neither local, nor global. Modules represent groups of networking nodes, and are related to the central concept of object grouping and classification. Modules of molecular networks often encode cellular functions. Moreover, the exploration of modular structure was proposed as a key factor to understand the complexity of biological systems. Therefore, module determination gained much attention in recent years. Modules of molecular networks are formed from nodes, which are more densely connected with each other than with their neighborhood (Girvan & Newman, 2002; Fortunato, 2010; Kovács et al., 2010; Koch, 2012; Szalay-Bekő et al., 2012). In Section 1.3 we introduced disease modules, i.e. modules of disease-related genes in protein–protein interaction networks (Goh et al., 2007; Oti & Brunner, 2007; Jiang et al., 2008; Suthram et al., 2010; Bauer-Mehren et al., 2011; Loscalzo & Barabasi, 2011; Nacher & Schwartz, 2012). These node-related properties influence the modular functions, making them attractive network drug-targets (Cho et al., 2012). However, the determination of network modules proved to be a notoriously difficult problem resulting in more than two hundred independent modularization methods (Fortunato, 2010; Kovács et al., 2010).

Modules of molecular networks have an extensive (often called pervasive) overlap, which was recently shown to be denser than the center of the modules in some social networks (Palla et al., 2005; Ahn et al., 2010; Kovács et al., 2010; Yang & Leskovec, 2012). This reflects the economy of our cells using a protein in more than one function. Modules with sparse edge structure also characterize protein–protein interaction networks (Srihari & Leong, 2012). Modules of real world networks were shown to form a 'very small world' having an average distance of 3 from each other (Li & Li, 2013). Inter-modular nodes are attractive drug targets.

Bridges connect two neighboring network modules (Fig. 8). Bridges may also be identified by k-shell analysis (Reppas & Lawyer, 2012). Bridges usually have fewer neighbors than hubs, and are independently regulated from the nodes belonging to both modules, which they connect. This makes them attractive as drug targets, since they may display lower toxicity, while the disruption of information flow between functional network modules could prove to be therapeutically effective (Hwang et al., 2008). Proteins involved in the aging process are often bridges (Wang et al., 2009). Proteins bridging disease modules may provide important points of interventions (Nguyen & Jordan, 2010; Nguyen et al., 2011).

Inter-modular hubs form a special class of inter-modular nodes (Fig. 8). Date hubs, i.e. hubs having only a single or few binding sites and frequently changing their protein partners, were shown to occupy an inter-modular position as opposed to party hubs residing mostly in modular cores (Han et al., 2004a; Kim et al., 2006; Komurov & White, 2007; Kovács et al., 2010). Party hubs tend to have higher affinity binding surfaces than date hubs (Kar et al., 2009). Inter-modular hubs usually have a regulatory role (Fox et al., 2011), and are mutated frequently in cancer (Taylor et al., 2009).

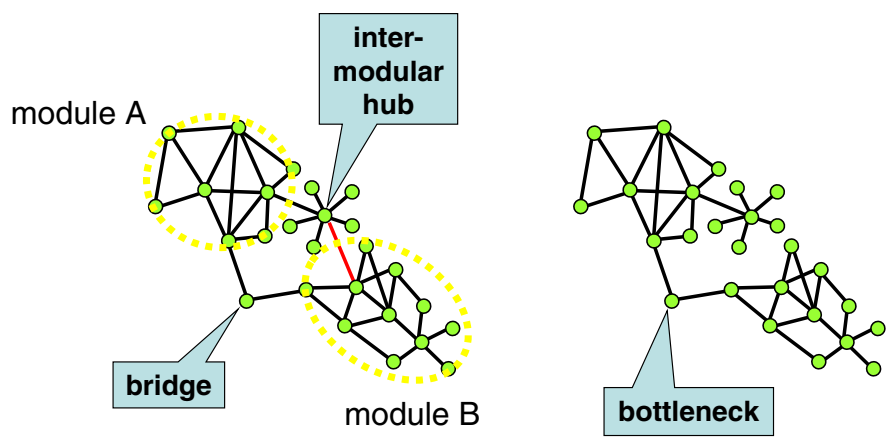

**Fig. 8.** Bridge, inter-modular hub and bottleneck. The network on the left side of the figure has two modules (modules A and B marked by the yellow dotted lines), which are connected by a bridge and by an inter-modular hub. By the removal of the red edge from the network on the left side, the former bridge obtains a unique and monopolistic role connecting modules A and B, and is therefore called as a bottleneck.

Nodes occupying a unique and monopolistic inter-modular position have been termed 'bottlenecks' (Fig. 8), because almost all information flowing through the network must pass through these nodes. This makes bottlenecks more effective drug targets than bridges (Yu et al., 2007b). In agreement with this concept, hub-bottlenecks were shown to be preferential targets of microRNAs (Wang et al., 2011c) and play an important role in cellular re-programming (Buganim et al., 2012). However, inhibition of bottlenecks often compromises network integrity, restricting their use as drug targets to anti-infectious and (in the case of cancer-specific bottlenecks) anti-cancer therapies (Yu et al., 2007b). In agreement with this proposition, cancer proteins tend to be inter-modular hubs of cancer-specific networks offering an important target option (Jonsson & Bates, 2006).

Nodes connecting more than two modules are in modular overlaps. Overlapping nodes occupy a network position, which can provide more subtle regulation than bridges or bottlenecks. Modular overlaps are primary transmitters of network perturbations, and are key determinants of network cooperation (Farkas et al., 2011). Overlapping nodes play a crucial role in cellular adaptation to stress. In fact, changes in the overlap of network modules were suggested to provide a general mechanism of adaptation of complex systems (Mihalik & Csermely, 2011; Csermely et al., 2012). Modular overlaps (called cross-talks between signaling pathways) are most prevalent in humans, if compared to *Caenorhabditis elegans* or *Drosophila* (Korcsmáros et al., 2010). All these make modular overlaps especially attractive drug targets (Farkas et al., 2011). As we described earlier, 'creative nodes' are in the overlap of multiple modules belonging roughly equally to each module. These nodes play a prominent role in regulating the adaptivity of complex networks, and are lucrative network targets (Csermely, 2008; Farkas et al., 2011).

Despite the important role of hierarchy in network structures (Ravasz et al., 2002; Liu et al., 2012; Mones et al., 2012), the exploration of network hierarchy is largely missing from network pharmacology. Ispolatov and Maslov (2008) published a useful program to remove feedback loops from regulatory or signaling networks, and reveal their remaining hierarchy (http://www.cmth.bnl.gov/~maslov/programs.htm). Hartsperger et al. (2010) developed HiNO using an improved, recursive approach to reveal network hierarchy (http://mips.helmholtz-muenchen.de/hino). The hierarchical map approach of Rosvall and

Bergstrom (2011) used the shortest multi-level description of a random walk (http://www.tp.umu.se/~rosvall/code.html). A special class of hierarchy-representation and visualization uses the hierarchical structure of modules, i.e. the concept that modules can be regarded as meta-nodes and re-modularized, until the whole network coalesces into a single meta-node. Methods like Pyramabs (http://140.113.166.165/pyramabs.php; Cheng & Hu, 2010) or the Cytoscape (Smoot et al., 2011) plug-in, ModuLand (http://linkgroup.hu/modules.php; Szalay-Bekő et al., 2012) are good examples of this powerful approach.

Network hierarchy has recently been involved as a key factor of network controllability (Liu et al., 2012; Mones et al., 2012), which will be discussed in Section 2.3.4 in detail. However, not all hierarchical networks are 'autocratic', where top nodes have an unparalleled influence. Horizontal contacts of middle-level regulators play a key role in gene regulatory networks. Moreover, such a 'democratic network character' increases markedly in human gene regulation (Bhardwaj et al., 2010).

Similarly, the discrimination between network core and periphery has been published quite a while ago (Guimera & Amaral, 2005) and was extended recently to network modules (Li & Li, 2013), but its applications are largely missing from the field of drug design. As an example of the possible benefits, choke points were identified as those peripheral nodes that either uniquely produce or consume a certain metabolite (including here signal transmitters and membrane lipids too). Efficient inhibition of choke points may cause either a lethal deficiency, or toxic accumulation of the metabolite (Yeh et al., 2004; Singh et al., 2007).

### 2.3.3. Network centrality, network skeleton, rich-club and onion-networks

Network centrality measures span the entire network topology from local to global. Centrality is related to the concept of importance. Central nodes may receive more information, and may have a larger influence on the networking community. Thus it is not surprising that dozens of network centrality measures have been defined. Several centrality measures are local, like the number of neighbors (the network degree), or related to the modular structure, like bridging centrality, community centrality, or subgraph centrality. Centrality measures, like betweenness centrality (the number of shortest paths traversing through the node), random walk related centralities (like the PageRank algorithm of Google), or network salience are based on more global network properties. Recently a number of centrality measures have been defined based on network dynamics (Freeman, 1978; Estrada & Rodríguez-Velázquez, 2005; Estrada, 2006; Hwang et al., 2008; Kovács et al., 2010; Du et al., 2012; Ghosh & Lerman, 2012; Grady et al., 2012; Grassler et al., 2012;

Joseph & Chen, 2012; Mantzaris et al., in press). Global network centrality calculations may be faster, assessing only network segments and using network compression (Sariyüce et al., 2012). Network module-based centralities are related to the determination of bridges and overlaps (Hwang et al., 2008; Kovács et al., 2010), while betweenness centrality is used for the definition of bottlenecks (Yu et al., 2007b). Both are important drug target candidates as we discussed in the previous section. As an additional example, high betweenness centrality hubs were shown to dominate the drug–target network of myocardial infarction (Azuaje et al., 2011).

The network skeleton is an interconnected subnetwork of high centrality nodes. Network skeletons may contain hubs (we call this a 'rich-club'; Colizza et al., 2006; Fig. 9), may consist of high betweenness centrality nodes (Guimera et al., 2003), or may comprise inter-connected centers of network modules (Kovács et al., 2010; Szalay-Bekő et al., 2012). Network skeletons may be densely interconnected forming an inner core of the network, or may be truly skeleton-like traversing the network like a highway. In both network skeleton representations nodes participating in the network skeleton form the 'elite' of the network, like the respective persons in social networks (Avin et al., 2011). Network skeleton nodes are attractive drug target candidates. As an example of this, Milenkovic et al. (2011) defined a dominating set of nodes as a connected network subgraph having all residual nodes as its neighbor. They showed that the dominating set (especially if combined with a network-module type centrality measure called as graphlet degree centrality measuring the summative degree of neighborhoods extending to 4 layers of neighbors) captures disease-related and drug target genes in a statistically significant manner. It will be interesting to see, whether the recently defined intra-modular dominating sets (Li & Li, 2013) also possess similar features. Nicosia et al. (2012) defined a subset of nodes (called controlling sets), which can assign any prescribed set of centrality values to all other nodes by cooperatively tuning the weights of their out-going edges. Nacher and Schwartz (2008) identified a rich-club of drugs serving as a core of the drug–therapy network composed of drugs and established classes of medical therapies.

Network assortativity characterizes the preferential attachment of nodes having similar degrees to each other. Network cores (such as rich-clubs, Fig. 9) may or may not be a part of an assortative network. In a disassortative network low degree, peripheral network nodes are connected to the network core and not to each other. These core–periphery networks have a nested structure (Fig. 9). If peripheral nodes are connected to each other and form consecutive rings around the core, we call the network 'onion-type' (Fig. 9). Nested networks were shown to characterize ecosystems and trade networks, while onion-networks are especially resistant against targeted attacks (Saavedra et al., 2011; Schneider et al., 2011; Wu & Holme, 2011). Despite the exciting features of nested and onion networks, these network characteristics have not been assessed yet in disease-related, or drug design related-studies.

### 2.3.4. Global network topology: small worlds, network percolation, integrity, reliability, essentiality and controllability

The global topology of most real world networks is characterized by the small world property first generalized in the landmark paper of Watts and Strogatz (1998). Nodes of small worlds are connected well—as it was popularized by the proverbial "six degrees of separation" meaning that members of the social network of Earth can reach each other using 6 consecutive contacts (edges) as an average. In fact, modern web-based social networks, like Facebook, are an even smaller world having an average shortest path of 4.74 edges (Blackstrom et al., 2011).

Percolation is a broader term of global network topology than small worldness, since it refers to the connectedness of network nodes, i.e. the presence of a connected, giant network component. Sequential attacks on network nodes can induce a progressive and dramatic decrease of network percolation. Despite being a sensitive measure, to date the concept of percolation has not been extended to characterize network modules and other non-global structures of molecular networks (Antal et al., 2009). Percolation is related to network integrity and network reliability; that is, related to how much of the network remains connected if a network node or edge fails. In the case of directed networks the connection of sources or sinks can be calculated separately (Gertsbakh & Shprungin, 2010). The network efficiency measure of Latora and Marchiori (2001) is a widely used criterion to judge the integrity of a network. As noted before, intentional attack of hubs can be deleterious to most real world networks (Albert et al., 2000). The effect of a single attack of the largest hub in gene transcription networks can be substituted by a surprisingly low number of partial attacks, which is making the multi-target approaches listed in Section 4.1.5 a viable option from the network point of view (Agoston et al., 2005; Csermely et al., 2005).

In the case of anti-infectious or anti-cancer agents we would like to destroy the network of the parasite or of the malignant cell. In other words we need to predict essential proteins as targets of these therapeutic approaches. This makes network integrity a key measure to judge the efficiency of drug target candidates in these fields. Prediction of essential proteins is also important to forecast the toxicity of other drugs. The number of neighbors in protein–protein interaction networks is an important network measure of essentiality (Jeong et al., 2001). Later more global network measures were also shown to contribute to the prediction of node essentiality (Chin & Samanta, 2003; Estrada, 2006; Yu et al., 2007b; Missiuro et al., 2009; Li et al., 2011a; Song & Singh, 2013). Edge weights and directions may also significantly alter the determination of attack efficiency (Dall'Astra et al., 2006; Yu et al., 2007b). Finally, the constraints of metabolic networks define different contexts of essentiality exemplified by choke points, i.e. proteins uniquely producing or consuming a certain metabolite (Yeh et al., 2004; Singh et al., 2007). We will describe metabolic network essentiality in Section 3.6.2 in detail.

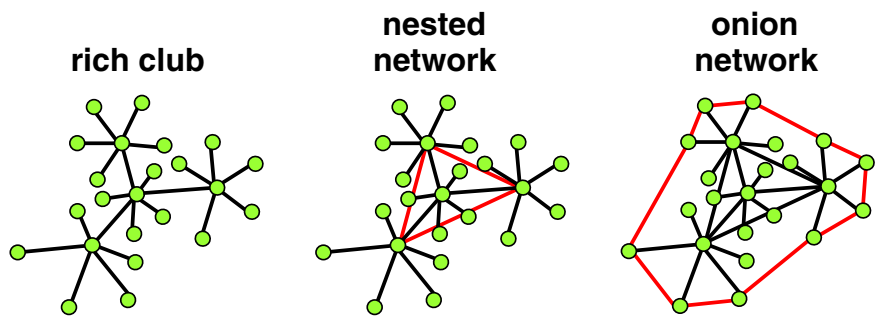

**Fig. 9.** Rich club, nested network and onion network. The figure illustrates the differences between a network having a rich club (left side), having a highly nested structure (middle) and developing an onion-type topology (right side). Note that the connected hubs of the rich club become even more connected by adding the 3 red edges on the middle panel. Connection of the peripheral nodes by an additional 10 red edges on the right panel turns the nested network to an onion network having a core and an outer layer. Note that the rich club network already has a nested structure, and both the nested network and the onion network have a rich club. Larger onion networks have multiple peripheral layers.

The most recent aspect of global network topology is similar to essentiality in the sense that it is also related to the influence of nodes on network behavior. However, here node influence is not judged on a 'yes/no scale', i.e. by whether the organism survives the malfunction of the node, but based on the more subtle scale of changing cell behavior. In this way node influence studies are closely related to network dynamics as we will detail in Section 2.5. Network centrality measures, or the dominating set of network nodes we mentioned before, are also related to the influence of selected nodes on others. Recent publications added network controllability, i.e. the ability to shift network behavior from an initial state to a desired state, to the repertoire of network-related measures of node influence. From these initial studies central nodes emerged as key players of network control (Cornelius et al., 2011; Liu et al., 2011, 2012; Banerjee & Roy, 2012; Cowan et al., 2012; Mones et al., 2012; Nepusz & Vicsek, 2012; Wang et al., 2012a; Pósfai et al., 2013). It is important to note that control here is a weak form of control, since we do not want to control how the system reaches the desired state (San Miguel et al., 2012). Despite of the clear applicability of network controllability to drug design (i.e. finding the nodes, which can shift molecular networks of the cell from a malignant state to a healthy state) there were only a few studies testing various aspects of this rich methodology in drug design (Xiong & Choe, 2008; Luni et al., 2010). Development of drug-related applications of network influence and control models is an important task of future studies.

### 2.4. Network comparison and similarity

As we summarized in Section 2.2, uncovering network similarities is useful to predict nodes and edges. Alignment of networks from various species identifies interologs corresponding to conserved interactions between a pair of proteins having interacting homologs in another organism, or the analogous regulogs in regulatory networks, signalogs in signal transduction networks and phenologs as disease associated-genes. Thus, network comparison may uncover novel protein functions and disease-specific changes. All these greatly help drug design (Yu et al., 2004b; Sharan et al., 2005; Leicht et al., 2006; Sharan & Ideker, 2006; Zhang et al., 2008; McGary et al., 2010; Korcsmáros et al., 2011). However, the great potential to uncover network similarities comes with a price: network comparison is computationally expensive, and remains one of the greatest challenges of the field.

Lovász (2009, 2012) gives an excellent summary of the network similarity problem including a number of network similarity measures such as edit distance (the number of edge changes required to get one network from another), sampling distance (measuring the similarity by an ensemble of random networks), cut distance and similarity distance. A later study also used an interesting combined distance metrics of the edit and spectral distances (Jurman et al., 2012b). Similarity measures based on the comparison of the top-k nodes were recently described (Amin et al., 2012; Lee et al., 2012a). Similarity indices may be local (comparing the closest neighborhood of selected nodes), mesoscopic (which are usually based on local walks), or global (often involving extensive, network-wide walks). Edge neighborhood may be compared by using the modular structure, hypergraphs, network hierarchy, a stochastic bloc model, or a probabilistic model. Comparison may also use an ensemble of random, scale-free or other model networks, and the distribution of the best fitting ensemble. Reviews of Sharan and Ideker (2006), Zhang et al. (2008) and Lü and Zhou (2011) give further details of the methodology used in the comparison of molecular networks.

A specific example of network comparison is the comparison of network descriptions of chemical structures, which we will summarize in Section 3.1. Table 4 summarizes a few major methods and related web-sites to compare molecular networks. Quite a few methods compare small subnetworks to larger ones. Sometimes the "small subnetwork" is small, containing only 3 to 5 nodes. This reduces the problem of finding a motif in a larger network (also called as network querying). Recent methods 1.) include an expansion process, which explores the network structure beyond the direct neighborhood; 2.) compress the network to meta-nodes, then align this representative network and finally refine the alignment; 3.) use k-hop network coloring to speed up the comparison of the traditional coloring techniques of neighboring nodes, or 4.) extend the comparison using multiple types of networks and functional information (Table 4; Ay et al., 2011; Ay et al., 2012; Berlingerio et al., 2012; Gulsoy et al., 2012). Despite the extensive progress in the field, additional work is needed to develop efficient comparison methods for large molecular networks and multiple network datasets. A widely used area of network comparison is the assessment of two time points, or a time series of a changing network, which will be discussed in the next section.

### 2.5. Network dynamics

In this section, which concludes the inventory of network analytical concepts and methods, we will summarize the approaches describing network dynamics. First we will list the methods describing the temporal changes of networks, then we describe the usefulness of network perturbation analysis in drug design, and finally we will draw attention to the potential use of spatial games to assess the influence of nodes on network cooperation. Description of network dynamics is a fast developing field of network science holding great promise to renew systems-based thinking in drug design.

#### 2.5.1. Network time series, network evolution

As we mentioned in Section 2.1 summarizing the key points of network definition, the time-window of observation is crucial for the detection of contacts between network nodes. The duration of observation becomes even more important, when describing the temporal changes of networks, which is also often called network evolution. (It is important to note that the concept of network evolution usually has no connection to the Darwinian concept of natural selection.) The order of network edge development has key consequences in directed networks which differ from network topology measures, like shortest path, or small world. As an interesting example of these changes, in the A → B → C connection pattern, A cannot influence C, if the B → C contact preceded the A → B contact. Such effects may slow down the propagation of signals by an order of magnitude (Tang et al., 2010; Pfitzner et al., 2012).

The description of temporal changes of network structures is related to the difficult concept and methodology of network comparison and similarity described in the preceding section. Following the early summary of Dorogovtsev and Mendes (2002) on network evolution, Holme and Saramäki (2011) had an excellent review on network time-series re-defining a number of static network parameters, such as connectivity, diameter, centrality, motifs and modules, to accommodate temporal changes. The prediction algorithms described in Section 2.2 can be used to predict edges that may appear at later time points in evolving networks (Lü & Zhou, 2011). Prediction may work backwards, to infer past structures of a current network identifying core-nodes around which the network was organized (Navlakha & Kingsford, 2011). Recently, a method to test the reversibility of changes in network time-series was published (Donges et al., 2012). However, most network time description studies have concentrated on neuronal or social networks offering many, albeit yet untested, possibilities for drug design.

Recently a number of centrality measures were introduced describing central nodes of dynamically changing networks (Joseph & Chen, 2012; Mantzaris et al., in press). The development of network modules gained considerable attention in network evolution studies, since this representation concentrates on the functionally most relevant changes in the network structure. Network modules may grow, contract, merge, split, be born or die. Some of the modules display a much larger stability than others. The intra-modular nodes of these modules bind to each other with high affinity and to nodes outside the module with low affinity. Interestingly, small modules (of say less than 10 nodes) seem to persist better, if they have a dense contact structure, while larger modules survive better, if they have a dynamic, fluctuating membership (Palla et al., 2007; Fortunato, 2010). Mucha et al. (2010) developed the technique of multislice networks, which monitor the module development of nodes with multiple types of edges. Taylor et al. (2009) showed that altered modularity of hubs had a prognostic value in breast cancer and suggested cancer-specific inter-modular hubs as drug targets in cancer therapies.

**Table 4**
Comparison methods of molecular networks.

| Name[a] | Network type(s)[b] | Description and website | References |
|---|---|---|---|
| AlignNemo | Protein–protein interaction networks | Uncovers subnetworks of proteins and uses an expansion process, which gradually explores the network beyond the direct neighborhood. http://sourceforge.net/p/alignnemo | Ciriello et al., 2012a |
| Differential dependency network analysis | Transcriptional networks | A set of conditional probabilities is proposed as a local dependency model, and a learning algorithm is developed to show the statistical significance of the local structures. http://www.cbil.ece.vt.edu/software.htm | Zhang et al., 2009 |
| Graphcrunch2, C-GRAAL, MI-GRAAL | Multiple networks | Compares networks with random networks. Additionally, clusters nodes based on their topological similarities in the compared networks. http://bio-nets.doc.ic.ac.uk/graphcrunch2; http://bio-nets.doc.ic.ac.uk/MI-GRAAL | Kuchaiev and Przulj, 2011; Kuchaiev et al., 2011; Memisevic & Przulj, 2012 |
| IsoRankN (IsoRank Nibble) | Metabolic networks | Uses spectral clustering on the induced graph of pair-wise alignment scores. http://isorank.csail.mit.edu | Liao et al., 2009 |
| MetaPathway-Hunter | Metabolic networks | Finds tree-like pathways in metabolic networks. http://www.cs.technion.ac.il/~olegro/metapathwayhunter | Pinter et al., 2005 |
| MNAligner | Molecular networks | Combines molecular and topological similarity using integer quadratic programming, enabling the comparison of weighted and directed networks and finding cycles beyond tree-like structures. http://doc.aporc.org/wiki/MNAligner | Li et al., 2007 |
| Module Preservation | Measures module preservation in different datasets | Uses several module comparison statistics based on the adjacency matrix, or on the basis of pair-wise correlations between numeric variables. http://www.genetics.ucla.edu/labs/horvath/CoexpressionNetwork/ModulePreservation | Langfelder et al., 2011 |
| NeMo | Gene co-expression networks | Detects frequent co-expression modules among gene co-expression networks across various conditions. http://zhoulab.usc.edu/NeMo | Yan et al., 2007b |
| NetAlign | Protein–protein interaction networks | Aligns conserved network substructures. http://netalign.ustc.edu.cn/NetAlign | Liang et al., 2006 |
| NetAligner | Protein–protein interaction networks + pathways | Compares whole interactomes, pathways and protein complexes of 7 organisms. http://netaligner.irbbarcelona.org | Pache et al., 2012 |
| NetMatch | Cytoscape plug-in for molecular networks | Finds subgraphs of the original network connected in the same way as the querying network. Can also handle multiple edges, multiple attributes per node and missing nodes. http://baderlab.org/Software/NetMatch | Ferro et al., 2007 |
| PathBLAST | Search of smaller linear pathways | Finds smaller linear pathways in protein–protein interaction networks. http://www.pathblast.org | Kelley et al., 2004 |
| PINALOG | Protein–protein interaction network | Combines information from protein sequence, function and network topology. http://www.sbg.bio.ic.ac.uk/~pinalog | Phan & Sternberg, 2012 |
| Rahnuma | metabolic networks | Represents metabolic networks as hypergraphs and computes all possible pathways between two or more metabolites. http://portal.stats.ox.ac.uk:8080/rahnuma | Mithani et al., 2009 |

[a] The summaries of Sharan and Ideker (2006) and Zhang et al. (2008) describe and compare some of the methods above.
[b] The network type is indicating the primary network, where the method has been tested. However, most methods are applicable to other types of molecular networks.

Detailed analyses identified change points, i.e. short periods where large changes of modular structure can be observed (Falkowski et al., 2006; Sun et al., 2007; Rosvall & Bergstrom, 2010). The alluvial diagram (applying the visualization technique of Sankey diagrams) introduced by Rosvall and Bergstom (2010; Fig. 10) illustrates the temporal changes of network modules particularly well. Recently the addition of routes between nodes of the network, called the accessibility graph, was also used successfully to describe network time series (Lentz et al., 2012).

Dramatic changes of network structure called "topological phase transitions" occur when the resources needed to the maintain network contacts diminish, or environmental stress becomes much larger. Networks may develop a hierarchy, a core or a central hub as the relative costs of edge-maintenance increase. Under extreme circumstances, the network may disintegrate to small subgraphs, which corresponds to the death of the complex organism encoded by the formerly connected network (Derényi et al., 2004; Csermely, 2009; Brede, 2010). Change points and topological phase transitions have not been assessed in disease, or in other therapeutically interesting situations showing an abrupt change, such as apoptosis, and thus provide an exciting field of future drug-related studies.

Going beyond the changes of system structure, network descriptions may also be applied to describe changes of systems-level emergent properties. In these descriptions nodes represent phenotypes of the complex system in the state-space, and edges are the transitions or similarities of these phenotypes. This approach is used in the network representations of energy landscapes (or fitness landscapes) resulting in transition networks, and in the recurrence-based time series analysis resulting in correlation networks, cycle networks, recurrence networks or visibility graphs (Doye, 2002; Rao & Caflisch, 2004; Donner et al., 2011). Recently a method to compare two visibility graphs, i.e. two network time series was published (Mehraban et al., 2013).

### 2.5.2. Network robustness and perturbations

In the network-related scientific literature perturbations often mean the complete deletion of a network node. However, in drug action the complete inhibition of a molecule is seldom achieved. Therefore, when summarizing network perturbations, we will concentrate on the transient changes of network-encoded complex systems. Transient perturbations play a major role in signaling and in the development of diseases. The action of drugs can be perceived as a network perturbation nudging pathophysiological networks back into their normal state (Gardner et al., 2003; di Bernardo et al., 2005; Ohlson, 2008; Antal et al., 2009; Huang et al., 2009; Lum et al., 2009; Baggs et al., 2010; del Sol et al., 2010; Chua & Roth, 2011). Therefore, studies addressing perturbation dynamics have a key importance in drug design.

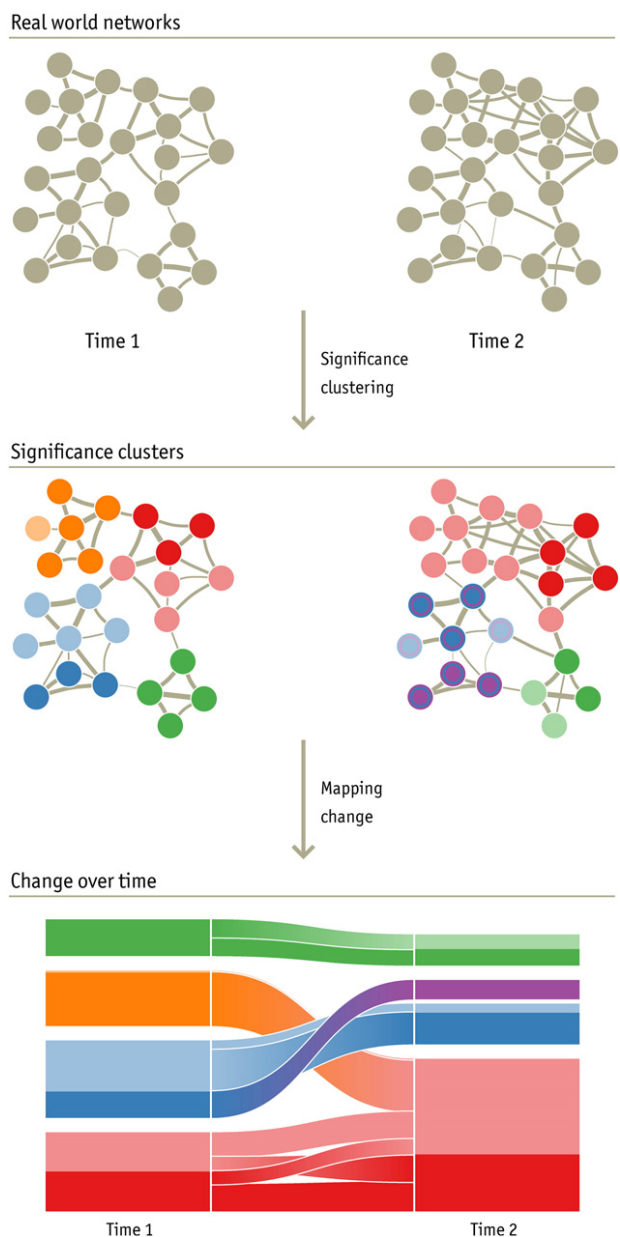

**Fig. 10.** Alluvial diagram illustrating the temporal changes of network communities. Each block represents a network module with a height corresponding to the module size. Modules are ordered by size (in case of a hierarchical structure within their supermodules). Darker colors indicate module cores. Modules having a non-significant difference are closer to each other. The height of the changing fields in the middle of the representation corresponds to the number of nodes participating in the change. To reduce the number of crossovers, changes are ordered by the order of connecting modules. To make the visualization more concise transients are passing through the midpoints of the entering and exiting modules and have a slim waist. Note the split of the blue module, and the merge of the orange and red modules.
Reproduced with permission from Rosvall and Bergstrom (2010).

Robustness is an intrinsic property of cellular networks that enables them to maintain their functions in spite of various perturbations. Enhanced robustness is a property of only a very small number of all possible network topologies. Cellular networks both in health and in disease belong to this extreme minority. Drug action often fails due to the robustness of disease-affected cells or parasites. In contrast, side-effects often indicate that the drug hits an unexpected point of fragility of the affected networks (Kitano, 2004a, 2004b, 2007; Ciliberti et al., 2007). Robustness analysis was used to reveal primary drug targets and to characterize drug action (Hallén et al., 2006; Moriya et al., 2006; Luni et al., 2010).

Cellular robustness can be caused by a number of mechanisms.

- Network edges with large weights often form negative or positive feedbacks helping the cell to return to the original state (attractor) or jump to another, respectively.
- Network edges with small weights provide alternative pathways, give flexible inter-modular connections disjoining network modules to block perturbations and buffer the changes by additional, yet unknown mechanisms. These 'weak links' grossly outnumber the 'strong links' participating in feedback mechanisms. Therefore, the two mechanisms have comparable effects at the systems level.
- Finally, robustness of molecular networks also depends on the robustness of their nodes, e.g. the stability of protein structures (Csermely, 2004, 2009; Kitano, 2004a, 2004b, 2007).

We summarize the possible mechanisms through which drugs can overcome cellular robustness in Fig. 11 (letters in the list correspond to symbols of the figure).

a. Drugs may activate a regulatory feedback helping disease-affected cells to return to the original equilibrium.
b. Drugs may activate a positive feedback and push disease-affected cells to a new state.

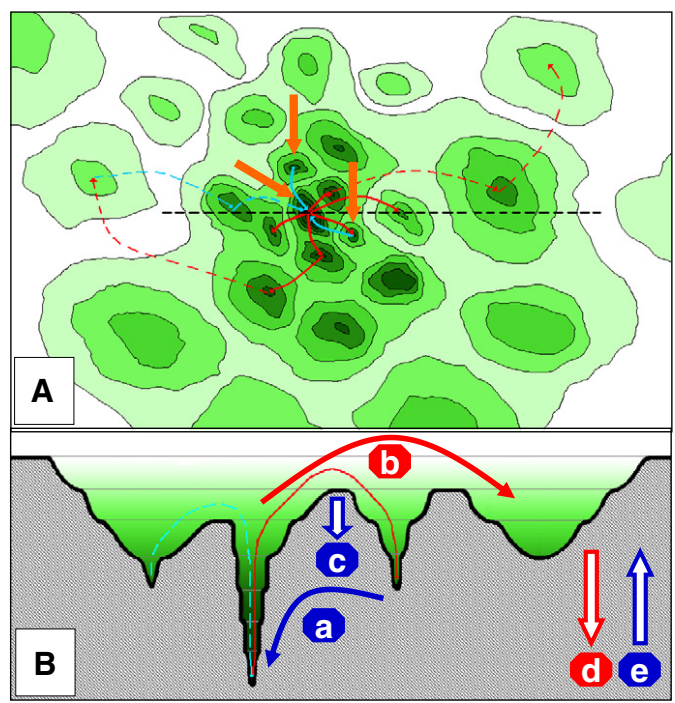

**Fig. 11.** Mechanisms of drug action changing cellular robustness. Panel **A** shows a 2-dimensional contour plot of the stability landscape of healthy and diseased phenotypes. Healthy states are represented by the central and the adjacent two minima marked with the large orange arrows, while all additional local minima are diseased states. Darker green colors refer to states with larger stability. Thin blue and red arrows mark shifts to healthy and diseased states, respectively. Dashed arrows refer to less probable changes. Panel **B** illustrates mechanisms of drug action on cellular robustness. The valleys and hills are a vertical representation of the stability-landscape shown on Panel **A** along the horizontal dashed black line. Blue symbols represent drug interactions with disease-prone or disease-affected cells, while red symbols refer to drug effects on cancer cells or parasites. (a) Counteracting regulatory feedback; (b) positive feedback pushing the diseased cell or parasite to another trajectory; (c) a transient decrease of a specific activation energy enabling a shift back to healthy state; (d) 'error-catastrophe': drug action diminishing many activation energies at the same time, causing cellular instability, which leads to cell death; (e) general increase in activation energies leading to the stabilization of healthy cells to prevent their shift to diseased phenotype.

c. Drugs may transiently lower a specific activation energy helping disease-affected cells to return to the healthy state.
d. Drugs may decrease activation energies and thus destabilize malignant or infectious cells causing an 'error catastrophe' and activating cell death.
e. Drugs may increase activation energies and thus stabilize healthy cells preventing their shift to the diseased phenotype (Csermely, 2004, 2009; Kitano, 2004a, 2004b, 2007).

If cellular robustness is conquered, critical transitions, i.e. large unexpected changes, may also occur. Critical transitions are often responsible for unexplained cases of excessive drug side-effects and toxicity. Lack of stabilizing negative feedbacks, excessive positive feedbacks, accumulating cascades may all lead to extreme events characterizing critical transitions (San Miguel et al., 2012).

Recently increasing attention focused on finding, predicting and influencing extreme events, i.e. outliers of common (e.g. scale-free) statistics, also called 'dragon kings' (Tajer & Poor, 2012; de S. Cavalcante et al., 2013). The detection of early warning signals of these critical transitions (such as a slower recovery after perturbations, increased self-similarity of the behavior, or increased occurrence of extreme behavior) was shown to characterize different complex systems, such as ecosystems, the market, climate change, or population of yeast cells (Scheffer et al., 2009; Farkas et al., 2011; Sornette & Osorio, 2011; Dai et al., 2012). Boettiger and Hastings (2012) emphasized the importance of using correct statistics in order not to 'over-examine' systems which are known to undergo critical transitions. Prediction and control of critical changes (delay/prevention in the case of normal cells and induction/acceleration in the case of malignant or infecting cells) may be an especially important area of future drug-related network studies.

The number of possible regulatory combinations for a given gene increases dramatically with an increase in input-complexity and network size. For example, with 100 genes and 3 inputs per gene there are a million input combinations for each gene in the network resulting in 10,600 different network wiring diagrams (Tegnér & Bjorkegren, 2007). The complexity of precise network perturbation models increases even more with system size. Therefore, it is not surprising that most studies of network dynamics described small networks with at most a few dozens of nodes. As an example of this, the Tide software analyzes the combined effects and optimal positions of drug-like inhibitors or activators using differential equations of reaction pathways up to 8 components (Schulz et al., 2009). Karlebach and Shamir (2010) presented an algorithm determining the smallest perturbations required for manipulating a network of 14 genes. Perturbations of Boolean networks, where nodes may only have an "on" or "off" mode, describe the dynamics of 20 to 50 nodes. These models often incorporate activating, inhibiting, or conditional edges, too (Huang, 2001; Shmulevich et al., 2002; Gong & Zhang, 2007; Abdi et al., 2008; Azuaje et al., 2010; Saadatpour et al., 2011; Wang & Albert, 2011; Garg et al., 2012). To help these studies a versatile, publicly available software library, BooleanNet (http://booleannet.googlecode.com) was developed by Albert et al. (2008). PATHLOGIC-S (http://sourceforge.net/projects/pathlogic/files/PATHLOGIC-S) offers a scalable Boolean framework for modeling cellular signaling (Fearnley & Nielsen, 2012).

Systems-level molecular networks have a size in the range of thousand to ten-thousand nodes. At this level of system complexity the optimal selection of the perturbation model becomes a key issue. At this system size the highly anisotropic perturbation propagation inside protein structures is usually neglected (we will detail the possibilities to construct atomic resolution interactomes in Section 4.1.6 on allo-network drugs; Nussinov et al., 2011). In current network perturbation models of larger systems delays, differences in individual dissipation patterns, effects of water or molecular crowding are also neglected (Antal et al., 2009).

We summarized an early promising approach of systems-level perturbation studies in Section 2.2.3 on reverse engineering. Here perturbations were assessed by systems-level mRNA expression profiles and the perturbed network was reconstructed from the output data (Liang et al., 1998a; Akutsu et al., 1999; Ideker et al., 2000; Kholodenko et al., 2002; Yeung et al., 2002; Segal et al., 2003; Tegnér et al., 2003; Friedman, 2004; Tegnér & Bjorkegren, 2007; Ahmed & Xing, 2009; Stokic et al., 2009; Marbach et al., 2010; Yip et al., 2010; Schaffter et al., 2011; Altay, 2012; Crombach et al., 2012; Kotera et al., 2012). Reverse engineering techniques were successfully applied to reconstruct drug-induced system perturbations (Gardner et al., 2003; di Bernardo et al., 2005; Chua & Roth, 2011).

Maslov and Ispolatov (2007) used the mass action law to calculate the effect of a two-fold increase in the expression of single protein on the free concentration of other proteins in the yeast interactome. Despite an exponential decay of changes, there were a few highly selective pathways, where concentration changes propagated to a larger distance (Maslov & Ispolatov, 2007). This and other models of network dynamics have been used in various publicly available algorithms including:

• the system dynamics modeling tool BIOCHAM using Boolean, differential, stochastic models and providing among others bifurcation diagrams (http://contraintes.inria.fr/biocham; Calzone et al., 2006);
• the random walk-based ITM-Probe, also available as a Cytoscape plug-in (http://www.ncbi.nlm.nih.gov/CBBresearch/Yu/mn/itm_probe/doc/cytoitmprobe.html; Stojmirović & Yu, 2009; Smoot et al., 2011);
• the mass action-based Cytoscape plug-in, PerturbationAnalyzer (http://chianti.ucsd.edu/cyto_web/plugins/displaypluginfo.php?name=PerturbationAnalyzer; Li et al., 2010a; Smoot et al., 2011);
• a user-friendly, Matlab-compatible, versatile network dynamics tool, Turbine supplying a communication vessels propagation model, but handling any user-defined dynamics, and enabling the user to simulate real world networks that include 1 million nodes and 10 million edges per GByte of free system memory, exporting and converting numerical data to a visual image using an inbuilt viewer function (www.linkgroup.hu/Turbine.php; Farkas et al., 2011);
• Conedy, a Python-interfaced C++ program capable to handle various dynamics including differential equations and oscillators (http://www.conedy.org; Rothkegel & Lehnertz, 2012).

Studying perturbations of larger networks Adilson Motter and colleagues developed an exciting model of compensatory perturbations showing that surprisingly, a debilitating effect can often be compensated by another inhibitory effect in a complex, cellular system (Motter et al., 2008; Motter, 2010; Cornelius et al., 2011). Perturbation dynamics of signaling networks was extensively analyzed including close to 10 thousand phosphorylation events in an experimental study of yeast cells (Bodenmiller et al., 2010). As we described in Section 2.2.3 on reverse engineering, perturbation studies are often used to reconstruct networks. As examples of this, the signaling network of T lymphocytes was reconstructed using single cell perturbations (Sachs et al., 2005), and the perturbations of 21 drug pairs were predicted from the reconstituted network of phospho-proteins and cell cycle markers of a human breast cancer cell line (Nelander et al., 2008). As another example, a perturbation amplitude scoring method was developed to test the biological impact of drug treatments, and was assessed using the transcriptome of colon cancer cells treated with the CDK cell cycle inhibitor, R547 (Martin et al., 2012).

Despite their complexity and robustness, cellular networks have their 'Achilles-heel'. Hitting it, a perturbation may cause dramatic changes in cell behavior. Stem cell reprogramming is a well-studied example of these network-reconfigurations (Huang et al., 2012a), where special bottleneck proteins may play a pivotal role (Buganim et al., 2012). As another example of 'streamlined' cellular responses, effects of multiple drug-combinations on protein levels can be quite accurately described by the linear superposition of drug-pair effects (Geva-Zatorsky et al., 2010).

Recent perturbation studies identified key nodes governing network dynamics. Central nodes, such as hubs, or inter-modular overlaps and bridges were shown to serve as highly efficient mediators of perturbations (Cornelius et al., 2011; Farkas et al., 2011). Network oscillations can be governed by a few central nodes forming a small network skeleton (Liao et al., 2011). Targets of viral proteins were shown to be major perturbators of human networks (de Chassey et al., 2008; Navratil et al., 2011). Perturbation mediators are often at cross-roads of cellular pathways. These key nodes bind multiple partners at shared binding sites. These shared binding sites can be identified as hot spot residues in protein structures (Ozbabacan et al., 2010). The fast-developing field of viral marketing identified influential spreaders of information at network cores and at other central network positions (Kitsak et al., 2010; Valente, 2012). Spreader proteins may be excellent targets of antiinfectious or anti-cancer therapies. Conversely, drugs against other diseases need to avoid these central proteins affecting a number of cellular functions. The identification of influential spreaders may provide important analogies of future drug target studies.

### 2.5.3. Network cooperation, spatial games

Spatial games, i.e. social dilemma games (such as the well known Prisoners' Dilemma, hawk–dove or ultimatum games) played between neighboring network nodes, provide a useful model of cooperation (Nowak, 2006). In a recent review Foster (2011) described the 'sociobiology of molecular systems' and provided convincing evidence how molecular networks determine social cooperation. We argue that cooperation of proteins and other macromolecules may offer an important description of cellular complexity. This view is based on the delicate dynamics of protein–protein interactions, which proceed via mutual selection of the binding-compatible conformations of the two protein partners. As the two proteins approach each other, they signal their status to the other via the hydrogen-bonded network of water molecules. Binding is achieved by a complex set of consecutive conformational adjustments. These concerted, conditional steps were called a 'protein dance', and can be perceived as rounds of a repeated game (Kovács et al., 2005; Csermely et al., 2010).

The stepwise encounter of protein molecules can be modeled as a series of rounds in common social dilemma games. In hawk–dove games the more rigid binding partner (corresponding to the drug) can be modeled as a hawk, while the more flexible binding partner (corresponding to the drug target) will be the dove. The hawk/dove encounter corresponds to an induced-fit-like scenario, where the conformational change of the dove is much larger than that of the hawk. The game is won by drug (hawk), since its enthalpy gain is not accompanied by an entropy cost. On the contrary, the flexible drug target loses several degrees of freedom during binding (Kovács et al., 2005; Chettaoui et al., 2007; Schuster et al., 2008; Antal et al., 2009; Csermely et al., 2010). In agreement with our previous description (Csermely et al., 2010) we note that induced-fit is conceived here as an extremity of an extended conformational selection model, where one of the partners (in this case the drug) is much more rigid than the other.

If we model drug binding with the ultimatum game, the drug and its target want to share the free energy decrease as a common resource. The drug proposes how to divide the sum between the two partners, and the target can either accept or reject this proposal, i.e. bind the drug or not (Kovács et al., 2005; Chettaoui et al., 2007; Schuster et al., 2008; Antal et al., 2009; Csermely et al., 2010).

Extending the above drug-binding scenario to the network level of the whole cell spatial game models are not only important to provide an estimate of systems-level cooperation, but are able to predict, which protein can most efficiently destroy the existing cooperation of the cell. This is a very helpful model of drug action in anti-infectious or anti-cancer therapies. Game models also identify those proteins, which are the most efficient to maintain cellular cooperation. This provides a useful model of drug efficiency in maintaining normal functions of diseased cells. Recently a versatile program, called NetworGame (www.linkgroup.hu/NetworGame.php) was made publicly available for simulating spatial games using any user-defined molecular networks and identifying the most influential nodes to establish, maintain or break cellular cooperation. Nodes having an exceptional influence in these cellular games may be promising targets of future drug development efforts (Farkas et al., 2011).

### 2.6. Limitations of network-related description and analysis methods

After completing a large inventory of network-related description and analysis methods here we list some of the major limitations of this approach.

- The first and foremost limitation of all network-related methods is data quality. Network description is a tool, which depends on the accuracy and coverage of input data. The entire dataset must have the same, well defined quality control. Currently we often lack these high-quality datasets especially on system dynamics (Henney & Superti-Furga, 2008; Prinz et al., 2011; Begley & Ellis, 2012; Landis et al., 2012).
- The definition of network nodes and edges often restricts network descriptions to well-defined connections. Hypergraph descriptions, which may overcome this problem are not wide-spread, and well documented yet. Despite the flexibility offered by hypergraphs, network descriptions often reduce the dimensions of the available information. When this is performed without a deep knowledge of the problem and the system, it may result in significant information loss. We described these difficulties of network construction in Sections 1.2 and 2.1.
- Visualization of networks improved over the years (see Table 1), but there is still room for development of 3D, large-capacity, zoom-in-type network visualization tools.
- Since the information spread is not only local in the cell, network analysis concentrating only local topological signatures, such as hubs or motifs may not use the full potency of network description. Definition of network modules or central nodes resulted in a large number of methods, but has reached a consensus neither in the applicability nor in overcoming the limitations of these methods. Additional studies on the identification of influential network nodes and edges are needed.
- Network-based prediction methods also need considerable improvement.
- Comparison of dynamically changing networks needs additional methods. Generally, the dynamics of complex systems is not adequately described by networks yet.

Applications of network description and analysis to molecular problems started less than 15 years ago. Therefore, several limitations arise from the current state-of-the-art of network related studies. Others may indeed pose limits for the use of network analysis. We need a number of well-based comparative studies in the future to understand the areas, where the use of network description and analysis gives the most efficient help in drug design. The following sections of our review try to help this clarification process.

## 3. The use of molecular networks in drug design

In this section we will describe molecular networks starting from networks of chemical substances, followed by protein structure networks (i.e. networks of amino acids forming 3D protein structures), protein–protein interaction networks, signaling networks, genetic interaction and chromatin networks (i.e. networks of chromatin segments forming the 3D structure of chromatin). We will conclude the section with the description of metabolic networks, i.e. networks of metabolites connected by enzyme reactions. The section will not give a detailed description of all studies on these networks, but will concentrate only on the most important aspects related to drug development.

Nodes of the networks above are connected either physically or conceptually. Chemical compound networks are often constructed by connecting two chemical compounds, if there is a chemical reaction to transform one of them to the other. This logic is very similar to that used in the construction of metabolic networks. In another form of chemical compound networks two drugs are considered similar, if they have a common binding protein. This is actually the inverse of drug target networks (where two drug targets are connected, if the same drug binds to them). Substrates and products also have a common binding protein, the enzyme, serving as the edge of metabolic networks. However, drug-related studies on metabolic networks often incorporate knowledge of protein–protein interaction and signaling networks. Therefore, we will summarize metabolic networks separately, at the end of the current section. Similarly, drug target networks often use the rich conceptual context of the drug development process. Therefore, we will re-assess the major features of drug target networks in Section 4.1.3. However, due to the unavoidable overlaps we encourage the Reader to compare the sections on chemical compound, metabolic and drug target networks.

### 3.1. Chemical compound networks

In this section we will summarize all networks which are related to chemical compounds: structural networks, reaction networks, and the large variety of chemical similarity networks. Of all these networks, especially the latter, chemical similarity networks can be used very well in lead optimization and selection of drug candidates. There is a very large variability in the names of these networks in the literature. Therefore, we selected the most discriminative name as the titles of sub-sections, and refer to some other network denotations in the text.

#### 3.1.1. Chemical structure networks

The structure of chemical compounds can be perceived as a network, where labeled (colored) nodes are the atoms constructing the molecule, and labeled (colored) edges are the covalent bonds binding the atoms together (Fig. 12). Chemical structure networks (also called chemical graphs) may use multiple edges representing multiple bonds. The core electron structure of the various atoms is often represented as a complete graph. Descriptors of this network structure, such as discrete invariants representing the chemical structure, connectivity indices, topological charge indices, electro-topological indices, shape indices and others are useful for quantitative structure/property and structure/activity (QSPR and QSAR) models (Bonchev & Buck, 2007; García-Domenech et al., 2008; Gonzalez-Diaz et al., 2010a). Molgen (http://molgen.de; Baricic & Mackov, 1995) and Modeslab (http://modeslab.com; Estrada & Uriarte, 2001) are widely used programs to draw and analyze chemical structure networks. SIMCOMP (http://www.genome.jp/tools/simcomp) and SUBCOMP (http://www.genome.jp/tools/subcomp) compare chemical structure networks and show the position of results in molecular pathways (Hattori et al., 2010).

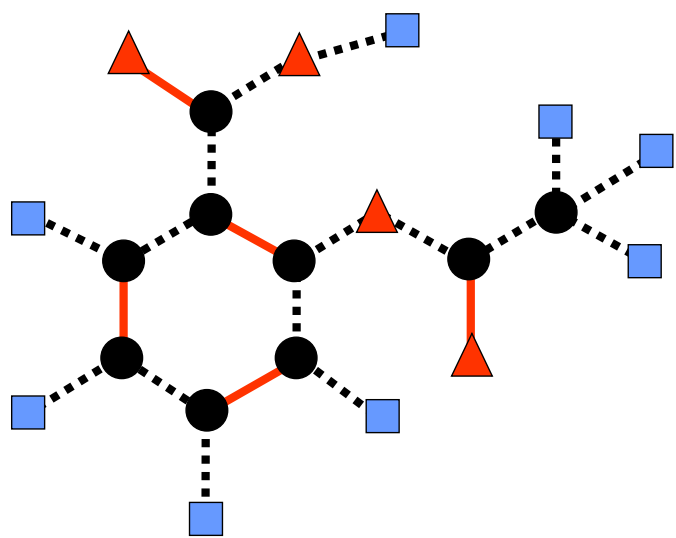

**Fig. 12.** Example of a chemical structure network. From the network point of view chemical structures are networks with differently labeled (colored) nodes representing different kinds of atoms and differently labeled (colored) edges related to different types of bonds. The chemical structure network representation of aspirin is shown. Black circles, red triangles and blue rectangles represent carbon, oxygen and hydrogen atoms, respectively. Dotted black edges stand for single bonds, while red solid edges represent double bonds.

#### 3.1.2. Chemical reaction networks

The mind-boggling set of 1060 chemical compounds that can be created by chemical reactions, defines the so-called chemical space (Kirkpatrick & Ellis, 2004). The size of drug-like chemical space is estimated to be larger than a million compounds (Drew et al., 2012). The increasing costs of experiments and the need for compounds with specific properties increased the efforts to apply new tools for chemical space discovery (Lipinski & Hopkins, 2004). Chemical reactions make the chemical space continuous. Therefore, their network representation serves as a promising tool. Nodes of chemical reaction networks are the chemical compounds and their edges are the reactions transforming them to one another (Christiansen, 1953; Temkin & Bonchev, 1992).

The chemical reaction network, comprising the whole synthetic knowledge of organic chemistry containing 7 million compounds in 2012, was first assembled by Fialkowski et al. (2005). Chemical reaction networks may only contain the participating compounds, or may be bipartite networks, where besides the participating compounds a different type of nodes represents the reactions (Fig. 13). The chemical reaction network is a small-world containing hubs, i.e. compounds, which can be formed and transformed to and from many other compounds. The chemical reaction network contains hubs. Importantly, hub compounds have a lower market price than chemicals involved in a low number of reactions. Moreover, hub molecules are more likely to be prepared via new methodologies, and may also be involved in the synthesis of many new compounds (Grzybowski et al., 2009). The chemical reaction network can be separated into a core, containing over 70% of the top 200 industrial chemicals, and a periphery, which has a tree-like structure, and can be easily synthesized from the core (Bishop et al., 2006). Chemical reaction networks offer help in the design of 'one-pot' reactions without the need for isolation, purification and characterization of intermediate structures, and without the production of much chemical waste. Gothard et al. (2012) used 8 filters of 86,000 chemical criteria to identify more than 1 million 'one-pot' reaction series. The number of possible synthetic pathways can be astronomical having 1019 routes of just 5 synthetic steps. Network analysis of Kowalik et al. (2012) identified optimal synthetic pathways of single and multiple-target syntheses using a simulated annealing-based network optimization. These optimizations help in the synthesis of drug candidate variants for lead selection.

#### 3.1.3. Similarity networks of chemical compounds: QSAR, chemoinformatics, chemical genomics

Molecular similarity can be viewed as the distance between molecules in a continuous high-dimensional space of numerical descriptors (Johnson & Maggiora, 1990; Bender & Glen, 2004; Eckert & Bajorath, 2007). This high dimensional similarity space is called the chemistry space, which constitutes an important part of chemoinformatics (Faulon & Bender, 2010; Krein & Sukumar, 2011; Varnek & Baskin, 2011). Nodes in similarity networks are most often chemical compounds, but may also be molecular fragments, or molecular scaffolds (Hu & Bajorath, 2010; Hu & Bajorath, 2011). Edge definition is a difficult task in similarity networks. Un-weighted networks can be constructed using a pre-determined similarity threshold, while the extent of similarity may also be used as edge-weight. From the large number of numerical descriptions of similarity listed in Table 5, we will first consider those networks, which are based on simple chemical similarity of the compounds involved using e.g. the Tanimoto-coefficient for the definition of edges (Rogers & Tanimoto, 1960; Tanaka et al., 2009; Bickerton et al., 2012). We will call these networks chemical similarity networks.

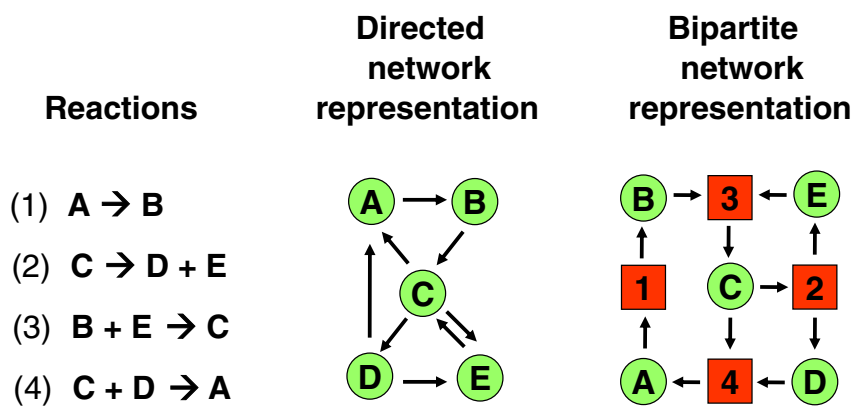

**Fig. 13.** Directed and bipartite network representations of chemical reaction networks. On the middle panel the 4 reactions of the left side of the figure are represented as a directed network of participating compounds. On the bipartite network of the right panel green circles represent chemicals, while red rectangles stand for reactions. Figures were adapted from Fialkowski et al. (2005) and from Grzybowski et al. (2009).

Chemical similarity networks are also small-worlds possessing hubs with a modular structure. Similarity hubs may be used as priority starting points in fragment-based drug design. If hubs become non-hits, many fragment-combinations can be excluded as candidates, under the assumption that molecules similar to non-hits are also non-hits. This strategy was shown to explore the chemistry space in much less trials than random selection or the selection of cluster centers (Tanaka et al., 2009). Well connected fragments can also be used in library design and in fragment-based database searches. The top 10% of most frequently occurring molecular segments accounts for the majority of overall fragment occurrences, thus, storing a relatively small number of fragments can cover a large portion of the searching space (Benz et al., 2008). Chemical similarity networks were shown to be a very useful description of the diversity and drug-likeness of bioactive compounds against various drug targets (Bickerton et al., 2012).

**Table 5**
Chemical compound similarity networks.

| Basis of chemical compound similarity | References |
|---|---|
| *Chemical compound similarity networks* | |
| Chemical similarity based on e.g. the Tanimoto-coefficient | Tanaka et al., 2009; Bickerton et al., 2012 |
| QSAR-related similarity networks (a freely available program to mine structure–activity and structure–selectivity relationship information in compound datasets, SARANEA) http://www.limes.uni-bonn.de/forschung/abteilungen/Bajorath/labwebsite/downloads/saranea/view) | Estrada et al., 2006; Gonzalez-Diaz & Prado-Prado, 2008; Hert et al., 2008; Prado-Prado et al., 2008; Wawer et al., 2008; Bajorath et al., 2009; Garcia et al., 2009;Prado-Prado et al., 2009; Gonzalez-Diaz et al., 2010a; Lounkine et al., 2010; Peltason et al., 2010; Prado-Prado et al., 2010; Wawer et al., 2010; Iyer et al., 2011a; Iyer et al., 2011b; Iyer et al., 2011c; Krein & Sukumar, 2011; Wawer & Bajorath, 2011a; Wawer & Bajorath, 2011b |
| BioAssay network: bioassay data of chemical compounds from PubChem | Zhang et al., 2011b |
| Similarity of protein binding sites | Paolini et al., 2006; Keiser et al., 2007; Hert et al., 2008; Park & Kim, 2008; Adams et al., 2009; Keiser et al., 2009; Hu et al., 2011 |
| Network of drug–receptor pairs with multitarget QSAR | Vina et al., 2009 |
| CARLSBAD: a Cytoscape plug-in for connecting common chemical patterns to biological targets via small molecules http://carlsbad.health.unm.edu | |
| Drug–target network combined with the chemical structure network of the drug and the protein structure network of its target giving quality-scores of drug–target networks | Riera-Fernandez et al., 2012 |
| Similarity of mRNA expression profiles extended with disease mRNA expression profiles: Connectivity Map http://www.broadinstitute.org/cmap | Lamb et al., 2006; Iorio et al., 2009; Huang et al., 2010a |
| Side-effect similarity of drugs | Campillos et al., 2008 |
| Protein–protein interaction network topology of the target neighborhood (a database of more than 700,000 chemicals, 30,000 proteins and their over 2 million interactions integrated to a human protein–protein interaction network having over 400,000 interactions, ChemProt: http://www.cbs.dtu.dk/services/ChemProt) | Hansen et al., 2009; Li et al., 2009a; Taboreau et al., 2011; Edberg et al., 2012 |
| *Integrated bio-entity relationship datasets and networks* | |
| • Structural similarity, QSAR, gene–disease interactions, biological processes, drug absorption, distribution, metabolism and excretion (ADME) data and toxicity mechanisms | Brennan et al., 2009 |
| • Integrated semantic network of chemogenomic repositories, Chem2Bio2RDF http://cheminfov.informatics.indiana.edu:8080 | Chen et al., 2010a |
| • Drug therapeutic and chemical similarity with protein–protein interaction network data: drugCIPHER | Zhao & Li, 2010 |
| • Protein–protein interactions, protein/gene regulations, protein–small molecule interactions, protein–Gene Ontology relationships, protein–pathway relationships and pathway–disease relationships: bio-entity network (IBN) | Bell et al., 2011 |
| • Phenotype/single-nucleotide polymorphism (SNP) associations, protein–protein interactions, disease–tissue, tissue–gene and drug–gene relationships: integrated Complex Traits Networks, iCTNet Cytoscape plug-in, http://flux.cs.queensu.ca/ictnet | Wang et al., 2011b |
| • Protein–protein interactions, protein–small molecule interactions, associations of interactions with pathways, species, diseases and Gene Ontology terms with the user-selected integration of manually curated and/or automatically extracted data: integrated molecular interaction database, IMID, http://integrativebiology.org | Balaji et al., 2012 |

Molecular similarity is particularly important in medicinal chemistry. This is due to the 'similar property principle' which states that similar molecules have similar biological activity (Johnson & Maggiora, 1990). This principle also serves as a basis of most quantitative structure–activity relationship (QSAR) modeling methods (note that we will use the term, QSAR to describe structure activity relationships in general). However, the relationship between chemical similarity and biological activity is not always straightforward (Martin et al., 2002), which necessitates the use of sophisticated approaches in drug design, such as the multi-component similarity networks listed in Table 5.

In QSAR-related similarity networks (also called network-like similarity graphs) nodes are often color-coded according to their biological action potency value ($pIC_{50}$ or $pK_i$), and scaled in size based on their contribution to the QSAR landscape features such as 'activity cliffs' or smooth regions. Near activity cliffs, small changes in molecular structure induce large changes in biological activity, while in smooth regions of the QSAR landscape changes in chemical structure only result in small or gradual changes in activity. QSAR-related similarity networks contain more information than chemical similarity networks. In contrast, chemical similarity networks were found to be topologically robust to the methods for representing and comparing chemical information. The choice of molecular representation (molecular descriptors) may change the interpretation of QSAR landscapes, where the appropriate selection of similarity (distance) cut-offs was proven to be crucial. If the cut-off value was too low, there were many isolated nodes; if the cut-off was too high, QSAR-related similarity networks became overcrowded and less useful for predictions. QSAR-related similarity networks are small worlds and contain hubs. Subsets of compounds related by different local QSARs are often organized in small communities (also called as clusters). High centrality nodes form 'chemical bridges' between various compound communities providing important QSAR information. These nodes can be used for 'hopping' between sub-networks having different chemical characteristics. Searching for nodes with high centrality and a closer look at their properties may contribute to the discovery of new drug candidates and uncover new directions for mechanistic-, scaffold- or target-hopping approaches (Gonzalez-Diaz & Prado-Prado, 2008; Hert et al., 2008; Prado-Prado et al., 2008, 2009, 2010; Wawer et al., 2008, 2010; Bajorath et al., 2009; Gonzalez-Diaz et al., 2010a; Peltason et al., 2010; Iyer et al., 2011a, 2011b, 2011c; Krein & Sukumar, 2011; Wawer & Bajorath, 2011a, 2011b). SARANEA (http://www.limes.uni-bonn.de/forschung/abteilungen/Bajorath/labwebsite/downloads/saranea/view) is a freely available program to mine structure–activity and structure–selectivity relationship information in compound datasets (Lounkine et al., 2010). Recently, group-based QSAR models were introduced, providing promising approaches in multi-target drug design (Ajmani & Kulkarni, 2012). Methods for the systematic comparison of molecular descriptors, such as that introduced by Bender et al. (2009), are very useful to guide future work—including network-related applications.

Dehmer et al. (2009) showed the usefulness of network complexity analysis in the determination of topological descriptor uniqueness. We demonstrate the usefulness of QSAR-related similarity network descriptors on chirality, since the different enantiomers of drug candidates can exhibit large differences in activity. Using complex networks, Garcia et al. (2009) investigated the drug–drug similarity relationship of more than 1600 experimentally unexplored, chiral 3-hydroxy-3-methyl-glutaryl coenzyme A inhibitor derivatives with a potential to lower serum cholesterol preventing cardiovascular disease. Inclusion of chirality in network description may guide synthesis efforts towards new chiral derivatives of potentially high activity. QSAR-related similarity networks including chiral information of G protein-coupled receptor ligands identified that opposing chiralities induced alterations in molecular mechanism (Iyer et al., 2011b).

Another important application of QSAR-related similarity networks is the molecular fragment network of human serum albumin binding defined by Estrada et al. (2006). The identification of polar 'emphatic' fragments anchoring chemicals to serum albumin and hydrophobic fragments determining albumin binding was an important step in network-related prediction of bioavailability.

Interestingly, a similar growth mechanism was found in the evolution of chemical reaction networks (Fialkowski et al., 2005; Grzybowski et al., 2009) and QSAR-related similarity networks (Iyer et al., 2011a). Growth was predominantly observed around a few hubs that emerged early in the growth process, and did not reach whole segments of the network until a very late phase of development. Analyzing evolving datasets can be very important to identify over-sampled regions containing redundant compound structure information, or yet unexplored regions in the chemical reaction network or QSAR-related similarity network.

The 'similar property principle' stating that similar molecules have similar biological activity (Johnson & Maggiora, 1990) can be reversed, and used for the construction of similarity networks, which means that compounds having a similar biological action are similar. Compounds or compound scaffolds can be connected using the similarity of their protein binding sites, as well as of the protein domains or the entire proteins harboring these sites. The emerging network defined the 'pharmacological space'. Hub ligands of this network were bridges between different ligand clusters. The network representation proved to be useful for identifying drug chemotypes, and for the probabilistic modeling of yet undiscovered biological effects of chemical compounds (Paolini et al., 2006; Keiser et al., 2007, 2009; Yildirim et al., 2007; Hert et al., 2008; Park & Kim, 2008; Yamanishi et al., 2008, 2011; Adams et al., 2009; Bleakley & Yamanishi, 2009; Hu et al., 2011; Tabei et al., 2012). Using the above datasets He et al. (2010) encoded chemical compounds with functional groups and proteins with biological features of 4 major drug target classes, and worked out a prediction of drug–target interactions using the maximum relevance minimum redundancy method. Riera-Fernandez et al. (2012) gave quality-scores of drug–target network edges using the combined information of the chemical structure network of the drug and the protein structure network of its target.

An important approach to compare the similarity of chemical compounds is to construct the network of drug–therapy interactions, where drugs are connected, if they are used in the same therapy class of the five hierarchical Anatomical Therapeutic Chemical (ATC) classification levels. Average paths in this drug–therapy network are shorter than 3 steps. Distant therapies are separated by a surprisingly low number of chemical compounds. Inter-modular, bridging and otherwise central drugs in the drug–therapy network may have more indications than currently known, thus drug–therapy network data may be useful for drug-repositioning (Nacher & Schwartz, 2008). Text mining may be an important method to enrich drug–therapy networks in the future (Ruan et al., 2004).

mRNA expression patterns were the first system-wide descriptors of drug effects enabling target clustering, target identification, and prediction of the mechanism of action of new compounds (Marton et al., 1998; Hughes et al., 2000; Lamb et al., 2006; Iorio et al., 2009; Chua & Roth, 2011). Huang et al. (2010a) connected mRNA expression profiles with a disease diagnosis database. Using a Bayesian learning algorithm they could query drug-treatment related mRNA expression profile and decipher drug similarity not only to each other, but also to specific disease and disease classes.

As we will discuss in detail in Sections 4.1.5 and 4.3.5, drugs seldom have a single effect. Based on this, the chemical similarity of drugs may be derived from their side-effects, describing a broader repertoire of drug action than the effect related to the original target. Campillos et al. (2008) connected drugs sharing a certain degree of side-effect similarity. This network uncovered shared targets of unrelated drugs and forms an important network method for drug repositioning.

Going one level further in systems-level abstraction, similarity of compounds can be measured by comparing the topological similarity of their target neighborhoods in protein–protein interaction networks (Hansen et al., 2009; Edberg et al., 2012). Li et al. (2009a) concluded from the investigation of an Alzheimer's Disease-related dataset, that the combination of curated drug–target databases and literature mining data outperformed both datasets when used alone. Systems-level inquiries are helped by ChemProt (http://www.cbs.dtu.dk/services/ChemProt), a database of more than 700,000 chemicals, 30,000 proteins and their over 2 million interactions integrated to a human protein–protein interaction network having over 400,000 interactions (Taboureau et al., 2011).

Baggs et al. (2010) encouraged the inclusion of network readouts (like transcriptome, proteome, phosphoproteome, metabolome and epigenetic system-wide datasets) in QSAR methods leading to QNSAR (quantitative network structure–activity relationships). In agreement with this suggestion, in recent years an increasing number of complex databases were published, where network reconstitution was used to predict biologically meaningful clusters of datasets, novel drug-candidate molecules, new drug applications, unexpected drug–drug interactions, drug side-effects and toxicity. We list these datasets in Table 5. As noted by Vina et al. (2009), increased reliance on indirect data similarities may compromise accuracy, but may also enable the exploration of those segments of the data association landscape, where no direct alignments were available. The aggregative assessment of multiple (and system-wide) datasets helps to pick up those similarities, which are the most relevant despite the many uncertainties of the individual data or their associations.

Utilizing the rich repertoire of the assessment of network topology and dynamics, listed in Section 2, will be helpful for predicting future directions in compound optimization, or redirecting research efforts to unexplored or more fruitful regions of chemical space. Moreover, detailed analysis of complex similarity networks is useful for predicting new targets of existing drugs, i.e. multi-target drug identification and drug repositioning. Finally, assessment of similarity networks can be used as an efficient predictor of drug specificity, efficacy, ADME, resistance, side-effects, drug–drug interactions and toxicity.

### 3.2. Protein structure networks

Proteins are the major targets of drug action, and therefore the description of their structure and dynamics has a crucial importance in the determination of drug binding sites, as well as in prediction of drug effects at the sub-molecular level. In this section we will show how protein structure networks help the characterization of disease-related proteins, the understanding of drug action mechanisms and drug targeting.

### 3.2.1. Definition and key residues of protein structure networks

In most protein structure network representations (also called amino acid networks, residue interaction networks, or protein meta-structures) nodes are the amino acid side-chains. Though occasionally protein structure network nodes are defined as the atoms of the protein, the side-chain representation is justified by the concerted movement of side-chain atoms. Edges of protein structure networks are defined using the physical distance between amino acid side-chains. Distances are usually measured between Cα or Cβ atoms, but in some representations the centers of mass of the side chains are calculated, and distances are measured between them. Edges of unweighted protein structure networks connect amino acids having a distance below a cut-off distance, which is usually between 4 and 8.5 Å (Artymiuk et al., 1990; Kannan & Vishveshwara, 1999; Greene & Higman, 2003; Bagler & Sinha, 2005; Böde et al., 2007; Krishnan et al., 2008; Vishveshwara et al., 2009; Doncheva et al., 2011, 2012b; Csermely et al., 2012; Di Paola et al., in press). A detailed study compared the effect of various Cα–Cα contact assessments, such as the atom distance criteria, the isotropic sphere chain and the anisotropic ellipsoid side-chain models, as well as of the selection of various cut-off distances. The study showed that the atom distance criteria model was the most accurate description having a moderate computational cost. The best amino acid pair specific cut-off distances varied between 3.9 and 6.5 Å (Sun & He, 2011). In protein structure networks with weighted edges, edge weight is usually inversely proportional to the distance between the two amino acid side-chains (Artymiuk et al., 1990; Kannan & Vishveshwara, 1999; Greene & Higman, 2003; Bagler & Sinha, 2005; Böde et al., 2007; Krishnan et al., 2008; Vishveshwara et al., 2009; Doncheva et al., 2011, 2012b; Csermely et al., 2012; Di Paola et al., in press).

Web-servers have been established to convert Protein Data Bank 3D protein structure files into protein structure networks, and to provide their network analysis. The RING server (http://protein.bio.unipd.it/ring) gives a set of physico-chemically validated amino acid contacts (Martin et al., 2011), and imports it to the widely used Cytoscape platform (Smoot et al., 2011) enabling their network analysis using the tool-inventory described in Section 2. Recently a specific, Cytoscape-linked (Smoot et al., 2011) tool-kit for protein structure network assessment, RINalyzer (http://www.rinalyzer.de) was published. The program is complemented by a protein structure determination module, called RINerator (http://rinalizer.de/rindata.php), which determines the protein structure networks and stores pre-determined protein structure networks of Protein Data Bank 3D protein structure files. The RINalyzer program was also linked to the NetworkAnalyzer software (http://med.bioinf.mpi-inf.mpg.de/netanalyzer; Assenov et al., 2008) allowing the comparison of protein structure networks and the extension of their analysis to protein–protein interaction networks (Doncheva et al., 2011, 2012b).

Protein structure networks are "small worlds". This is very important for the fast transmission of drug-induced conformational changes, since in the small-world of protein structure networks all amino acids can communicate with each other by taking only a few steps. Path-length analysis of individual amino acid side-chains was shown to be effective in predicting whether the protein, or its segment, is disordered. In protein structure networks we may find considerably less large hubs than in other networks. However, the existing smaller hubs still play an important role in protein structures, since these 'micro-hubs' were shown to increase the thermodynamic stability of proteins (Kannan & Vishveshwara, 1999; Greene & Higman, 2003; Atilgan et al., 2004; Bagler & Sinha, 2005; Brinda & Vishveshwara, 2005; Del Sol et al., 2006, 2007; Alves & Martinez, 2007; Krishnan et al., 2008; Konrat, 2009; Morita & Takano, 2009; Estrada, 2010; Csermely et al., 2012). Protein structure networks possess a rich club structure with the exception of membrane proteins, where hubs form disconnected, multiple clusters (Pabuwal & Li, 2009).

Protein structure networks have modules, which often encode protein domains (Xu et al., 2000; Guo et al., 2003; Delvenne et al., 2010; Delmotte et al., 2011; Szalay-Bekő et al., 2012). High-centrality segments of protein structure networks (i.e. hubs, or nodes with high closeness or betweenness centralities) having a low clustering coefficient participate in hem-binding (Liu & Hu, 2011). High-centrality, inter-modular bridges play a key role in the transmission of allosteric changes as we will describe in the next section.

Evolutionary conservation patterns of amino acids in related protein structures identified protein sectors (Halabi et al., 2009; McLaughlin et al., 2012). A similar concept has been published by Jeon et al. (2011), who determined that co-evolving amino acid pairs are often clustered in flexible protein regions. Protein sectors are sparse networks of amino acids spanning a large segment of the protein. Protein sectors are collective systems operating rather independently from each other. Segments of protein sectors are correlated with protein movements related to enzyme catalysis, and sector-connected surface sites are often places of allosteric regulation (Reynolds et al., 2011).

### 3.2.2. Key network residues determining protein dynamics

Understanding protein dynamics is a key step in the prediction of drug-induced changes and the identification of novel types of drug binding sites. Several questions related to protein dynamics, such as the mechanism of allosteric changes gained much attention in the last century (Fischer, 1894; Koshland, 1958; Straub & Szabolcsi, 1964; Závodszky et al., 1966; Tsai et al., 1999; Goodey & Benkovic, 2008; Csermely et al., 2010; Szilágyi et al., in press), but have not been completely elucidated yet.

Our current understanding indicates that an allosteric change extends across a spectrum: at one end there is a switch-type conformational change, where signaling appears focused on a small number of amino acids (Fig. 14A). At the other end, allosteric signaling may involve a large number of amino acids. In both cases allosteric signals propagate through multiple trajectories, with different distributions and weights. Pathways often converge at inter-domain boundaries (Fig. 14B). While protein segments involved in switch-type allosteric changes may be more rigid, protein segments harboring multiple trajectories may be more flexible. Convergence points in these latter proteins may mark the more flexible inter-domain regions. Switch-type mechanism is typical of multidomain proteins, and is often expressed in all-or-none observable consequences.

Disordered protein regions serve the need for large flexibility, and therefore are often used, especially in the human proteome (Csermely et al., 2012; Tompa, 2012). Protein structure network description may be a useful method to describe the complexity of protein structures, which are neither very rigid, nor very flexible. However, protein structure networks may not adequately describe the dynamics of very rigid and extremely flexible (e.g. disordered) protein structures (Szilágyi et al., in press).

Hinges connecting relatively rigid protein segments often play a decisive role in switch-type changes. Hinges may be co-localized with independent dynamic segments, which are situated in the stiffest parts of the protein, and harbor spatially localized vibrations of non-linear origin, like those of discrete breathers. Independent dynamic segments exchange their energy largely via a direct energy transfer, which is in agreement with a switch-type behavior (Daily et al., 2008; Piazza & Sanejouand, 2008, 2009; Csermely et al., 2010, 2012).

In contrast, in allosteric systems, where signaling involves a large number of amino acids, signals propagate using multiple trajectories (Fig. 14B). These multiple trajectories often converge at inter-modular residues of protein structure networks (Pan et al., 2000; Chennubhotla & Bahar, 2006; Ghosh & Vishveshwara, 2007, 2008; Tang et al., 2007; Daily et al., 2008; Sethi et al., 2009; Tehver et al., 2009; Vishveshwara et al., 2009; Csermely et al., 2012; Gasper et al., 2012).

A given protein may have a spectrum of the above mechanisms for the propagation of conformational changes. In agreement with this behavior, discrete breathers were shown to be present at the interface between monoatomic and diatomic granular chain models (Hoogeboom et al., 2010). If certain protein segments become more rigid, the mechanism may shift towards the first, switch-type signal transduction mechanism. This can be conceptualized as the propagation of a 'rigidity-front', which we recently proposed as a mechanism of allosteric signaling (Csermely et al., 2012). Panel C of Fig. 14 shows an illustrative mechanism of rigidity front propagation. Consecutive 'rigidization' of protein segments both induces similar changes in the neighboring segment, and accelerates the propagation of the allosteric change within the rigid segment. Rigidity front propagation may use sequential energy transfers (illustrated by the violet arrows; Piazza & Sanejouand, 2009; Csermely et al., 2010), and thus may increase the speed of the allosteric change (Csermely et al., 2012). The rigidity front propagation model combines elements of 'rigidity propagation' (Jacobs et al., 2001, 2003; Rader & Brown, 2011) with the 'frustration front' concept of Zhuravlev and Papoian (2010), with the dynamic pre-stress of proteins as large as a few 100 pN (Edwards et al., 2012), and is in agreement with the recent proposal of Dixit and Verkhivker (2012) suggesting an interactive network of minimally frustrated (rigid) anchor sites (hot spots) and locally frustrated (flexible) proximal recognition sites to play a key role in allosteric signaling. We will describe the use of allosteric signal propagation mechanisms to design allo-network drugs (Nussinov et al., 2011) in Section 4.1.6.

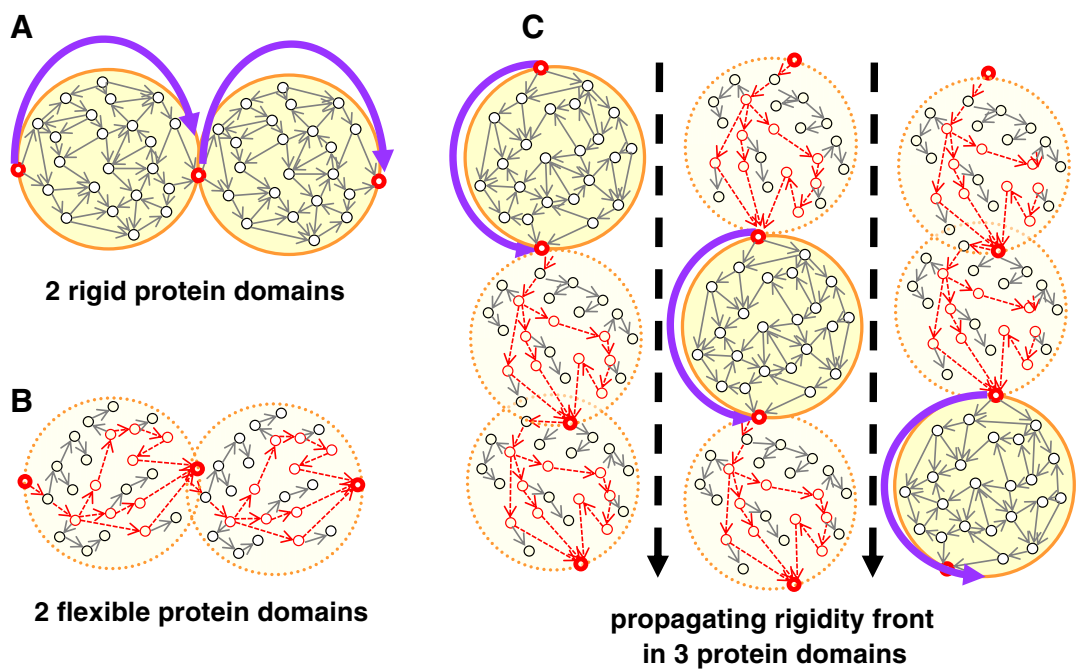

**Fig. 14.** Saltatoric signal transduction along a propagating rigidity-front: a possible mechanism of allosteric action in protein structures. Panel **A** shows two rigid modules of protein structure networks (corresponding to protein segments or domains). Such modules have little overlap, behave like billiard balls, and transmit signals 'instantaneously' (illustrated with the violet arrows). Panel **B** shows two flexible modules. These modules have a larger overlap, and transmit signals via a slower mechanism using multiple trajectories, which converge at key, bridging amino acids situated in modular boundaries. Panel **C** combines rigid and flexible modules in a hypothetical model of rigidity front propagation of the allosteric conformational change. In the 3 snapshots of this illustration of protein dynamics (organized from left to right) the 3 protein segments become rigid from top to bottom. Consecutive 'rigidization' of protein segments both induces similar changes in the neighboring segment, and accelerates the propagation of the allosteric change within the rigid segment. Rigidity front propagation may use sequential energy transfers (illustrated by the violet arrows), and may increase the speed of the allosteric change approaching that of an instantaneous process (Piazza & Sanejouand, 2009; Csermely et al., 2010, 2012).

Protein structure networks may be efficiently used to identify key amino acids involved in intra-protein signal transmission. In these studies topological network analysis was often combined with the assessment of evolutionary conservation, elastic network models and/or normal mode analysis. Inter-modular nodes, hinges, loops and hubs were particularly important in information transmission (Chennubhotla & Bahar, 2006, 2007; Zheng et al., 2007; Chennubhotla et al., 2008; Tehver et al., 2009; Liu & Bahar, 2010; Liu et al., 2010a; Park & Kim, 2011; Su et al., 2011; Dixit & Verkhivker, 2012; Ma et al., 2012a; Pandini et al., 2012). The examination of a hierarchical representation of protein structure networks showed a key function of top level, 'superhubs' in allosteric signaling (Ma et al., 2012a). xPyder provides an interface between the widely employed molecular graphics system, PyMOL and the analysis of dynamical cross-correlation matrices (http://linux.btbs.unimib.it/xpyder; Pasi et al., 2012).

The incorporation of novel network centrality measures (described in Section 2.3.3) and network dynamics (described in Section 2.5) will enrich our knowledge of the mechanism of conformational changes (including allosterism).

### 3.2.3. Disease-associated nodes of protein structure networks

Proteins related to more frequently occurring diseases tend to be longer than average (Lopez-Bigas et al., 2005). Disease-related proteins have a smaller 'designability'; that is, their folds can be built up from fewer variants than the average. In other words: disease-related proteins have a constrained structure, which may explain, why they have debilitating mutations (Wong & Frishman, 2006). Disease-associated mutations (single-nucleotide polymorphisms) often occur at sites having a high local or global centrality in the protein structure network, and are enriched by 3-fold at the interaction interfaces of proteins associated with the disorder (Akula et al., 2011; Li et al., 2011b; Wang et al., 2012b). Recently, a machine learning method has been developed to predict the disease-association of a single-nucleotide polymorphism using the network neighborhood of the mutation site (Li et al., 2011b).

### 3.2.4. Prediction of hot spots and drug binding sites using protein structure networks

Key functional residues are very useful in the identification of drug binding sites as we will discuss in Section 4.2. Usual drug binding sites are cavity-type, and overlap with substrate or allosteric ligand binding. High centrality residues of protein structure networks were shown to participate in ligand binding (Liu & Hu, 2011). Protein structure network position-based scores improved the rigid-body docking algorithm of pyDock (Pons et al., 2011). Protein structure comparison can also be used for the identification of chemical scaffolds of potential drug candidates (Konrat, 2009).

Binding sites of edgetic drugs modifying protein–protein interactions (see Section 4.1.2) are large and flat, and have been considered as non-druggable for a long time. Hot spots are those residues of these alternative drug binding sites, which provide a key contribution (>2 kcal/mol) to the decrease in binding free energy. Hot spots tend to cluster in tightly packed, relatively rigid hydrophobic regions of the protein–protein interface also called hot regions. Hot spots and hot regions are very helpful; they aid drug design, since 1.) they constitute small focal points of drug binding, which can be predicted within the large and flat binding-interface; 2.) these focal points are relatively rigid, helping rigid docking and molecular dynamics simulations (Clackson & Wells, 1995; Bogan & Thorn, 1998; Keskin et al., 2005, 2007; Ozbabacan et al., 2010). Hot spots can be predicted as central nodes of protein structure networks (del Sol & O'Meara, 2005; Liu & Hu, 2011; Grosdidier & Fernández-Recio, 2012).

The elastic network model-guided molecular dynamics simulation of Isin et al. (2012) showed that different ligands of the β2-adrenergic receptor prefer different predicted conformers of the receptor. This model predicted a novel allosteric binding site for larger drugs, such as salmeterol. Since different conformations participate in different metabolic and signaling pathways, such conformational modeling will be a powerful tool in the determination of novel drug binding sites and in the analysis of refined drug action mechanisms.

Despite these advances, the use of the predictive power of protein structure networks is surprisingly low in the determination of drug binding sites. We believe that the arsenal of network analytical and network dynamics methods we listed in Sections 2.3 and 2.5 and their application to protein structure networks will provide a much greater help in the identification of drug binding sites in the future.

### 3.3. Protein–protein interaction networks (network proteomics)

Protein–protein interaction networks are one of the most promising network types to predict drug action or identify new drug target candidates. In this section we will summarize the major properties of protein–protein interaction networks and will assess their use in the characterization and prediction of disease-related proteins and drug targets.

#### 3.3.1. Definition and general
properties of protein–protein interaction networks

Protein–protein interaction networks (PPI-networks) are often called interactomes—especially if they cover genome-wide data. Nodes of protein–protein interaction networks are proteins, and network edges are their direct, physical interactions. Protein–protein interaction networks are probability-type networks; that is, the edge weights reflect the probability of the actual interaction. Interactome edge weights are often calculated as confidence scores. Interaction probability includes protein abundance, interaction affinity, and also co-expression levels, co-localization in subcellular compartments, etc. (De Las Rivas & Fontanillo, 2010; Jessulat et al., 2011; Sardiu & Washburn, 2011; Seebacher & Gavin, 2011). Table 6 summarizes a number of major protein–protein interaction datasets concentrating on publicly available, human interactome data. There are several types of protein–protein interaction networks, which we list below.

- Even though protein–protein interaction datasets usually cover multiple species, the derived networks, that is, the interactomes are usually species-specific. Interactome subnetworks may be restricted to cell type, to cellular sub-compartment, or to certain temporal segments of cellular life, such as a part of the cell cycle, cell differentiation, malignant transformation, etc. These specializations may be direct, where the interactions of proteins are experimentally measured in the given species, cell type, cellular compartment, or

**Table 6**
Protein–protein interaction network resources.

| Name | Content | Website | References |
|---|---|---|---|
| 3did | Domain–domain interaction with 3D data | http://3did.irbbarcelona.org | Stein et al., 2011 |
| APID | Interactome exploration | http://bioinfow.dep.usal.es/apid/index.htm | Prieto & De Las Rivas, 2006 |
| atBioNet | Integration of 7 interactomes, protein complex identification | http://www.fda.gov/ScienceResearch/BioinformaticsTools/ucm285284.htm | Ding et al., 2012 |
| BioGRID | Integrated protein–protein interaction data | http://thebiogrid.org | Stark et al., 2011 |
| BioProfiling | Inference of network data from expression patterns | http://www.bioprofiling.de | Antonov et al., 2009 |
| DIP | Experimental protein–protein interaction data | http://dip.doe-mbi.ucla.edu | Salwinski et al., 2004 |
| DomainGraph | Cytoscape plug-in for domain–domain interaction analysis | http://domaingraph.bioinf.mpi-inf.mpg.de | Emig et al., 2008 |
| DOMINE | Domain–domain interaction data | http://domine.utdallas.edu | Yellaboina et al., 2011 |
| Estrella | Detection of mutually exclusive protein–protein interactions | http://bl210.caspur.it/ESTRELLA/help.php | Sanchez Claros & Tramontano, 2012 |
| HAPPI | Human protein–protein interaction data | http://discern.uits.iu.edu:8340/HAPPI | Chen et al., 2009c |
| HINT | High quality human protein–protein interaction data | http://hint.yulab.org | Das & Yu, 2012 |
| HPRD | Human protein–protein interaction data | http://www.hprd.org | Goel et al., 2012 |
| Hubba | Identification of hubs (potentially essential proteins) | http://hub.iis.sinica.edu.tw/Hubba | Lin et al., 2008 |
| IntAct | Curated protein–protein interaction data | http://www.ebi.ac.uk/intact/main.xhtml | Kerrien et al., 2012 |
| IntNetDB | Human protein–protein interaction data | http://hanlab.genetics.ac.cn/sys | Xia et al., 2006 |
| IRView | Protein interacting regions | http://ir.hgc.jp | Fujimori et al., 2012 |
| MiMI | Protein interaction information | http://mimi.ncibi.org | Gao et al., 2009 |
| MINT | Protein–protein interactions in refereed journals | http://mint.bio.uniroma2.it/mint | Licata et al., 2012 |
| NeAT (Network Analysis Tools) | Interactome analysis | http://rsat.ulb.ac.be/neat/ | Brohee et al., 2008 |
| NetAligner | Interactome comparison | http://netaligner.irbbarcelona.org | Pache et al., 2012 |
| PanGIA | A Cytoscape plug-in for integration of physical and genetic interactions into hierarchical module maps | http://prosecco.ucsd.edu/PanGIA | Srivas et al., 2011 |
| Pathwaylinker | Combines protein–protein interaction and signaling data | http://PathwayLinker.org | Farkas et al., 2012 |
| PINA | Interactome analysis | http://cbg.garvan.unsw.edu.au/pina | Wu et al., 2009b; Cowley et al., 2012 |
| PIPs | Human protein–protein interaction prediction | http://www.compbio.dundee.ac.uk/www-pips | McDowall et al., 2009 |
| PPISearch | Search of homologous protein–protein interactions across many species | http://gemdock.life.nctu.edu.tw/ppisearch | Chen et al., 2009d |
| STRING | Integrated protein–protein interaction data | http://string.embl.de | Szklarczyk et al., 2011 |
| UniHI | Integrated protein–protein interaction and drug target data | http://www.unihi.org | Chaurasia & Futschik, 2012 |

The table is focused on recently available public databases or web-servers applicable to human protein–protein interaction data and/or to drug design. Network visualization tools were listed in Table 1. A collection of protein–protein interaction network analysis web-tools can be found in recent reviews (Ma'ayan, 2008; Moschopoulos et al., 2011; Ma & Gao, 2012; Sanz-Pamplona et al., 2012). The Reader may find a more extensive list of web-sites in recent collections (http://ppi.fli-leibniz.de; Seebacher & Gavin, 2011).

condition. In many cases the specializations are indirect, where the presence of the actual proteins and/or the intensity of the protein–protein interactions are estimated from mRNA expression levels. Disease-specific or drug treatment-related interactomes hold promise for future drug development efforts (De Las Rivas & Fontanillo, 2010; Jessulat et al., 2011; Sardiu & Washburn, 2011; Seebacher & Gavin, 2011).

• Protein–protein interaction networks may be refined to networks of interacting protein domains, called domain networks or DDI-networks. Domain networks can be a better representation of drug action, deciphering domain-specific inhibition or activation (Fig. 15). Current lists of possible domain–domain interactions predict millions of novel, potential protein–protein interactions. However, not all domain–domain interactions may occur in the cellular context due to hindrances, binding competition or subcellular localization. Domain–domain interactions and their networks were used both to score protein–protein interactions (bottom–up approach) and to predict domain composition and interactions from interactome data (top–down approach) (Deng et al., 2002; Ng et al., 2003; Moon et al., 2005; Santonico et al., 2005; Emig et al., 2008; Prieto & De Las Rivas, 2010; Stein et al., 2011; Yellaboina et al., 2011).

• Atomic resolution interactomes expand protein–protein interaction networks with the protein structure networks of each interacting node aiming to construct the 3D structure of the whole interactome, and discriminating between parallel and sequential interactions (Kim et al., 2006; Prieto & De Las Rivas, 2010; Bhardwaj et al., 2011; Clarke et al., 2012; Pache & Aloy, 2012; Sanchez Claros & Tramontano, 2012). It is important to note that atomic level interactomes will never reach the real 3D complexity of the cell, since protein–protein interactions are probabilistic, reflecting an average of the possible interactions.

Protein–protein interaction data can be obtained using various high-throughput methods (such as protein fragment complementation assays, or affinity purification combined with mass spectrometry), text mining or prediction techniques. For details on the increasing number of methodologies the Reader is referred to recent reviews on the subject (Chautard et al., 2009; De Las Rivas & Fontanillo, 2010, 2012; Jessulat et al., 2011; Sardiu & Washburn, 2011; Seebacher & Gavin, 2011; Gonzalez & Kann, 2012). Prediction methods were also summarized in Sections 1.3.3 and 2.2.2, as well as in Tables 2 and 3. Data-quality is a major problem of interactomes. Sampling bias, missing interactions and false positives are all important factors influencing the robustness of interactome results. Evolutionary conservation rates of interactions are often low (Lewis et al., 2012). High-quality data are more reliable, but are not necessarily representative of whole interactomes. Some of these problems may be circumvented by using confidence scores calculated by various methods, such as the summative, network topology-based or Bayesian network-based models. Since different methods have different biases, composite scores taking multiple data-types into account perform better (Hakes et al., 2008; Sanchez Claros & Tramontano, 2012). The size of the human interactome has been estimated to have 650,000 interactions (Stumpf et al., 2008). Though a recent report (Havugimana et al., 2012) added 14 thousand high-confidence interactions to the growing list of human interactome edges, currently we are still far from deciphering the full complexity of this richness.

Table 6 lists a number of web-resources used for interactome analysis. A collection of protein–protein interaction network analysis web-tools can be found in recent reviews (Ma'ayan, 2008; Moschopoulos et al., 2011; Ma & Gao, 2012; Sanz-Pamplona et al., 2012). Protein–protein interaction networks are small worlds, have hubs and a well developed, hierarchical modular structure. These interactomes do not possess such an extensive rich-club as the social elite, i.e. hubs do not form dense clusters with each other (Maslov & Sneppen, 2002; Colizza et al., 2006; De Las Rivas & Fontanillo, 2010; Sardiu & Washburn, 2011). Soluble proteins tend to possess more connections than membrane proteins (Yu et al., 2004a). Steric hindrances severely limit the maximum number of simultaneous interactions. Tsai et al. (2009) warned that large interactome hubs may often be a result of aggregated data not taking into account protein conformations, posttranslational modifications, isoforms, expression differences and localizations. Another possibility to increase binding partners is sequential binding, which results in the formation of date hubs (as opposed of party hubs binding their partners simultaneously). Date hubs are often singlish-interface hubs as opposed to party-hubs, which are multi-interface hubs. Multi-interface hubs display a greater degree of conformational change than singlish-interface hubs (Han et al., 2004a; Kim et al., 2006; Bhardwaj et al., 2011). Interestingly, natural product drugs were shown to target proteins having a higher number of neighbors than targets of synthetic drugs (Dancik et al., 2010).

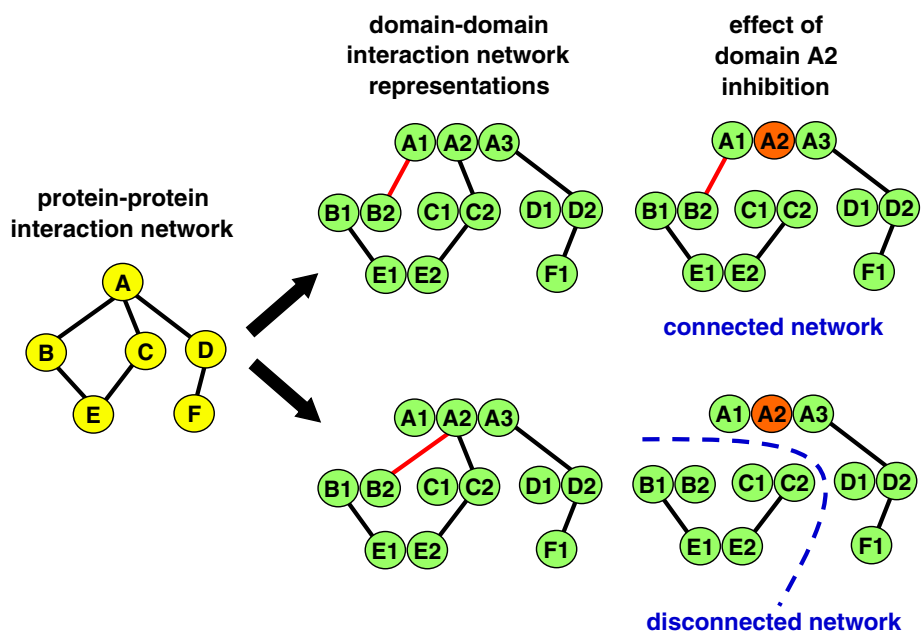

**Fig. 15.** The effect of more detailed representation of protein–protein interaction networks in representation of drug mechanism action. The left side of the figure shows a hypothetical protein–protein interaction network (yellow nodes). The middle panels show two representations of the very same network as a domain–domain interaction network (green nodes). Note that on the middle top panel the edge marked with red connects domains A1 and B2 while on the middle bottom panel the same edge connects domains A2 and B2. Note that these two representations cannot be discriminated at the protein–protein interaction level (shown on the left side marked with green nodes). If domain A2 (highlighted with red) is inhibited by a drug (and there is limited domain–domain interaction in protein A), this single edge-change leaves the sub-interactome in the right top panel intact. On the contrary, in the right bottom panel, the inhibition of domain A2 leads to the dissociation of the sub-network. An atomic level resolution of the interactome can discriminate even more subtle changes as we will discuss in Section 4.1.6 on allo-network drugs (Nussinov et al., 2011).
The figure is re-drawn from Figure 2 of Santonico et al. (2005) with permission.

Interactome modules overlap with each other, since most proteins are members of multiple protein complexes. Modules often correspond to major cellular functions (Palla et al., 2005). Refined modularization methods define modular cores containing only a few proteins, which occupy a central position of the interactome module. Major function of core proteins often reflects a consensus function of the whole module (Kovács et al., 2010; Szalay-Bekő et al., 2012; Srihari & Leong, 2013). Importantly, protein complexes may also form sparsely interconnected network modules, which often escape traditional detection methods (Srihari & Leong, 2012). Date hubs occupy inter-modular positions as opposed to party-hubs, which are in modular centers. Multi-component hubs (which, similarly to date-hubs, bridge multiple dense local network components) were enriched in regulatory proteins. Bridges and other inter-modular nodes play a key role in drug action (Han et al., 2004a; Komurov & White, 2007; Kovács et al., 2010; Fox et al., 2011; Szalay-Bekő et al., 2012).

As we discussed in Section 2.3.4, interactome hubs were shown to be an important predictor of essentiality (Jeong et al., 2001). Hub Objects Analyzer (Hubba) is a web-based service for exploring potentially essential nodes of interactomes assessing the maximum neighborhood component (Lin et al., 2008). Single-component hubs (i.e. hubs in the middle of a stable network neighborhood) were shown to be more essential than multi-component hubs, i.e. hubs connecting multiple dense network regions (Fox et al., 2011). Essential proteins associate with each other more closely than the average, and tend to be more promiscuous in their function. Many of these essential genes are housekeeping genes with high and less fluctuating expression levels (Jeong et al., 2001; Yu et al., 2004c). Later more global network measures, such as bottlenecks or more globally central proteins were also shown to contribute to the determination of essential nodes (Chin & Samanta, 2003; Estrada, 2006; Yu et al., 2007b; Missiuro et al., 2009; Li et al., 2011a).

The recent work of Hamp and Rost (2012) uncovered that variability of protein–protein interactions is much more frequent than previously thought. Besides single-nucleotide polymorphisms, alternative splicing, addition of N- or C-terminal tags, partial proteolysis and other post-translational modifications (such as phosphorylation), changes in protein expression patterns may dramatically re-configure protein complexes. Dynamic changes of protein–protein interactions, such as co-expression based clustering are key determinants of the disease state as we discuss in the next section. Importantly, interactome analysis has not been adequately extended to the assessment of interactome dynamics, and currently the application of the tools listed in Section 2.5 is largely missing. The human proteome is enriched in disordered proteins causing dynamically fluctuating, 'fuzzy' interaction patterns (Tompa, 2012). As an initial example of these studies the yeast interactome was shown to develop more condensed and more separated modules after heat shock and other types of stresses than under optimal growth conditions. Importantly, yeast cells preserved a few inter-modular bridges during stress and developed novel, stress-specific bridges containing key proteins in cell survival (Mihalik & Csermely, 2011).

### 3.3.2. Protein–protein interaction networks and disease

Most human diseases are oligogenic or polygenic affecting a whole set of proteins and their interactions. In the last decade several genome-wide datasets became available to characterize disease-related patho-mechanisms. mRNA expression patterns, genome-wide association studies (GWAS) of disease-associated single-nucleotide polymorphisms (SNPs) and disease-related changes in posttranslational modifications (such as the phospho-proteome) are just three of the most widely used datasets, which may also include system-wide changes of subcellular localization. All this information can be incorporated in protein–protein interaction networks as changes in edge configuration and weights (Zanzoni et al., 2009; Coulombe, 2011).

As we described in Section 1.3, disease-associated proteins do not generally act as interactome hubs, with the important exception of somatic mutations, such as those occurring in cancer, where disease-associated multi-interface hubs form an inter-connected rich club (Jonsson & Bates, 2006; Goh et al., 2007; Feldman et al., 2008; Kar et al., 2009; Barabási et al., 2011; Zhang et al., 2011a). Disease-related proteins have a smaller clustering coefficient than average, which was used for the prediction of novel disease-related genes (Feldman et al., 2008; Sharma et al., 2010a).

Disease-related proteins form overlapping disease modules. Suthram et al. (2010) identified 59 core modules out of the 4620 modules of the human interactome, which were affected by mRNA changes in more than half of the 54 diseases examined. These core modules were often targeted by drugs, and drugs affecting the core modules were more often multi-target drugs than those acting on 'peripheral' modules, which changed their mRNA levels only in a few specific diseases. Bridges and additional types of overlaps between disease-related interactome modules may provide important points of interventions (Nguyen & Jordan, 2010; Nguyen et al., 2011).

### 3.3.3. The use of protein–protein interaction networks in drug design

Uncovering the estimated ~650,000 interactions of the human interactome (Stumpf et al., 2008) is an ongoing, key step in network-related drug design efforts (see Rual et al., 2005; Stelzl et al., 2005; Chautard et al., 2009; Burkard et al., 2011; De Las Rivas & Prieto, 2012; Havugimana et al., 2012 and databases of Table 6). Databases like ChemProt: http://www.cbs.dtu.dk/services/ChemProt including 700,000 chemicals and 2 millions of interactions of their target proteins in various species (Taboureau et al., 2011) provide a great help in this process. However, interactome complexity goes much beyond the inventory of contacts and binding partners, and includes expression level-induced, posttranslational modification-induced (such as phosphorylation-dependent), cellular environment-induced (such as calcium-dependent) and protein domain-dependent variations (Santonico et al., 2005). We illustrate the latter in Fig. 15.

Drug targets have a generally larger number of neighbors than average. In agreement with assumptions related to disease-associated protein contacts described in the previous section, the larger number of neighbors comes mostly from the contribution of middle-degree nodes; but not hubs. Drug targets in cancer are exceptions having a more defined hub-structure. Drug target proteins have a lower clustering coefficient than other proteins. Drug targets often occupy a central position in the human interactome bridging two or more modules. Nodes having an intermediate number of neighbors have an extensive contact structure. Targeting these non-hub nodes (with the exception of infectious diseases and cancer) is crucial to avoid unwanted side-effects. As opposed to targets of withdrawn drugs having a too large network influence, drug target proteins perturb the interactome in a controlled manner (Hase et al., 2009; Zhu et al., 2009; Bultinck et al., 2012; Yu & Huang, 2012).

Properties of the interactome topology were used to predict and score novel drug target candidates using mainly machine learning techniques (Zhu et al., 2009; Zhang & Huan, 2010; Yu & Huang, 2012). Network neighborhood similarity to the drug targets test-set proved to be a good predictor of additional targets (Zhang & Huan, 2010). This feature may actually show the limits of machine learning-based approaches: since current drug targets are often similar to each other (Cokol et al., 2005; Yildirim et al., 2007; Iyer et al., 2011a), machine learning techniques may not be useful to extend the current drug target inventory to surprisingly novel hits.

Modulation of specific protein–protein interactions provides a much higher specificity to restore disease pathology to the normal state than targeting a whole protein. We will describe methods for the design of such 'edgetic drugs' in Section 4.1.2. Conceptually, it is much easier to develop inhibitors of protein–protein interactions than agents for increasing binding affinity or stability. The latter option together with the inclusion of interactome dynamics using the tools listed in Section 2.5 is a very promising future trend of the field. As a recent advance to explore drug-induced interactome dynamics, Schlecht et al. (2012) investigated the changes in the yeast interactome upon the addition of 80 diverse small molecules. Their method could identify novel protein–protein contacts specifically disrupted by the addition of drugs such as the immunosuppressant, FK506.

### 3.4. Signaling, microRNA and transcriptional networks

"Representations of signaling networks in many textbooks and even in some of the most recent review articles have in their simplicity a striking similarity to children's drawings" (Lewitzky et al., 2012; Fig. 16). The complex representation of a signaling network is constructed by upstream and downstream subnetworks. The upstream subnetwork contains the intertwined network of signaling pathways, while the downstream, regulatory part contains DNA transcription factor binding sites and microRNAs (Fig. 17). As we will show in the following subsections, both subnetworks are highly structured, are linked to each other, and are very important in drug discovery. The systems-level exploration and understanding of signaling networks significantly facilitate drug target identification, target selection in pathological networks and the avoidance of unwanted side-effects. At the end of the section we will also point out important network features that make signaling-related drug discovery a challenging task.

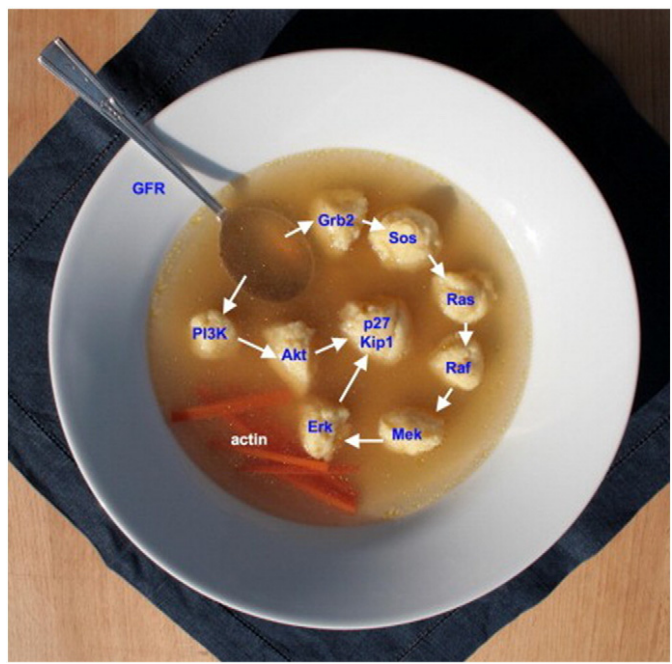

**Fig. 16.** The dumpling soup representation of growth factor initiated signaling. Reproduced with permission from Lewitzky et al. (2012).

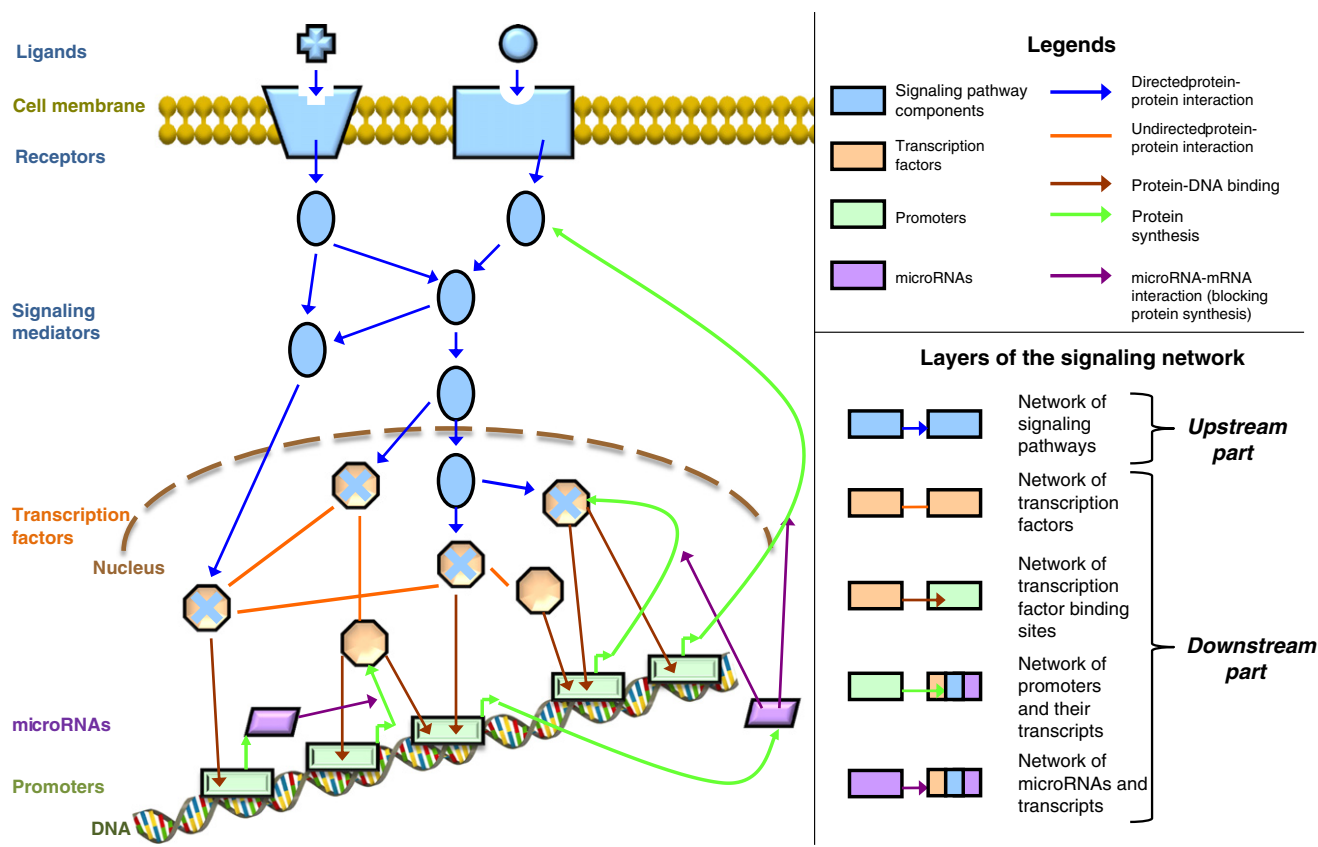

**Fig. 17.** Structure of the signaling network. The figure illustrates the major components of the signaling network including the upstream part of the signaling pathways and their cross-talks and the downstream part of gene regulation network. The gene regulation network contains the subnetworks of transcription factors, their DNA-binding sites and regulating microRNAs. Directed protein–protein interactions may encode enzyme reactions, such as phosphorylation events, while undirected protein–protein interactions participate (among others) in formation of scaffold and adaptor complexes.

### 3.4.1. Organization and analysis signaling networks

Signaling pathways, the functional building blocks of intracellular signaling, transmit extracellular information from ligands through receptors and mediators to transcription factors, which induce specific gene expression changes. Signaling pathways constitute the upstream part of signaling networks (Fig. 17). Over the past decade, it has been realized that signaling pathways are highly structured, and are rich in cross-talks, where cross-talk was defined as a directed physical interaction between pathways (Papin et al., 2005; Fraser & Germain, 2009). As the number of input signals (ligands/receptors) and output components (transcription factors) is limited, cross-talks between pathways can create novel input/output combinations contributing to the functional diversity and plasticity of the signaling network (Kitano, 2004a). However, cross-talks have to be precisely regulated to maintain output specificity (meaning that inputs preferentially activate their own output) and input fidelity (meaning that outputs preferentially respond to their own input). Regulation of cross-talks to prevent 'leaking' or 'spillover' can be achieved using different mediation mechanisms, such as scaffolding proteins, cross-pathway inhibitions, kinetic insulation, and the spatial and temporal expression patterns of proteins (Freeman, 2000; Bhattacharyya et al., 2006; Kholodenko, 2006; Behar et al., 2007; Haney et al., 2010; Lewitzky et al., 2012).

The regulatory subnetwork (gene regulatory network) constitutes the downstream part of the signaling pathway-network (Lin et al., 2012). The gene regulatory network can be separated into the transcriptional and the post-transcriptional levels. At the transcriptional level, transcription factors bind specific regions of DNA sequences (called transcription factor binding sites, or response elements) regulating their mRNA expression. Horizontal contacts of middle-level regulators play a key role in gene regulatory networks, especially in human cells. The human transcription factor regulatory network has a basic architecture, which is independent from the cell type and is complemented by cell type specific segments (Bhardwaj et al., 2010; Gerstein et al., 2012; Neph et al., 2012).

MicroRNAs (miRNAs or miRs) are key players of gene regulatory networks, and regulate gene expression by binding to complementary sequences (i.e. microRNA binding-sites) on target mRNAs. MicroRNA binding may suspend or permanently repress the translation of given transcripts (Doench & Sharp, 2004; Guo et al., 2010). In the last decade, it became evident that nearly all human genes can be controlled by at least one microRNA (Lewis et al., 2003), and that mutations in microRNA coding genes often have pathological consequences (Calin & Croce, 2006). Interactome hubs, bottleneck proteins and downstream signaling components, such as transcription factors are regulated by more microRNAs than other nodes (Cui et al., 2006; Liang & Li, 2007; Hsu et al., 2008).

Besides biochemical and molecular biological approaches, reverse engineering of genome-wide transcriptional changes proved to be very efficient for determining signaling networks as we detailed in Section 2.2.3. Signaling networks are small-worlds and possess signaling hubs. Networks (partially due to their pathway structures) have modules, and cross-talking proteins may often be considered as bridges between these modules. In the last decade several resources have been developed to provide signaling pathways, transcription factor and transcription factor binding site information, as well as microRNA networks. We summarize some of these signaling network resources in Table 7. A list of several other pathway databases can be found at PathGuide (http://pathguide.org; Bader et al., 2006). A compendium of human transcription factors have been collected and analyzed by Vaquerizas et al. (2009). Experimentally validated microRNA–mRNA interactions are available from TarBase (Vergoulis et al., 2012), while predicted interactions can be accessed at TargetScan and PicTar (Krek et al., 2005; Lewis et al., 2005). To examine the signaling network in a unified fashion, recently a few integrated resources, like IntegromeDB, TranscriptomeBrowser 3.0 and SignaLink 2.0, have been developed allowing the examination of all layers from signaling pathways to microRNAs through transcription factors (Korcsmáros et al., 2010; Baitaluk et al., 2012; Lepoivre et al., 2012; Fazekas et al., 2013).

**Table 7**
Signaling network resources.

| Name | Content | Website | References |
| --- | --- | --- | --- |
| IPAVS | Signaling pathway resources | http://ipavs.cidms.org | Sreenivasaiah et al., 2012 |
| Reactome | | http://reactome.org | Croft et al., 2011 |
| NCI Nature–Pathway Interaction Database | | http://pid.nci.nih.gov | Schaefer et al., 2009 |
| NetPath | | http://netpath.org | Kandasamy et al., 2010 |
| JASPAR | Transcription factor–transcription factor binding information | http://jaspar.genereg.net | Portales-Casamar et al., 2010 |
| HTRIdb | | http://www.lbbc.ibb.unesp.br/htri | Bovolenta et al., 2012 |
| MPromDB | | http://mpromdb.wistar.upenn.edu | Gupta et al., 2011 |
| PAZAR | | http://pazar.info | Portales-Casamar et al., 2007 |
| OregAnno | | http://oreganno.org | Griffith et al., 2008 |
| Expander | Transcription factor and microRNA target prediction form gene expression data | http://acgt.cs.tau.ac.il/expander | Ulitsky et al., 2010 |
| TarBase | mRNA–microRNA target information | http://diana.cslab.ece.ntua.gr/tarbase | Vergoulis et al., 2012 |
| TargetScan | | http://www.targetscan.org | Lewis et al., 2005 |
| PicTar | | http://pictar.mdc-berlin.de | Krek et al., 2005 |
| miRecords | Integrated resource of microRNA target information | http://mirecords.biolead.org | Xiao et al., 2009 |
| miRGen | | http://diana.cslab.ece.ntua.gr/mirgen | Alexiou et al., 2010 |
| TransMir | Regulatory information of microRNAs | http://202.38.126.151/hmdd/mirna/tf | Wang et al., 2010a |
| PutMir | | http://www.isical.ac.in/~bioinfo_miu/TF-miRNA.php | Bandyopadhyay & Bhattacharyya, 2010 |
| MiRandola | Extracellular microRNA database | http://atlas.dmi.unict.it/mirandola | Russo et al., 2012 |
| IntegromeDB | Integrated signaling network resources | http://integromedb.org | Baitaluk et al., 2012 |
| SignaLink 2.0 | | http://signalink.org | Korcsmáros et al., 2010; Fazekas et al., 2013 |
| TranscriptomeBrowser 3.0 | | http://tagc.univ-mrs.fr/tbrowser | Lepoivre et al., 2012 |

There was considerable progress in defining algorithms to identify the downstream components of a signaling network affected by the inhibition of a specific protein or protein set. Such methods identify targets, which inhibit certain outputs of the signaling network, while leaving others intact, redirecting the signal flow in the network (Dasika et al., 2006; Ruths et al., 2006; Pawson & Linding, 2008). This recuperates the output specificity and input fidelity in a drug–target context, where output specificity corresponds to the minimization of side-effects, while input fidelity represents drug efficiency at the signaling network level.

The dynamics of signaling networks is regulated by changes in the abundance of their components, by complex formation, by macromolecular crowding, and subcellular localization (Lewitzky et al., 2012). The assessment of signaling network kinetics is helped by perturbation analysis, differential equation models, constrained fuzzy logic models and Boolean methods. In the latter, the activity of signaling components is represented by 0:1 states connected by directed and conditional edges as we summarized in Section 2.5.2 on network perturbations (Kauffman et al., 2003; Shmulevich & Kauffman, 2004; Berg et al., 2005; Antal et al., 2009; Farkas et al., 2011). There are several excellent methods for the analysis of Boolean networks.

• BooleanNet (http://booleannet.googlecode.com) is a versatile, publicly available software library to describe signaling network dynamics using the Boolean description (Albert et al., 2008).
• PATHLOGIC-S (http://sourceforge.net/projects/pathlogic/files/PATHLOGIC-S) offers a scalable Boolean framework for modeling cellular signaling (Fearnley & Nielsen, 2012).

• PathwayOracle (http://old-bioinfo.cs.rice.edu/pathwayoracle) is a fast simulation program of large signaling networks taking into account their topology (Ruths et al., 2008a, 2008b).
• Changes in memory effects (i.e. specific decay times of gene products) greatly affected Boolean network behavior (Graudenzi et al., 2011a, 2011b).
• Boolean models can be trained by high-throughput data such as by phosphoproteomics (Videla et al., 2012).
• Recently the CellNOpt algorithm (http://www.ebi.ac.uk/saezrodriguez/cno) was introduced, which uses either Boolean logic or constrained fuzzy logic to generate and analyze cell-specific signaling networks (Terfve et al., 2012; Morris et al., 2013).
• The method of Chen et al. (2011) is able to identify sub-pathways and principal components of sub-pathways affected by a drug or disease.

The elucidation of signaling network dynamics can be greatly helped by quantitative phosphoproteomics (White, 2008). Reaction network analysis pointed out receptor-related events and not kinase-related events as rate-limiting factors in IL-12 signaling (Klinke & Finley, 2012). In an interesting study on signaling dynamics, Cheong et al. (2011) assessed the amount of information transduced by the TNF-related signaling network in the presence of cellular noise. They found that signaling bottlenecks may have a crucial influence on signaling capacity in the presence of noise. Negative feedback can reduce noise, increasing signal capacity in the short term (30 min after stimulation). However, negative feedback behaves as a double-edged sword, and also reduces the dynamic range of the signal input reducing the capacity in the longer run (4 h after stimulation). Perturbation modeling suggested that cancer-related proteins may have a larger than usual signaling capacity (Serra-Musach et al., 2012). Signal transmission capacity analysis has many unsolved questions. Currently, we do not understand the decision making limits of the vast majority of signaling systems. Information-loss, and information-integration all affect the information handling capacity, which is presumed to be minimally sufficient (Brennan et al., 2012). Future, network-related analysis of signaling dynamics faces the important task of finding an optimal ratio of large-scale signaling network topology and refined kinetic details to find answers to these questions.

### 3.4.2. Drug targets in signaling networks

Understanding the structure and dynamics of signaling networks is used more and more often in drug discovery (Pawson & Linding,

2008). Drugs having a similar pharmacological profile reach similarly discrete positions in signaling networks (Fliri et al., 2009). In signaling networks of healthy cells a distinctive role was suggested for proteins in the junctions of signaling pathways. These proteins were termed 'critical nodes' by Taniguchi et al. (2006) as exemplified by the PI3 kinase, AKT and IRS isoforms in insulin signaling. Proteins forming a bridge between signaling modules (e.g., SHC, SRC and JAK2) have a track record as targets of drug action (Hwang et al., 2008; Gardino & Yaffe, 2011).

Studies on pathologically altered signaling networks can uncover possible drug targets, whose malfunction is involved in the etiology of the disease. For example, driver mutations of tumorigenesis affect a limited number of central pathways (Tomlinson et al., 1996; Ali & Sjoblom, 2009). Targeting of these specific pathways may prevent tumor growth. However, the development of aggressive tumor cells causes a systems-level change of the signaling network causing the appearance of angiogenetic and metastatic capabilities, as well as the deregulation of cellular metabolism and avoidance of the immune system (Hanahan & Weinberg, 2000, 2011; Hornberg et al., 2006; Papatsoris et al., 2007). Changes of cross-talking (i.e. multi-pathway) proteins are key steps in disease-induced rewiring of the signaling network, e.g. transforming a 'death' signal into a 'survival' signal (Hanahan & Weinberg, 2000; Hornberg et al., 2006; Kim et al., 2007; Torkamani & Schork, 2009; Mimeault & Batra, 2010; Farkas et al., 2011). Multipathway proteins show a significant change in their expression level in hepatocellular carcinoma (Korcsmáros et al., 2010).

We have more than 500 types of post-translational modifications offering a rich repertoire for signaling (Lewitzky et al., 2012). Traditionally, among these, protein kinases are the most targeted proteins of the cellular signaling network (Pawson & Linding, 2008). However, the similarity of ATP-binding pockets poses a significant challenge in kinase targeting. Kinase domains and their target motifs (i.e., specific amino acid sequences in the substrate proteins) can be accessed in resources such as Phosphosite (Hornbeck et al., 2012), NetworKIN and NetPhorest (Linding et al., 2008; Miller et al., 2008). Kinase regulatory domain associations, kinase-associated scaffold proteins and multisite docking proteins often direct subcellular localizations, and play a key role determining signaling kinetics and substrate specificity (Remenyi et al., 2006; Pálfy et al., 2012). Scaffolding proteins are flexible; and they, and multi-site docking proteins may increase network flexibility (by allowing integrative cross-talks; Lewitzky et al., 2012). However, our systems-level knowledge on these undirected protein–protein interactions is rather limited. Disruption of kinase-centered sub-interactomes and/or remodeling of kinase-centered protein complexes are focus areas of drug design (Brehme et al., 2009; Bandyopadhyay et al., 2010; Li et al., 2012a).

Protein phosphatases play a dominant role in determining the spatio-temporal behavior of protein phosphorylation systems (Herzog et al., 2012; Sacco et al., 2012; Nguyen et al., 2013). Despite their promising effect, only a few protein tyrosine phosphatases are currently used as therapeutic targets (Alonso et al., 2004). The development of phosphatase-related drugs is more complicated than that of kinase targeting drugs, since i.) the high-level of homology between phosphatase domains limits the development of selective compounds; ii.) contrary to kinases, phosphatase substrate specificity is achieved through docking of the phosphatase complex at a site distant from the dephosphorylated amino acid (Roy & Cyert, 2009; Shi, 2009); (iii) the targeted sequences are highly charged, and many of the interacting compounds are not hydrophobic enough to cross the membrane (Barr, 2010). Despite these difficulties, phosphatase-targeting holds great promise in rational drug design.

In the last decade, microRNAs have been recognized as highly promising intervention points of the signaling network. Though the use of antisense nucleotides comes with great challenges in pharmacological availability, microRNA targeting affects mRNA clusters having a rather specific effect at the transcriptome level (Gambari et al., 2011). Down- or up-regulation of microRNAs is implicated in more than 270 diseases according to the Human MicroRNA Disease Database (http://202.38.126.151/hmdd/mirna/md; Lu et al., 2008) including cardiovascular, neurodegenerative diseases, viral infections and various types of cancer (McDermott et al., 2011). Identification of new microRNA targets may be helped by the microRNA clusters associated with the same disease (Lu et al., 2008), or with expression modules (Bonnet et al., 2010).

MicroRNA targeting is typically a systems-level endeavor. Most microRNAs have a number of targets, and are in an intensive crosstalk with transcription factors (Lin et al., 2012) forming a highly cross-reacting, and cross-regulated network. The microRNA network has hierarchical layers, and hundreds of 'target hubs', each potentially subject to massive regulation by dozens of microRNAs (Shalgi et al., 2007). Cancer cells, as opposed to normal cells, have disjoint microRNA networks, where major hubs of normal cells are down-regulated, and cancer-specific novel microRNA hubs emerge (Volinia et al., 2010). MicroRNA-regulated drug targets were shown to preferentially interact with each-other, and tend to form hub-bottlenecks of the human interactome (Wang et al., 2011c). Unwanted, network-level side-effects of microRNA targeting may be predicted using databases, such as SIDER (http://sideeffects.embl.de; Kuhn et al., 2010), web-services, such as PathwayLinker (http://PathwayLinker.org; Farkas et al., 2012) or the other integrated resources listed in Table 7. Miravirsen, a locked nucleic acid-modified antisense oligonucleotide, targets the liver-expressed microRNA-122 and is in phase-II clinical trial for treatment of hepatitis C virus infection (Lindow & Kauppinen, 2012).

### 3.4.3. Challenges of signaling network targeting

Signal transduction is highly context-specific: it depends on the gene expression patterns, mRNA stability, protein synthesis, and degradation conditions. Certain signaling modules (such as regulation of apoptosis) seem to share evolutionary traits with others, while other signaling proteins developed more independently. These variabilities necessitate a precise knowledge of the actual status of the signaling network in the disease condition and in the affected patient population (Cui et al., 2009; Davis et al., 2012; Hamp & Rost, 2012; Kirouac et al., 2012).

As we have shown in the preceding section, cross-talking (i.e., multi-pathway, bridge) proteins are in a critical position of the signaling network providing a very efficient set of potential drug targets (Korcsmáros et al., 2007, 2010; Hwang et al., 2008; Kumar et al., 2008; Spiró et al., 2008). However, numerous drug developmental failures were caused by undiscovered or underestimated cross-talk effects (Rajasethupathy et al., 2005; Jia et al., 2009). Cross-talking proteins may have opposite roles in healthy and diseased (such as in malignant) cells. Moreover, targeting of cross-talking proteins may significantly affect the systems-level stability (robustness) of healthy or diseased cells (Kitano, 2004b, 2007). Targeting proteins in negative feedback loops may suppress the inhibitory effect of the feedback loop, and thereby activate the targeted pathway (Sergina et al., 2007). Feedback loops are not always direct, and can exist at multiple levels of a pathway. In conclusion, targeting multi-pathway and feedback loop proteins requires a particularly detailed knowledge of signaling network responses (Berger & Iyengar, 2009; Barabási et al., 2011).

Systems-level properties are also needed to assess the development of drug resistance and drug toxicity.

- Development of drug resistance is often a result of a systems-level response of signaling networks involving mutation of key signaling proteins (such as multi-pathway proteins), or of the activation of alternative pathways due to system-robustness (Kitano, 2004a; Logue & Morrison, 2012). As a specific form of drug resistance, many anticancer drugs induce stress response/survival pathways directly, or indirectly, by producing a stressful environment (Tomida

& Tsuruo, 1999; Chen et al., 2006b; Tiligada, 2006). Thus, systems-level approaches that combine anti-tumor drugs and stress response targeting may increase therapeutic efficiency (Rocha et al., 2011; Tentner et al., 2012).

- Hepatotoxicity is a major cause of drug development failures in the pre-clinical, clinical and post-approval stages (Kaplowitz, 2001). Hepatic cytotoxicity responses are regulated by a multi-pathway signaling network balance of intertwined pro-survival (AKT) and pro-death (MAPK) pathways. Importantly, therapeutic modulation of cross-talks between these pathways as well as specific pathway inhibitors could antagonize drug-induced hepatotoxicity (Cosgrove et al., 2010).

We will address network-based assessment of drug toxicity and drug resistance in Sections 4.3.3 and 4.3.6 in more detail.

### 3.5. Genetic interaction and chromatin networks

In this section we will describe the drug-related aspects of genetic interaction networks. Genetic interaction networks are related to gene regulatory networks. However, here gene–gene interactions are often indirect. Chromatin networks encode 3D interactions between distant DNA-segments of the chromatin structure, and may be regarded as a specific representation of genetic interaction networks. While genetic interaction networks already helped drug design, chromatin interaction networks are recent developments holding a great promise for future studies.

#### 3.5.1. Definition and structure of genetic interaction networks

The most stringent (and traditional) description of a genetic interaction comes from comparing the phenotypes of the individual single mutants with the phenotype of the double mutant. We can distinguish between negative and positive genetic interactions: if the fitness of the double mutants is worse than the additive effect of the two single mutants, then the genes have a negative interaction. Conversely, if the fitness of the double mutants is better than expected, the two genes interact positively. A severe type of negative interactions is synthetic lethality, when the two single mutants are viable, but their double mutant becomes lethal. Genes of negative (i.e. aggravating) interactions may operate in parallel processes, while those of positive (i.e., alleviating or epistatic) interactions may function in the same process (Guarente, 1993; Hartman et al., 2001; Dixon et al., 2009). The complexity of the genetic interaction network is illustrated well by compensatory perturbations, where a debilitating effect can be compensated by another inhibitory effect (Motter, 2010; Cornelius et al., 2011).

Most comprehensive genome-wide studies were performed in inbred model systems, such as yeast and worm, as well as in isogenic populations of cultured cells derived from fruit flies and mammals. It is plausible that many genetic interactions identified in these unicellular organisms can be relevant for all other eukaryotes (Tong et al., 2004; Dixon et al., 2008, 2009; Roguev et al., 2008; Costanzo et al., 2010). However, comparison between orthologous genes of yeast and worm found less than 5% of synthetic lethal genetic interactions to be conserved (Byrne et al., 2007). Furthermore, most of the human disease genes are metazoan-specific. Despite the widespread specificity, there are some genetic interactions (such as those of DNA repair enzymes, which are commonly mutated in cancer), which are conserved from yeast to humans (McManus et al., 2009).

Besides mutational studies, system-wide assessments of output signals, such as transcriptomes, allowed the phenotype analysis of thousands of perturbations inferring a genetic interaction network. We reviewed the reverse engineering methods allowing network inference in Section 2.2.3. It appears that no single inference method performs optimally across all datasets. Therefore, the integration of predictions from multiple inference methods seems to give the best results (Marbach et al., 2012). The genetic interaction network obtained by direct mutational studies, or by reverse engineering methods exhibited dense local neighborhoods, while highly correlated profiles delineated specific pathways defining gene function, and were used for pre-clinical drug prioritization (Xiong et al., 2010). Recently, an algorithm called HotNet was introduced to identify genetic interaction network clusters (http://compbio.cs.brown.edu/software.html; Vandin et al., 2012).

Mapping of genetic interactions to protein–protein interaction or to signaling networks may uncover the underlying mechanisms. Consequently, we can define 'between-pathway', 'within-pathway' and 'indirect' types of genetic interactions. With this approach, ~40% of the yeast synthetic lethal genetic interactions were mapped to physical pathway models identifying 360 between-pathway and 91 within-pathway models (Kelley & Ideker, 2005). Synthetic lethal gene pairs were found mostly close to each other (often within the same modules), while rescuing genes were often in alternative pathways and/or modules (Hintze & Adami, 2008). Combinations of genetic interactions with mRNA expression patterns, interactome data, gene–drug interactions, or with chemical compound similarity measures (for details see the compendium of Table 5) offered a great help in the identification of drug targets and drug-affected genes (Parsons et al., 2006; Hansen et al., 2009; Gosline et al., 2012).

Measurement of time-series of genome-wide mRNA expression patterns after drug treatment of 95 genotyped yeast strains led to the identification of novel genetic interaction network relationships including novel feedback loops and transcription factor binding sites (Yeung et al., 2011). BioLayout Express gives an integrative network visualization and analysis of gene expression data (http://www.biolayout.org; Freeman et al., 2007; Theocharidis et al., 2009). SteinerNet provides integrated transcriptional, proteomic and interactome data to assess regulatory networks (http://fraenkel.mit.edu/steinernet/; Tuncbag et al., 2012). Genetic interaction networks may also be defined in a more general manner, where any types of interactions, such as correlated expression levels, interacting protein products, or co-participation in a disease etiology or drug action, may form an edge between two genes serving as nodes of the network (Schadt et al., 2009). Genome-wide association studies (GWAS) identified single-nucleotide polymorphism (SNP) derived gene–gene association networks revealing novel between-pathway models (Cowper-Sal lari et al., 2011; Fang et al., 2011; Hu et al., 2011; Li et al., 2012b).

#### 3.5.2. Chromatin networks and network epigenomics

An underlying molecular mechanism establishing genetic interaction networks is the network of long-range interactions of the 3D chromatin structure. Recent methodologies based on proximity ligation with next generation sequencing (abbreviated as Hi-C or ChIA-PET) enabled the construction of a functionally associated, long-range contact network of the human chromatin structure. This chromatin network contains functional modules, and has a rich club of hub–hub interactions (Fullwood et al., 2009; Lieberman-Aiden et al., 2009; Dixon et al., 2012; Li et al., 2012c; Sandhu et al., 2012).

The chromatin network determines cancer-associated chromosomal alterations (Fudenberg et al., 2011). Moreover, the chromatin network configuration was shown to be grossly altered by the overexpression of ERG, an oncogenic transcription factor activated primarily in prostate cancers (Rickman et al., 2012). The structure of the chromatin network is largely determined by inheritable epigenetic factors, such as histone posttranslational modifications, DNA silencers and nascent RNA scaffolds (Schreiber & Bernstein, 2002; Moazed, 2011; Pujadas & Feinberg, 2012). Chromatin networks are an exciting and fast-developing area of network-studies, which will provide promising tools to predict drug–drug interactions, drug side-effects and system-wide effects of anti-cancer and other drugs inducing chromatin reprogramming.

### 3.5.3. Genetic interaction networks as models for drug discovery

The genetic interaction network of yeast can be used as a system for rational ranking of potential new antifungal targets; it may also shed light on human drug mechanisms of action, since several human drugs specifically inhibit the orthologous proteins in yeast (Hartwell et al., 1997; Cardenas et al., 1999; Hughes, 2002). The identification of 16 genes, whose inactivation suppressed the defects in the retinoblastoma tumor suppressor pathway in another widely used model system, *C. elegans*, could point out potential targets for pharmaceutical intervention or prevention of human retinoblastoma-linked tumors (Lee et al., 2008b). Extending this methodology, McGary et al. (2010) defined orthologous phenotypes, or 'phenologs', which can be regarded as evolutionarily conserved outputs that arise from the disruption of a set of genes. The phenolog approach identified non-obvious equivalences between mutant phenotypes in different species, establishing a yeast model for angiogenesis defects, a worm model for breast cancer, mouse models of autism, and a plant model for the neural crest defects associated with the Waardenburg syndrome (McGary et al., 2010).

Many pharmacologically interesting genes, such as nuclear hormone receptors and GPCRs occur in large families containing paralogs, i.e. duplicated homologous genes. Though model organisms can significantly help us to understand how human genes interact with each other, it is important to keep in mind that paralogs often do not have the same function. This problem can be circumvented by targeting paralog-sets, which makes the identification of paralogs and their functions a key point in multi-target drug design (Searls, 2003).

Wang et al. (2012c) gave an interesting example for the use of genetic interaction networks in the assessment of the effects of drug combinations. They showed that drug combinations have significantly shorter effect radius than random combinations. Drug combinations against diseases affecting the cardiovascular and nervous systems have a more concentrated effect radius than immuno-modulatory or anti-cancer agents.

### 3.6. Metabolic networks

In this section we will describe metabolic networks, i.e. networks of major metabolites connected by enzyme reactions, which transform them to each other. Metabolic networks are the biochemically constrained aspects of the chemical reaction networks we summarized in Section 3.1.2. After the description of the structure and properties of metabolic networks we will summarize their use in drug targeting with special reference to the identification of essential reactions as potential drug targets in infectious diseases and in cancer.

### 3.6.1. Definition and structure of metabolic networks

In a metabolic network, each node represents a metabolite. Two nodes are connected, if there is a biochemical reaction that can transform one into the other. Edges of metabolic networks represent both reactions and the enzymes that catalyze them. (We note that metabolic networks may also have another projection, where nodes are the enzymes and edges are the metabolites connecting them, but this projection is seldom used, since it is less relevant to biological processes.) Metabolic processes may be represented as hypergraphs, where edges connect multiple nodes. An edge may correspond to multiple reactions both in the forward direction and in the opposite direction. Moreover, some reactions occur spontaneously, and therefore have no associated enzymes. Most metabolites are fairly general, but the biochemical reaction structure connecting them is often rather special to the given organism (Guimera et al., 2007b; Ma & Goryanin, 2008; Chavali et al., 2012).

Reconstruction of metabolic networks became a highly integrative process, which applies genome sequences, enzyme databases, and specifies the network using transcriptome and proteome data (Kell, 2006; Ma & Goryanin, 2008). In the last decade several metabolic networks, such as those of *Escherichia coli*, yeast and humans have been assembled. Moreover, recently bacterium-, strain-, tissue- and disease-specific metabolic networks were reconstituted. However, it should be kept in mind that metabolic network data are often still incomplete, and often reflect optimal growth conditions (Edwards & Palsson, 2000a; Forster et al., 2003; Duarte et al., 2007; Ma et al., 2007; Shlomi et al., 2008, 2009; Folger et al., 2011; Holme, 2011; Chavali et al., 2012; Szalay-Bekő et al., 2012).

Metabolic networks have a small-world character, possess hubs, and display a hierarchical bow-tie structure similar to other directed networks, such as the world-wide-web. Metabolic networks have a hierarchical modular structure (Jeong et al., 2001; Wagner & Fell, 2001; Ravasz et al., 2002; Ma & Zeng, 2003; Ma et al., 2004; Guimera & Amaral, 2005; Zhao et al., 2006). Correlated reaction sets (Co-sets) are representations of metabolic network modules encoding reaction-groups with similar fluxes. Hard-coupled reaction sets (HCR-sets) are those subgroups of Co-sets, where consumption/production rates of participating metabolites are 1:1. Since all reactions of a HCR-set changes, if any of its reactions is targeted, HCR-sets help in prioritizing potential drug target lists (Papin et al., 2004; Jamshidi & Palsson, 2007; Xi et al., 2011). Metabolic networks have a core and a periphery (Almaas et al., 2004, 2005; Guimera & Amaral, 2005; Guimera et al., 2007b). Core and periphery may also be discriminated in the non-topological sense that genes and gene pairs of the 'core' are essential under many environmental conditions, while those of the 'periphery' are needed under some environmental conditions (Papp et al., 2004; Pál et al., 2006; Harrison et al., 2007).

Metabolic control analysis (MCA) is good for smaller networks where kinetic parameters are known, while flux balance analysis (FBA), flux-variability analysis (FVA) and elementary flux mode analysis are very useful methods to characterize systems-level metabolic responses (Fell, 1998; Cascante et al., 2002; Klamt & Gilles, 2004; Chavali et al., 2012). Resendis-Antonio (2009) integrated high throughput metabolome data describing transient perturbations in a red blood cell metabolic network model. This approach may be applicable for the modeling and metabolome-wide understanding of drug-induced metabolic changes (Fan et al., 2012). In Table 8 we list resources to define and analyze metabolic networks.

### 3.6.2. Essential enzymes of metabolic networks as drug targets in infectious diseases and in cancer

Metabolic networks help in the identification essential proteins. This requires a systems-level approach, since an essential metabolite might be produced by several pathways (Palumbo et al., 2013). As an example of metabolic robustness, the early work of Edwards and Palsson (2000b) showed that the flux of even the tricarboxylic acid cycle can be reduced to 19% of its optimal value without significantly influencing the growth of *E. coli*. When designing a drug against a metabolic network of an infectious organism or against cancer cells, many parameters should be kept in mind. We list a few of them here.

- Network topology analysis is not enough to predict essential enzymes, since it does not indicate whether the topologically important enzymes are active under specific conditions. Moreover, metabolic fluxes are determined by gene expression levels (representative to the affected tissue, or cell status) and by signaling- or interaction-related activation/inhibition of pathway enzymes. Genes that are not identified correctly as essential genes are usually connected to fewer reactions and to less over-coupled metabolites, and/or their associated reactions are not carrying flux in the given condition (Becker & Palsson, 2008; Chavali et al., 2012; Kim et al., 2012).
- Infectious organisms often take advantage of the metabolism of the host requiring the analysis of integrated parasite–host metabolic networks (Fatumo et al., 2011).
- Essentiality is not a yes/no variable: essentiality of a given reaction

**Table 8**
Metabolic network resources.

| Name | Content | Website | References |
| --- | --- | --- | --- |
| KEGG | Metabolic pathway resource | http://kegg.jp | Kanehisa et al., 2012 |
| MetaCyc | | http://metacyc.org | Caspi et al., 2012 |
| HumanCyc | | http://humancyc.org | Romero et al., 2005 |
| SMPDB | Small Molecule (e.g., drug) Pathway Database | http://smpdb.ca | Frolkis et al., 2010 |
| HMDB | Human Metabolome Database | http://hmdb.ca | Wishart et al., 2009 |
| BRENDA | Comprehensive enzyme data resource | http://brenda-enzymes.info | Scheer et al., 2011 |
| YeastNet | Yeast metabolic network | http://comp-sys-bio.org/yeastnet | Herrgard et al., 2008 |
| iMAT another metabolic network construction and analysis tools | Several metabolic network construction and analysis tools | http://www.cs.technion.ac.il/~tomersh/methods.html | Shlomi et al., 2008; Zur et al., 2010 |
| ModelSEED and its Cytoscape plug-in, CytoSEED | Genome level metabolic network reconstruction and analysis | https://github.com/ModelSEED, http://sourceforge.net/projects/cytoseed | Henry et al., 2010; DeJongh et al., 2012 |
| Markov Chain Monte Carlo modeling | Bayesian inference method to uncover perturbation sites in metabolic pathways | ftp://anonymous@dbkweb.mib.man.ac.uk/pub/Bioinformatics_BJ.zip | Jayawardhana et al., 2008 |
| PyNetMet | Python library of tools for the analysis of metabolic models and networks | http://pypi.python.org/pypi/PyNetMet | Gamermann et al., 2012 |

depends on the environment of the infectious organism, or cancer cells. Therefore, metabolic network-based drug-design should incorporate environment interactions and stressor effects (Guimera et al., 2007b; Jamshidi & Palsson, 2007; Ma & Goryanin, 2008; Kim et al., 2012).

- A promising current trend, metabolic interactions of bacterial communities, such as the gut microbiome, are also important factors to consider (Chavali et al., 2012; Kim et al., 2012).
- Drug targets against infectious organisms or against cancer should be specific for the target itself or for its drug binding site, or for its network-related consequences of targeting (Guimera et al., 2007b; Ma & Goryanin, 2008; Chavali et al., 2012). This highlights the importance of comparing metabolic network pairs.
- Finally, network-analysis offers a great help to predict side-effects (Guimera et al., 2007b). We will detail network-methods of side-effect prediction in Section 4.3.5.

Enzymes catalyzing a single chemical reaction on one particular substrate are frequently essential (Nam et al., 2012). Through the analysis of metabolic network structure, choke points were identified as reactions that either uniquely produce or consume a certain metabolite. Efficient inhibition of choke points may cause either a lethal deficiency, or toxic accumulation of metabolites in infectious organisms (Yeh et al., 2004; Singh et al., 2007). Later choke point analysis was combined with load point analysis (identification of nodes with a high ratio of k-shortest paths to the number of nearest neighbor edges providing many alternative metabolic pathways) and with comparison of the metabolic networks of pathogenic and related non-pathogenic strains. Such methods can test multiple knock-outs on a high throughput manner predicting effective drug combinations (Fatumo et al., 2009, 2011; Perumal et al., 2009).

Guimera et al. (2007b) developed a network modularity-based method for target selection in metabolic networks. They systematically analyzed the effect of removing edges from the metabolic networks of E. coli and Heliobacter pylori quantifying the effect by the difference in growth rate. In both bacteria, essential reactions (edges) mostly involve satellite connector metabolites that participate in a small number of biochemical reactions, and serve as bridges between several different modules (Guimera et al., 2007b).

Essential and non-essential genes propagate their deletion effects via distinct routes. Flux selectivity of a deletion of a metabolic reaction was used to design the appropriate type and concentration of the inhibitor (Gerber et al., 2008). Recently, several iterative methods have been constructed, sequentially identifying a set of enzymes whose inhibition can produce the expected inhibition of targets with reduced side-effects in human and E. coli metabolic networks (Lemke et al., 2004; Sridhar et al., 2007, 2008; Song et al., 2009). Ma et al. (2012b) assembled 'damage lists' of reactions affected by deleting other reactions using flux-balance analysis. They showed that the knockout of an essential gene mainly affects other essential genes, whereas the knockout of a non-essential gene only interrupts other non-essential genes. Genes sharing the same 'damage list' tend to have the same level of essentiality.

A subset of genes and gene pairs may be essential under various environmental conditions, while most genes are essential only under a certain environmental condition. In yeast environmental condition-specificity accounts for 37–68% of dispensable genes, while compensation by duplication and network-flux reorganization is responsible for 15–28 and 4–17% of yeast dispensable genes, respectively (Papp et al., 2004; Blank et al., 2005). Almaas et al. (2005) suggested the use of metabolic network cores to identify drug targets. Barve et al. (2012) identified a set of 124 superessential reactions required in all metabolic networks under all conditions. They also assigned a superessentiality index for thousands of reactions. Superessentiality of the 37 reactions catalyzed by enzymes having a very low homology to human genes (Becker et al., 2006; Aditya Barve & Andreas Wagner, personal communication) can provide substantial help in drug target selection, since the index is not highly sensitive to the chemical environment of the pathogen.

An interesting approach to narrow metabolic networks to essential components is to identify essential metabolites. As examples of this process, in two pathogenic organisms a total of 221 or 765 metabolites were narrowed to 9 or 5 essential metabolites, respectively, after the removal of the currency metabolites, i.e. those present in the human metabolic network and those participating in reactions catalyzed by enzymes having human homologs. Enzymes that catalyze reactions involved in the production or consumption of these essential metabolites may be considered as drug targets. Moreover, structural analogs of essential metabolites may be considered as drug candidates for experimental evaluation (Kim et al., 2010, 2011a, 2011b).

Using a comparison of metabolic networks, Shen et al. (2010) provided a blueprint of strain-specific drug selection combining metabolic network analysis with atomistic level modeling. They deduced common antibiotics against E. coli and Staphylococcus aureus, and ranked more than a million small molecules identifying potential antimicrobial scaffolds against the identified target enzymes.

The analysis of disease-specific metabolic networks is a key step to find 'differentially essential' genes (Murabito et al., 2011). Analysis of cancer-specific human metabolic networks led Folger et al. (2011) to predict 52 cytostatic drug targets, of which 40% were targeted by known anticancer drugs, and the rest were new target-candidates. Their method also predicted combinations of synthetic lethal drug targets and potentially selective treatments for specific cancers. We will describe network-related anti-infection and anti-cancer strategies in more detail in Sections 5.1 and 5.2.

### 3.6.3. Metabolic network targets in human diseases

Many human diseases cause a metabolic deficiency rather than overproduction making the recovery of a specific metabolic reaction a widely used drug-development strategy (Ma & Goryanin, 2008). Systems-level assessment may lead to the development of successful combined-therapies, such as the combination of Niacin, an inhibitor of cholesterol transportation, with Lovastatin, an inhibitor of the cholesterol synthesis pathway to reduce blood cholesterol level (Gupta & Ito, 2002). As a very interesting approach, a flux-balance analysis model was developed to predict compensatory deletions (also called as synthetically viable gene pairs, or synthetic rescues), where a debilitating effect can be compensated by another inhibitory effect (Motter et al., 2008; Motter, 2010; Cornelius et al., 2011). Since inhibition is often a pharmacologically more feasible intervention than activation, this approach opens novel possibilities for drug design to restore disease-induced malfunctions. The approach of Jamshidi and Palsson (2008) to describe temporal changes of metabolic networks is an example of what seems to be a very promising future direction to study the process of disease progression, and to design disease-stage specific drug treatment protocols.

## 4. Areas of drug design: an assessment of network-related added-value

In this section we will highlight the added-value of network related methods in major steps of the drug design process. Fig. 18 illustrates various stages of drug development starting with target identification, followed by hit finding, lead selection and optimization including various methods of chemoinformatics, drug efficiency optimization, ADMET (drug absorption, distribution, metabolism, excretion and toxicity) studies, as well as optimization of drug–drug interactions, side-effects and resistance. Table 9 summarizes a few major data-sources and web-services, which can be used efficiently in network-related drug design studies.

### 4.1. Drug target prioritization, identification and validation

Network-based drug target prioritization and identification are essentially a top–down approach, where system-wide effects of putative targets are modeled to help in the identification of novel network drug targets. These network drug targets are non-obvious from a traditional magic-bullet type analysis aiming to find the single most important cause of a given disease. Network node-based drug target prediction may highlight non-obvious hits, and edge-targeting may make these hits even more specific. Drug target networks allow us to see the system-wide target landscape and, combined with other network methods, help drug repositioning. Multi-target drug design needs the integration of drug effects at the system level. The new concept of allo-network drugs may identify non-obvious drug targets, which specifically influence the major targets causing fewer side-effects than direct targeting. Finally, treating the whole cellular network (or its segment) as a drug target, gives a conceptual synthesis of network description and analysis in drug design.

### 4.1.1. Two strategies of network-based drug targeting: the central hit and the network influence strategies

Here we propose that our current knowledge discriminates two network-based drug identification strategies. We name the first strategy the central hit strategy. This strategy is useful to find drug target candidates in anti-infectious and in anti-cancer therapies. The second strategy is named network influence strategy. This strategy uses systems-level knowledge to find drug target candidates in therapies of polygenic, complex diseases (Fig. 19). In the central hit strategy our aim is to damage the network integrity of the infectious agent or of the malignant cell in a selective manner. For this, detailed knowledge of the structural differences of host/parasite or healthy/malignant networks can help. In the network influence strategy we would like to shift back the malfunctioning network to its normal state. For this, an understanding of network dynamics both in healthy and diseased states is required. Knowledge of the existing drug targets of the particular disease also helps.

System destruction of the central hit strategy finds hubs and central nodes of various networks (the latter are called load-points in metabolic networks), and uses the methods listed in Section 3.6.2 to find essential enzymes of metabolic networks (Jeong et al., 2001; Chin & Samanta, 2003; Agoston et al., 2005; Estrada, 2006; Guimera et al., 2007b; Yu et al., 2007b; Fatumo et al., 2009, 2011; Missiuro et al., 2009; Perumal et al., 2009; Li et al., 2011a). In addition, choke points of metabolic networks, i.e. proteins uniquely producing or consuming a certain metabolite are also excellent targets in anti-infectious therapies (Yeh et al., 2004; Singh et al., 2007). In the case of directed, hierarchical networks nodes at the top of the hierarchy should be attacked by the central hit strategy. Liu et al. (2012) suggested the random upstream attack to find high position nodes in hierarchical networks. Recent work on connections of essential reactions and on superessential reactions (where the latter are needed in all organisms) suggests that essential reactions form a core of metabolic networks (Barve et al., 2012; Ma et al., 2012b). Cytostatic drug targets have also been identified through analysis of cancer-specific human metabolic networks (Folger et al., 2011). Recent anti-cancer strategies mostly use the cancer-specific targeting of signaling networks as we will describe in detail in Section 5.2.

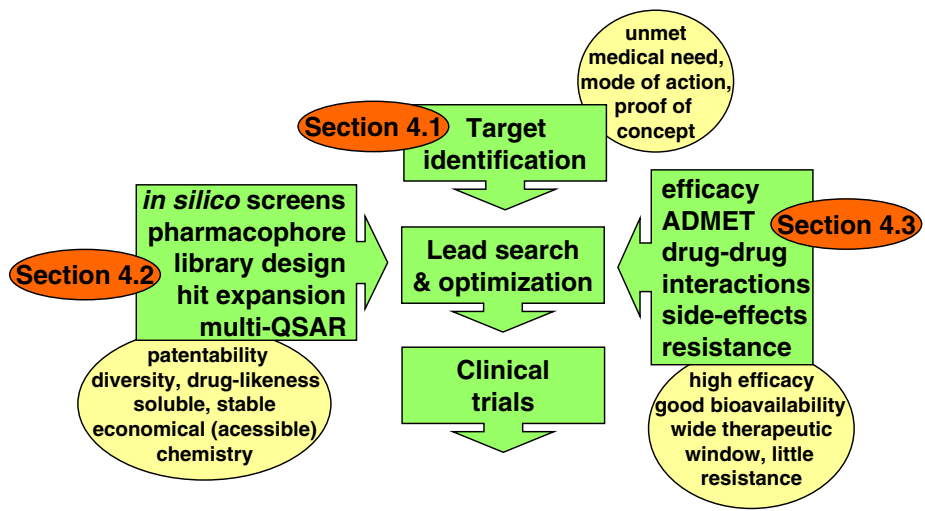

**Fig. 18.** The drug development process. Green boxes illustrate the major stages of the drug development process starting with target identification, followed by hit finding, hit confirmation and hit expansion leading to lead selection/optimization and concluded by clinical trials. Lead search and lead optimization are helped by various methods of chemoinformatics (left side), drug efficiency optimization, ADMET (drug absorption, distribution, metabolism, excretion and toxicity) studies, as well as optimization of drug–drug interactions, side-effects and resistance (right side). Yellow ellipses summarize a few major optimization criteria, while orange ellipses refer to the subsections of Section 4 discussing the given drug development stage.

**Table 9**
Drug-design related resources.

| Name | Content | Website | References |
|---|---|---|---|
| *Section 4.1. Drug target prioritization, identification and validation* | | | |
| Pubchem | Repository of small molecule biological activities | http://pubchem.ncbi.nlm.nih.gov | Wang et al., 2010b |
| chEMBLdb | Chemical properties and biological activities of drug-like molecules | https://www.ebi.ac.uk/chembldb | Gaulton et al., 2012 |
| DailyMed | Drug package insert texts | http://dailymed.nlm.nih.gov | de Leon, 2011 |
| DrugBank | Integrated drug and drug target information resource | http://drugbank.ca | Knox et al., 2011 |
| PharmGKB | Integrated drug and drug target information resource | http://pharmgkb.org | Thorn et al., 2010 |
| Therapeutic Target Database | | http://bidd.nus.edu.sg/group/cjttd/TTD_HOME.asp | Zhu et al., 2012b |
| MATADOR | | http://matador.embl.de | Günther et al., 2008 |
| Supertarget | | http://insilico.charite.de/supertarget | Hecker et al., 2012 |
| KEGG DRUG | | http://genome.jp/kegg/drug | Kanehisa et al., 2012 |
| TDR | Drug targets of neglected tropical diseases | http://tdrtargets.org | Agüero et al., 2008 |
| PDTD—Potential Drug Target Database | Information on drug targets | http://dddc.ac.cn/pdtd | Gao et al., 2008 |
| DTome | Drug–target network construction tool | http://bioinfo.mc.vanderbilt.edu/DTome | Sun et al., 2012 |
| My-DTome | Myocardial infarction-related drug target interactome | http://my-dtome.lu | Azuaje et al., 2011 |
| PROMISCUOUS | Interactome-based database for drug-repurposing | http://bioinformatics.charite.de/promiscuous | von Eichborn et al., 2011 |
| MANTRA | mRNA expression profile-based server for drug-repurposing | http://mantra.tigem.it | Iorio et al., 2010 |
| CDA | Combinatorial drug assembler and drug repositioner (mRNA expression profiles, signaling networks) | http://cda.i-pharm.org | Lee et al., 2012c |
| *Section 4.2.1. Hit finding for ligand binding sites* | | | |
| STITCH 3 | Integrated network resource of chemical–protein interactions | http://stitch.embl.de | Kuhn et al., 2012 |
| CARLSBAD | Cytoscape plug-in connecting common chemical patterns to biological targets via small molecules | http://carlsbad.health.unm.edu | |
| BindingDB | Binding affinity data for almost a million protein–ligand pairs | http://bindingdb.org | Liu et al., 2007a |
| BioDrugScreen | Structural protein ligand interactome + scoring system | http://biodrugscreen.org | Li et al., 2010c |
| CREDO | a protein–ligand interaction database including a wide range of structural information | http://www-cryst.bioc.cam.ac.uk/databases/credo | Schreyer & Blundell, 2009 |
| *Section 4.2.2. Hit finding for protein–protein interaction hot spots* | | | |
| TIMBAL | A curated database of ligands inhibiting protein–protein interactions | http://www-cryst.bioc.cam.ac.uk/databases/timbal | Higueruelo et al., 2009 |
| Dr. PIAS—Druggable Protein-protein Interaction Assessment System | Machine learning-based web-server to judge, if a protein–protein interaction is druggable | http://drpias.net | Sugaya & Furuya, 2011; Sugaya et al., 2012 |
| *Section 4.3. Drug efficiency, ADMET, drug–drug interactions, side-effects and resistance* | | | |
| Supertarget | Drug metabolism information | http://insilico.charite.de/supertarget | Hecker et al., 2012 |
| KEGG DRUG | | http://genome.jp/kegg/drug | Kanehisa et al., 2012 |
| ACToR | Integrated toxicity resource | http://actor.epa.gov | Judson et al., 2012 |
| DITOP | Drug-induced toxicity related protein database | http://bioinf.xmu.edu.cn:8080/databases/DITOP/index.html | Zhang et al., 2007 |
| DCDB | Drug combination database | http://www.cls.zju.edu.cn/dcdb | Liu et al., 2010b |
| DTome | Adverse drug–drug interactions | http://bioinfo.mc.vanderbilt.edu/DTome | Sun et al., 2012 |
| KEGG DRUG | | http://genome.jp/kegg/drug | Kanehisa et al., 2012 |
| SIDER | Drug side-effect resource | http://sideeffects.embl.de | Kuhn et al., 2010 |
| DRAR-CPI | Drug-binding structural similarity based server for adverse drug reaction and drug repositioning | http://cpi.bio-x.cn/drar | Luo et al., 2011 |
| DvD | An R/Cytoscape plug-in assessing system-wide gene expression data to predict drug side effects and drug repositioning | http://www.ebi.ac.uk/saezrodriguez/DVD | Pacini et al., 2013 |
| NPC | Approved and experimental drugs useful for drug-repositioning | http://tripod.nih.gov/npc | Huang et al., 2011b |
| SePreSA | Binding pocket polymorphism-based serious adverse drug reaction predictor | http://sepresa.bio-x.cn | Yang et al., 2009a; Yang et al., 2009b |
| *Section 4.3. Drug efficiency, ADMET, drug–drug interactions, side-effects and resistance (continuation)* | | | |
| SADR-Gengle | PubMed record text mining-based data on 6 serious adverse drug reactions | http://gengle.bio-x.cn/SADR | Yang et al., 2009c |

At the protein structure level the central hit strategy may mostly target active sites. Targeting allosteric regulatory sites may be shared by both strategies. These cavity-like binding sites are easier to target than the flat, 'hot-spot'-type protein–protein interfaces mostly involved in the network influence strategy (Keskin et al., 2007;

Ozbabacan et al., 2010). We will discuss the network-based identification of ligand binding sites in Sections 4.2.1 and 4.2.2.

Network-based methods of the network influence strategy are much less developed than those of the central hit strategy. Using the network influence strategy we need to conquer system robustness to push the cell back from the attractor of the diseased state to that of the healthy state, which is a difficult task—as we summarized in Section 2.5.2 on network dynamics. Nodes with intermediate connection numbers located in vulnerable points of disease-related networks (such as in inter-modular, bridging positions) driving disease-specific network traffic are preferred targets of the network influence strategy (Kitano, 2004a, 2004b, 2007; Ciliberti et al., 2007;

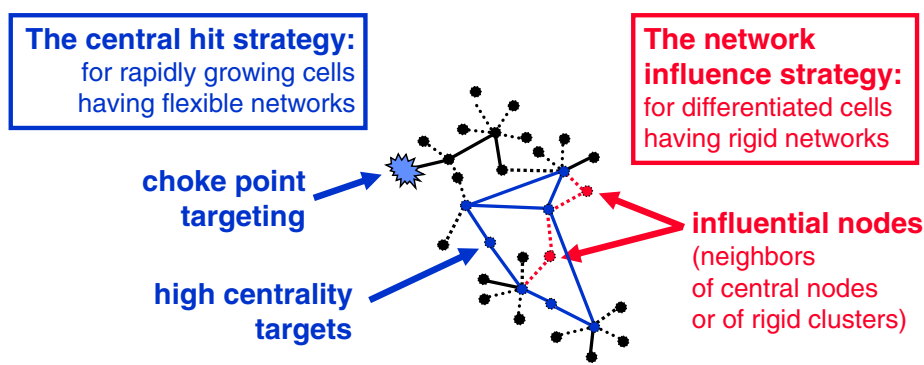

**Fig. 19.** Illustrative figure on the two major strategies to find network nodes as drug targets. The central hit strategy (represented by dark blue symbols) is useful to find drug targets against infectious agents or in anti-cancer therapies. These cells are presumed to have flexible networks. The central hit strategy targets central nodes (often forming a core of the network) or 'choke points', which are peripheral nodes uniquely producing or consuming a cellular metabolite. The network influence strategy (represented by red symbols) is needed to use the systems-level knowledge to find the targets in therapies of polygenic, complex diseases. The differentiated cells of these diseases are presumed to have rigid networks. Targeting central nodes here may cause an 'over-excitement' of the system leading to side-effects and toxicity. Thus the network influence strategy targets nodes, which are neither hubs nor otherwise central nodes themselves, but occupying strategically important disease-specific network positions able to influence central nodes. (In a typical case the network influence strategy targets are neighbors of central nodes exerting an indirect influence on the central nodes often representing the 'real targets'.) Solid lines represent network edges with high weight, while dashed lines represent network edges with low weight.

Antal et al., 2009; Hase et al., 2009; Zanzoni et al., 2009; Fliri et al., 2010; Cornelius et al., 2011; Farkas et al., 2011; Yu & Huang, 2012). In signaling networks preferred nodes of the network influence strategy inhibit certain outputs of the signaling network, while leaving others intact redirecting the signal flow in the network (Dasika et al., 2006; Ruths et al., 2006; Pawson & Linding, 2008). The network influence strategy often targets network segments, e.g. disease-modules (Cho et al., 2012), as we will discuss in Section 4.1.7. Edgetic drugs (Section 4.1.2.), multi-target drugs (Section 4.1.5.) and allo-network drugs (Section 4.1.6.) are all very promising forms of the network influence strategy.

It is important to note that the central hit strategy is applied against rapidly growing cells, such as those of infectious agents or cancer cells. Networks of these cells are in an 'exploratory phase', have larger entropy (West et al., 2012), and are presumably much more flexible, than the networks of differentiated cells. Efficient targeting of flexible systems having a high plasticity needs the targeting of their central nodes/edges. This is the major mode through which the central hit strategy operates.

In contrast, the network influence strategy is applied to differentiated cells. Networks of these cells are in an 'optimized phase', where the optimum was re-set for a diseased cell to a different attractor of its state-space than that of healthy cells. Networks of differentiated cells are presumably much more rigid than the networks of rapidly growing, undifferentiated or dedifferentiated cells. This is true all the more, since diseases are typically recognized by their late-appearing manifestations (Loscalzo and Barabasi, 2011), where the corresponding network already reached the disease-specific, rigid state. Targeting the central nodes/edges of systems having a low plasticity may easily 'over-saturate' the system leading to a change, which becomes too large to be selective, and causes side-effects and toxicity. Therefore, the network influence strategy often needs an indirect approach, where e.g. neighbors of the real target are targeted (allo-network drugs), or multiple targets are targeted 'mildly' (multi-target drugs), and their indirect and/or superposing effects lead to the reconfiguration of diseased network state back to normal.

In general, the strategy-pair we described here for drug action appears valid, and is a key for efficient modification of molecular structures, as well as neuronal and social systems at their different states besides the cellular networks discussed here. Thus, the best targeting strategy depends on the extent of network rigidity. Flexible, plastic systems are 'under-defined', and dissipate the perturbations well. These plastic systems may generally require a well-defined attack in the form of targeting of their central nodes/edges. Note that central nodes form the most rigid segments of these flexible, plastic networks, which transmit the attack with the best efficiency. On the other hand, rigid systems are 'well-defined', and transmit (but not dissipate) the perturbations well. Thus optimal modification of rigid systems may be achieved by an indirect, 'under-defined' attack of the neighbors of their central nodes or rigid clusters. Note that nodes connecting rigid clusters (or being neighbors of central nodes) constitute the most flexible segments of these rigid networks, which dissipate the perturbation best, thus their attack does not over-excite the network. Moreover, rigid systems often contain multiple rigid clusters, and are often multi-modular (Mihalik & Csermely, 2011; Gáspár & Csermely, 2012). A single central hit may not optimally target a multi-modular, multi-centered system, which makes the network influence strategy even more efficient.

Network effects of existing drugs (e.g. in the form of drug target networks detailed in Section 4.1.3) may help in finding disease-specific network control-points. Drugs showing an mRNA expression profile that is strongly anticorrelated with a disease expression profile might actually reverse some of the disease effects and can be used for drug-repurposing. On the contrary, positively correlated profiles may reveal side effects (Dudley et al., 2011; Sirota et al., 2011; Iskar et al., 2012; Pacini et al., 2013). Reverse-engineering methods finding the underlying network structure from complex dynamic system output data (such as genome-wide mRNA expression patterns, signaling network or metabolome, see Section 2.2.3), as well as discriminating the primary targets from secondarily affected network nodes help in identifying control nodes directing network dynamics (Gardner et al., 2003; di Bernardo et al., 2005; Hallén et al., 2006; Lamb et al., 2006; Xing & Gardner, 2006; Lehár et al., 2007; Madhamshettiwar et al., 2012). The identification of disease-specific control-points of network dynamics will be an exciting task in the near future.

Disease-specificity may well be hierarchical. Suthram et al. (2010) identified 59 modules out of the 4620 modules of the human interactome, which are dysregulated in at least half of the 54 diseases tested, and were enriched in known drug targets. Influence-cores of the interactome, signaling, metabolic and other networks may be involved in the regulation of many more diseases than the connection-core (e.g. hub containing rich club) or periphery of these networks.

Potential methods to find influential nodes redirecting perturbations, affecting cellular cooperation or asserting network control have been described in Sections 2.3.4, 2.5.2 and 2.5.3 (Xiong & Choe, 2008; Antal et al., 2009; Kitsak et al., 2010; Luni et al., 2010; Farkas et al., 2011; Liu et al., 2011, 2012; Banerjee & Roy, 2012; Cowan et al., 2012; Mones et al., 2012; Nepusz & Vicsek, 2012; Valente, 2012; Wang et al., 2012a). Influential nodes may have a hidden influence, like those highly unpredictable, 'creative' nodes, which may delay critical transitions of diseased cells (see Sections 2.2.2 and 2.5.2 for more details; Csermely, 2008; Scheffer et al., 2009; Farkas et al., 2011; Sornette & Osorio, 2011; Dai et al., 2012). Finding sets of influence-core nodes with fewer side-effects, or periphery nodes specifically influencing an influence-core or connection-core node, will be the subject of Sections 4.1.5 and 4.1.6 on multi-target drugs and allo-network drugs (Nussinov et al., 2011), respectively.

### 4.1.2. Edgetic drugs: edges as targets

Perturbations of selected network edges give a grossly different result than the partial inhibition (or deletion) of the whole node. Development of drugs targeting network edges (recently called: edgetic drugs) has a number of advantages (Arkin & Wells, 2004; Keskin et al., 2007; Sugaya et al., 2007; Dreze et al., 2009; Zhong et al., 2009; Schlecht et al., 2012; Wang et al., 2012b).

- Many disease-associated proteins, e.g. p53, were considered nontractable for small-molecule therapeutics, since they do not have an enzyme activity. In these cases edgetic drugs may offer a solution.
- Edgetic drugs are advantageous, since targeting network edges, i.e. protein–protein interaction, signaling or other molecular networks, is more specific than node targeting. This becomes particularly useful, when a protein simultaneously participates in two complexes having different functions, where only one of these functions is disease-related, like in the case of the mammalian target of rapamycin, mTOR (Huang et al., 2004; Agoston et al., 2005; Ruffner et al., 2007; Zhong et al., 2009; Wang et al., 2012b).
- Due to its larger selectivity, edge targeting may provide an efficient solution in targeting networks of multigenic diseases described as the network influence strategy in the preceding section. Edge targeting may also be used in the central hit strategy (targeting whole network-encoded systems) in the case of cancer, where selectivity may be more limited than in targeting of infectious agents. Importantly, the selectivity of edgetic drugs is not unlimited: hitting frequent interface motifs in a network may be as destructive as eliminating hubs. However, "interface-attack" may affect functional changes better than the attack of single proteins (Engin et al., 2012).

Edgetic drug development has inherent challenges. Interacting surfaces lack small, natural ligands, which may offer a starting point for drug design. Moreover, protein–protein binding sites involve large, flat surfaces, which are difficult to target. However, these flat surfaces often contain hot spots, which cluster to hot regions corresponding to a smaller set of key residues, which may be efficiently targeted by a drug of around 500 Da (Keskin et al., 2007; Wells & McClendon, 2007; Ozbabacan et al., 2010). We showed the usefulness of protein structure networks in finding hot spots in Section 3.2.4, and will summarize the possibilities to define edgetic drug binding sites in Section 4.2.2.

In one of the few systematic studies on edgetic drugs, Schlecht et al. (2012) constructed an assay to identify changes in the yeast interactome in response to 80 diverse small molecules, including the immunosuppressant FK506, which specifically inhibited the interaction between aspartate kinase and the peptidyl-prolyl-cis–trans isomerase, Fpr1. Sugaya et al. (2007) provided an in silico screening method to identify human protein–protein interaction targets. Edgetic perturbation of a *C. elegans* Bcl-2 ortholog, CED-9, resulted in the identification of a new potential functional link between apoptosis and centrosomes (Dreze et al., 2009). The TIMBAL database is a hand curated assembly of small molecules inhibiting protein–protein interactions (http://www-cryst.bioc.cam.ac.uk/databases/timbal; Higueruelo et al., 2009).

The Dr. PIAS server offers a machine learning-based assessment if a protein–protein interaction is druggable (http://drpias.net; Sugaya & Furuya, 2011; Sugaya et al., 2012).

Current development of edgetic drugs is mostly concentrated on protein–protein interaction networks. (We note here that most metabolic network-related drugs are by definition 'edgetic drugs', since in these networks target-enzymes constitute the edges between metabolites.) Signaling networks and gene interaction networks (including chromatin interaction networks) are promising fields of edgetic drug development. Scaffolding proteins and signaling mediators are particularly attractive targets of edgetic drug design efforts (Klussmann & Scott, 2008). In conclusion of this section, we list a few other future aspects of edgetic drug design.

- To date, the preferential topology of edge-targets in the human interactome has not been systematically addressed. Thus, currently we do not know, if indeed such a preference exists. Similarly, little attention has been paid to systematic studies of edge-weights, i.e. binding affinity-related drug target preference. Low-affinity binding is easier to disrupt, but interventions may not be that efficient. Disruption of high-affinity interactions may be more challenging (Keskin et al., 2007).
- Those interactions of intrinsically disordered proteins, which couple binding with folding, display a large decrease in conformational entropy, which provides a high specificity and low affinity. This pair of features is useful for regulation of protein–protein interactions and signaling, and this mechanism is widely used in human cells. Coupled binding and folding interactions often involve localized, small hydrophobic interaction surfaces, which provide a feasible targeting option in edgetic drug design (Cheng et al., 2006).
- Both low-probability interactions and interactions of intrinsically disordered proteins involve transient binding complexes. Modulation of these transient edges by 'interfacial inhibition' (Pommier & Cherfils, 2005; Keskin et al., 2007) may be an option in future edgetic drug design.
- Edgetic drugs are usually inhibiting interactions (Gordo & Giralt, 2009). Stabilization of specific interactions is an area of great promise in drug design as we will discuss in Section 4.1.6 on allo-network drugs (Nussinov et al., 2011). Since the changed cellular environment in diseases often induces protein unfolding, general stabilizers of protein–protein interactions in normal cells, such as chemical chaperones, or chaperone inducers and co-inducers (Vígh et al., 1997; Sőti et al., 2005; Papp & Csermely, 2006; Crul et al., 2013) offer an exciting therapeutic area of network-wide restoration of protein–protein interactions.

### 4.1.3. Drug target networks

A broader representation of drug target networks are the protein-binding site similarity networks, where network edges between two proteins are defined by not only common, FDA-approved drugs, but also by a wide variety of common natural ligands and chemical compounds, as well as by binding site structural similarity measures. We list a few approaches to construct such protein binding site similarity networks below.

- Protein binding site networks can be constructed by large-scale experimental studies. One of these systematic studies examined naturally binding hydrophobic molecule profiles of kinases and proteins of the ergosterol biosynthesis in yeast using mass spectrometry. Hydrophobic molecules, such as ergosterol turned out to be potential regulators of many unrelated proteins, such as protein kinases (Li et al., 2010b).
- Protein binding site similarity networks may be constructed using a simplified representation of binding sites as geometric patterns, or numerical fingerprints. Here similarities are ranked by similarity scores based on the number of aligned features (Kellenberger et al., 2008).

- Pocket frameworks encoding binding pocket similarities were also used to create protein binding site similarity networks (Weisel et al., 2010). Pocket frameworks are reduced, graph-based representations of pocket geometries generated by the software PocketGraph using a growing neural gas approach. Another pocket comparison method, SMAP-WS combines a pocket finding shape descriptor with the profile-alignment algorithm, SOIPPA (Ren et al., 2010).
- Enzyme substrate and ligand binding sites have been compared using cavity alignment. Clustering of cavity space resembles most the structure of chemical ligand space and less that of sequence and fold spaces. Unexpected links of consensus cavities between remote targets indicated possible cross-reactivity of ligands, suggested putative side-effects and offered possibilities for drug repositioning (Zhang & Grigorov, 2006; Liu et al., 2008a; Weskamp et al., 2009).
- Andersson et al. (2010) proposed a method avoiding geometric alignment of binding pockets and using structural and physicochemical descriptors to compare cavities. This approach is similar to QSAR models of comparison detailed in Section 3.1.3.

Identifying clusters of proteins with similar binding sites may help drug repositioning, and could be a starting point for designing multi-target drugs as we will describe in the following two sections. Binding site similarities help in finding appropriate chemical molecules for new drug target candidates as described in Section 4.2. However, designing drugs for a group of targets with similar binding sites is challenging due to low specificity as exemplified by the drug design efforts against the ATP binding sites of protein kinases. Construction and analysis of protein binding site similarity networks in these cases can be helpful to identify proteins, whose active sites are different enough to be targeted selectively. Using 491 human protein kinase sequences, Huang et al. (2010b) constructed similarity networks of kinase ATP binding sites. The recent tyrosine kinase target, EphB4 belonged to a small, separated cluster of the similarity network supporting the experimental results of selective EhpB4 inhibition.

Signaling components, particularly membrane receptors and transcription factors form a major segment of drug target networks. Drug target networks are bipartite networks having drugs and their targets as nodes, and drug–target interactions as edges. These networks can be projected as drug similarity networks (where two drugs are connected, if they share a target). We summarized these projections as similarity networks in Section 3.1.3. In the other projection of drug–target networks, nodes are the drug targets, which are connected, if they both bind the same drug (Keiser et al., 2007, 2009; Ma'ayan et al., 2007; Yildirim et al., 2007; Hert et al., 2008; Yamanishi et al., 2008; Bleakley & Yamanishi, 2009; van Laarhoven et al., 2011). We describe the drug development applications of this projection in the remaining part of this section.

Drug target networks are particularly useful for comparisons of drug target proteins, since such a network comparison can be more informative pharmacologically than comparing protein sequences or protein structures. Drug target networks are modular: many drug targets are clustered by ligand similarity even though the targets themselves have minimal sequence similarity. This is a major reason, why drug target networks were successfully used to predict and experimentally verify novel drug actions (Keiser et al., 2007, 2009; Ma'ayan et al., 2007; Yildirim et al., 2007; Hert et al., 2008; Yamanishi et al., 2008, 2010; Bleakley & Yamanishi, 2009; van Laarhoven et al., 2011; Nacher & Schwartz, 2012; Mei et al., 2013).

Chen et al. (2012a) merged protein–protein similarity, drug similarity and drug–target networks and applied random walk-based prediction on this meta-network to predict drug–target interactions. Riera-Fernandez et al. (2012) developed a Markov–Shannon entropy-based numerical quality score to measure connectivity quality of drug–target networks extended by both the chemical structure networks of the drugs and the protein structure networks of their targets. As we will detail in Section 4.1.6 on allo-network drugs

(Nussinov et al., 2011), the integration of protein structure networks and protein–protein interaction networks may significantly enhance the success-rate of drug target network-based predictions of novel drug target candidates. Importantly, many drugs do not target the actual disease-associated proteins but proteins in their network-neighborhood (Yildirim et al., 2007; Keiser et al., 2009). Drugs having a target less than 3 or more than 4 steps from a disease-associated protein in human signaling networks have significantly more side-effects, and fail more often (Wang et al., 2012c). This substantiates the importance of the targeting of 'silent', 'by-stander' proteins further, which may influence the disease-associated targets in a selective manner (Section 4.1.6; Nussinov et al., 2011).

We listed a number of drug target databases and resources useful for the construction of drug–target networks in Table 9 at the beginning of Section 4. However, a refined representation of a drug target network should also include protein conformations (Fig. 20). Drugs may favor or disfavor certain protein conformations, and therefore this information is important for a more detailed understanding of drug action (Isin et al., 2012).

Indirect drug target networks may also be constructed using available data on human diseases, patients, their symptoms, therapies, or the systems-level effects of drug-induced perturbations (see Fig. 6 in Section 1.3.1; Spiró et al., 2008). Recently, several approaches extended drug/target datasets. Vina et al. (2009) assessed drug/target interaction pairs in a multi-target QSAR analysis enriching the dataset with chemical descriptors of targets and affinity scores of drug–target interactions. Wang et al. (2011b) assembled the Cytoscape (Smoot et al., 2011) plug-in of the integrated Complex Traits Networks (iCTNet, http://flux.cs.queensu.ca/ictnet) including phenotype/single-nucleotide polymorphism (SNP) associations, protein–protein interactions, disease–tissue, tissue–gene and drug–gene relationships. Balaji et al. (2012) compiled the integrated molecular interaction database (IMID, http://integrativebiology.org) containing protein–protein interactions, protein–small molecule interactions, associations of interactions with pathways, species, diseases and Gene Ontology terms with user-selected integration of manually curated and/or automatically extracted data. These and other complex approaches to drug target networks (Yamanishi et al., 2010, 2011; Savino et al., 2012; Tabei et al., 2012; Takarabe et al.;, 2012) will lead to the development of prediction techniques of novel drug targets, and improve drug efficiency, as well as ADMET, drug–drug interaction, side-effect and resistance profiles.

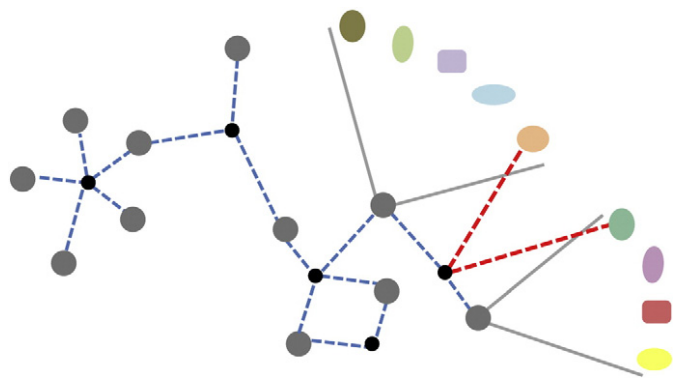

**Fig. 20.** A refined representation of a drug target network includes protein conformations. In current drug target network representations drug targets (gray circles) are interpreted as single entities connected through drugs (black circles). In these representations protein conformations preferred or dispreferred by a certain drug are ignored. For a more complete understanding of the interactions of drugs to their targets a target should be represented by its different functionally relevant conformations (differently colored shapes within gray line enclosed areas). Drug targets that are represented by single structures are connected to drugs by blue dashed lines. The target conformations that preferentially bind a drug are connected by red dashed lines.
Reproduced by permission from Isin et al. (2012).

### 4.1.4. Network-based drug repositioning

Drug repositioning (or drug repurposing) aims to find a new therapeutic modality for an existing drug, and thus provides a cost-efficient way to enrich the number of available drugs for a certain therapeutic purpose. Drug repurposing uses a compound having a well-established safety and bioavailability profile, and a proven formulation and manufacturing process, as well as a well-characterized pharmacology. Most drug repositioning efforts use large screens of existing drugs against a multitude of novel targets (Chong & Sullivan, 2007). The pharmacological network approach asks, given a pattern of chemistry in the ligands, which targets a particular drug may bind (Kolb et al., 2009)? Here we list network-based methods mobilizing and efficiently using systems-level knowledge for rational drug repositioning.

- Analysis of common segments of protein–protein interaction and signaling networks affected by different drugs or participating in different diseases may reveal unexpected cross-reactions suggesting novel options for drug repurposing (Bromberg et al., 2008; Kotelnikova et al., 2010; Ye et al., 2012). As an example of these efforts, PROMISCUOUS (http://bioinformatics.charite.de/promiscuous) offers a web-tool for protein–protein interaction network-based drug-repositioning (von Eichborn et al., 2011).

- As an extension of the above approach, analysis of the complex drug similarity networks, by modularization, edge-prediction or by machine learning methods, described in Section 3.1.3 (see Table 5 there), may show unexpected links between remote drug targets indicating possible cross-reactivity of existing drugs with novel targets (Zhang & Grigorov, 2006; Liu et al., 2008a; Weskamp et al., 2009; Zhao & Li, 2010; Gottlieb et al., 2011a; Chen et al., 2012a; Cheng et al., 2012a, 2012b; Lee et al., 2012b). Network-based comparison of drug-induced changes in gene expression profiles (combined with disease-induced gene expression changes, disease–drug associations, interactomes, or signaling networks), was used to suggest unexpected, novel uses of existing drugs (Hu & Agarwal, 2009; Iorio et al., 2010, MANTRA server, http://mantra.tigem.it; Kotelnikova et al., 2010; Suthram et al., 2010; Gottlieb et al., 2011a; Luo et al., 2011, DRAR-CPI server, http://cpi.bio-x.cn/drar; Jin et al., 2012; Lee et al., 2012c, CDA server: http://cda.i-pharm.org; Pacini et al., 2013, DvD server as a Cytoscape plug-in: http://www.ebi.ac.uk/saezrodriguez/DVD).

- Genome-wide association studies (GWAS) may also be used to construct drug-related networks helping drug repositioning including in a personalized manner (Zanzoni et al., 2009; Coulombe, 2011; Cowper-Sal lari et al., 2011; Fang et al., 2011; Hu et al., 2011; Li et al., 2012b; Sanseau et al., 2012). Important future applications may use the comparison of phosphoproteome and metabolome data to reveal further drug repositioning option, including personalized drug application protocols.

- Drug target networks (including drug-binding site similarity networks and drug–target–disease networks) summarized in the preceding section help in drug repositioning. Modularization or edge prediction of these networks may reveal novel applications of existing drugs (Keiser et al., 2007, 2009; Ma'ayan et al., 2007; Yildirim et al., 2007; Hert et al., 2008; Yamanishi et al., 2008; Bleakley & Yamanishi, 2009; Kinnings et al., 2010; Mathur & Dinakarpandian, 2011; van Laarhoven et al., 2011; Daminelli et al., 2012; Nacher & Schwartz, 2012).

- Central drugs of drug–therapy networks, where two drugs are connected, if they share a therapeutic application (Nacher & Schwartz, 2008), such as inter-modular drugs connecting two otherwise distant therapies, may reveal novel drug indications. Drug–disease networks have also been constructed and used for this purpose (Yildirim et al., 2007; Qu et al., 2009). Moreover, disease–disease networks (Goh et al., 2007; Rzhetsky et al., 2007; Feldman et al., 2008; Spiró et al., 2008; Hidalgo et al., 2009; Barabási et al., 2011; Zhang et al., 2011a) and the other disease and drug-related network representations we listed in Section 1.3.1 (see Fig. 6 there; Spiró et al., 2008) may also be used for drug repositioning. Edge prediction methods (detailed in Section 2.2.2) and network-based machine learning methods may also be applied to these networks to uncover novel drug–therapy associations.

- Tightly interacting modules of drug–drug interaction networks (Yeh et al., 2006; Lehár et al., 2007) may also reveal unexpected, novel therapeutic applications.

- Side-effects of drugs, summarized in Section 4.3.5, may often reveal novel therapeutic areas. Shortest path, random walk and modularity analysis of side-effect similarity networks offers a number of novel options for network-based drug repositioning (Campillos et al., 2008; Yamanishi et al., 2010; Oprea et al., 2011; Takarabe et al., 2012).

Network-related datasets and methods to reveal drug–drug interactions (Section 4.3.4), or drug side-effects (Section 4.3.5) may all give clues for drug re-positioning. Drug repositioning also has challenges, such as validation of the drug candidate from incomplete and outdated data, and the development of novel types of clinical trials (Mei et al., 2012). However, most network-based methods helping drug repositioning may also be used to predict multi-target drugs, an area we will summarize in the next section.

### 4.1.5. Network polypharmacology: multi-target drugs

Robustness of molecular networks may often counteract drug action on single targets thus preventing major changes in systems-level outputs despite the dramatic changes in the target itself (see Section 2.5.2; Kitano, 2004a; Kitano, 2004b; Papp et al., 2004; Pál et al., 2006; Kitano, 2007; Tun et al., 2011). Moreover, most cellular proteins belong to multiple network modules in the human interactome, signaling or metabolic networks (Palla et al., 2005; Kovács et al., 2010; Wang et al., 2012d). As a consequence, efficient targeting of a single protein may influence many cellular functions at the same time. In contrast, efficient restoration of a particular cellular function to that of the healthy state (or efficient cell damage in anti-cancer strategies) can often be accomplished only by a simultaneous attack on many proteins, wherein the targeting efficiency on each protein may only be partial. These target sets preferentially contain proteins with intermediate number of neighbors having an intermediate level of influence of their own (Hase et al., 2009; Wang et al., 2012d).

The above systems-level considerations explain the success of polypharmacology, also called as modular pharmacology, i.e. the development and use of multi-target drugs (Fig. 21; Ginsburg, 1999; Csermely et al., 2005; Mencher & Wang, 2005; Millan, 2006; Hopkins, 2008; Wang et al., 2012d). The goal of polypharmacology is "to identify a compound with a desired biological profile across multiple targets whose combined modulation will perturb a disease state" (Hopkins, 2008). Multiple targeting is a well-established strategy. Snake or spider venoms, part defense strategies are all using multi-component systems. Traditional medicaments and remedies often contain multi-component extracts of natural products. Combinatorial therapies are used with great success to treat many types of diseases, including AIDS, atherosclerosis, cancer and depression (Borisy et al., 2003; Keith et al., 2005; Dancey & Chen, 2006; Millan, 2006; Yeh et al., 2006; Lehár et al., 2007). Importantly, more than 20% of the approved drugs are multi-target drugs (Ma'ayan et al., 2007; Yildirim et al., 2007; Nacher & Schwartz, 2008). Many drugs interact with more than one transporter, which increases the complexity of polypharmacology (Kell et al., 2013). Moreover, multi-target drugs have an increasing market-value (Lu et al., 2012). Multi-target drugs possess a number of beneficial network-related properties, which we list below.

- Multi-target drugs can be designed to act on a selected set of primary targets influencing a set of key, therapeutically relevant secondary targets.

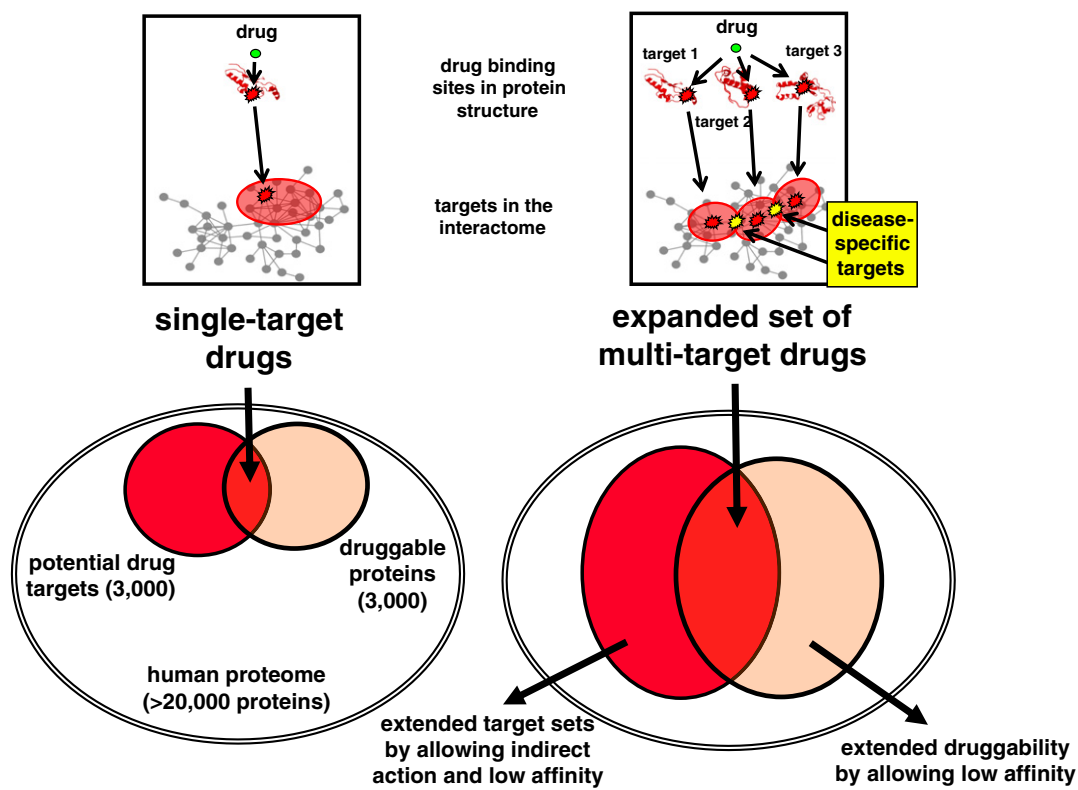

**Fig. 21.** Multi-target drugs are target multipliers. The top left panel and the red circle of the bottom left part of the figure show the targets of single-target drugs situated in pharmacologically interesting pathways and the hits of chemical proteomics, which represent those proteins, which can interact with druggable molecules. (The numbers are only approximate, and in case of the human proteome contain only the non-redundant proteins.) The overlap between the two sets constitutes the 'sweet spot' of drug discovery (Brown & Superti-Furga, 2003). On the right side of the figure the expansion of the 'sweet spot' is shown by multi-target drugs. The top left part illustrates the action of multi-target drugs. Yellow asterisks highlight the indirect targets, where the changes initiated by the multiple primary targets are superposed. It is a significant advantage, if these targets are disease-specific. On the bottom left part the indirect targets of multi-target action and the allowed low affinity binding of multi-target drugs both expand the number of pharmacologically relevant targets, while low-affinity binding enlarges the number of druggable proteins. The overlap of the two groups (the 'sweet spot') displays a dramatic increase.

- Multiple targeting may need a compromise in binding affinity. However, even low-affinity binding multi-target drugs are efficient: in our earlier study a 50% efficient, partial, but multiple attack on a few sites of *E. coli* or yeast genetic regulatory networks caused more damage than the complete inhibition of a single node (Agoston et al., 2005; Csermely et al., 2005).
- Via the above, 'indirect' targeting, and via their low affinity binding, multi-target drugs may avoid the presently common dual-trap of drug-resistance and toxicity (Lipton, 2004; Csermely et al., 2005; Lehár et al., 2007; Zimmermann et al., 2007; Ohlson, 2008; Savino et al., 2012).
- Due to their low affinity binding, multi-target drugs may often stabilize diseased cells, which sometimes may be at least as beneficial as their primary therapeutic effect (Csermely et al., 2005; Korcsmáros et al., 2007; Csermely, 2009; Farkas et al., 2011).

In summary, multi-target drugs offer a magnification of the 'sweet spot' of drug discovery, where the 'sweet spot' represents those few hundred proteins, which are both parts of pharmacologically important pathways, and are druggable (Brown & Superti-Furga, 2003). The resulting beneficial effects have two reasons. First, both indirect and partial targeting by multi-target drugs expand the number of possible targets. Second, low-affinity binding eases druggability constraints, and allows the targeting of partially hydrophilic molecules by orally-deliverable, hydrophobic molecules. These two effects cause a remarkable increase of the drug targets situated in the overlap region of the potential target and druggable pools. Thus, multi-target drugs are, in fact, target multipliers (Fig. 21; Keith & Zimmermann, 2004; Csermely et al., 2005; Korcsmáros et al., 2007).

We list a number of network-related methods below to find target-sets of drugs by systems-level, rational multi-target design.

- Network efficiency (Latora & Marchiori, 2001), or critical node detection (Boginski & Commander, 2009) may serve as a starting measure to judge network integrity after multi-target action (Agoston et al., 2005; Csermely et al., 2005; Li et al., 2011c). Pathway analysis of molecular networks gives a more complex picture, and may reveal multiple intervention points affecting pathway-encoded functions, utilizing pathway cross-talks, or switching off compensatory circuits of network robustness. Network methods allow the identification of target sets, which disconnect signaling ligands from their downstream effectors with the simultaneous preservation of desired pathways (Dasika et al., 2006; Ruths et al., 2006; Lehár et al., 2007; Jia et al., 2009; Hormozdiari et al., 2010; Kotelnikova et al., 2010; Pujol et al., 2010). Deconvolution of network dynamics showing interrelated dynamics modules, such as those of elementary signaling modes (Wang & Albert, 2011), is a promising approach for future multi-drug design efforts.
- Experimental testing of drug combinations may uncover unexpected effects in drug–interactions, which may be used for selection of multi-target sets (Borisy et al., 2003; Keith et al., 2005; Dancey & Chen, 2006; Yeh et al., 2006; Lehár et al., 2007; Jia et al., 2009; Liu et al., 2010b). Combination therapies may also be designed using network methods, such as the minimal hitting set method (Vazquez, 2009), or a complex method taking into account adjacent network position and action-similarity (Li et al., 2011d). Recently, several iterative algorithms were developed to find optimal target combinations restricting the search to a few combinations out of the potential search space of several millions to billions of combinations (Calzolari et al., 2008; Wong et al., 2008; Small et al., 2011; Yoon, 2011; Zhao et al., 2011a). Pritchard et al. (2013) demonstrated the existence of simple and predictable combination mechanisms using RNA interference signatures. Network-based search algorithms may improve this search efficiency even further in the future. Drug combinations against diseases affecting the cardiovascular and nervous systems have a more concentrated effect radius in the human genetic interaction network than that of immuno-modulatory or anti-cancer agents (Wang et al., 2012e). Network methods were applied to predict and avoid unwanted drug–drug interaction effects and the emergence of multi-drug resistance as we will describe in Sections 4.3.4 and 4.3.6, respectively.

- Side-effect networks connect drugs by the similarity of their side-effects. Shortest path and random walk analysis, as well as the identification of tight clusters, bridges and bottlenecks of these networks (Campillos et al., 2008; Yamanishi et al., 2010; Oprea et al., 2011; Takarabe et al., 2012) combined with the selective optimization of side activities (Wermuth, 2006) may be used to design multi-target drugs.

- The combined similarity networks of chemical molecules including drug targets, various molecular networks (such as interactomes or signaling networks), system-wide biological data (such as mRNA expression patterns) and medical knowledge (such as disease characterization) listed in Tables 5 and 9 (Lamb et al., 2006; Paolini et al., 2006; Brennan et al., 2009; Hansen et al., 2009; Iorio et al., 2009; Li et al., 2009a; Huang et al., 2010a; Zhao & Li, 2010; Azuaje et al., 2011; Bell et al., 2011; Taboureau et al., 2011; Wang et al., 2011b; Balaji et al., 2012; Edberg et al., 2012) may all be used for multi-target drug design using modularization method-, similarity score-, network inference-, Bayesian network- or machine learning-based clustering (Hopkins et al., 2006; Hopkins, 2008; Chen et al., 2009e; Xiong et al., 2010; Yang et al., 2010; Hu et al., 2011; Takigawa et al., 2011; Yabuuchi et al., 2011; Cheng et al., 2012a, 2012b; Lee et al., 2012c; Nacher & Schwartz, 2012; Yu et al., 2012). We note that some of the above methods describe and interpret multi-target action and thus need further development for multi-target prediction.

- Multiple perturbations of interactomes, signaling networks or metabolic networks may uncover alternative target sets causing a similar systems-level perturbation to that of the original target set. Differential analysis of networks in healthy and diseased states may enable an even more efficient prediction (Antal et al., 2009; Farkas et al., 2011). Such perturbation studies were successfully applied to smaller, well-defined networks before using differential equation sets and disease-state specific Monte Carlo simulated annealing (Yang et al., 2008). Assessment of network oscillations may reveal central node sets governing the dynamic behavior (Liao et al., 2011).

- Recent advances in establishing the controllability conditions of large networks and in defining complex network hierarchy measures (Cornelius et al., 2011; Liu et al., 2011, 2012; Banerjee & Roy, 2012; Cowan et al., 2012; Mones et al., 2012; Nepusz & Vicsek, 2012; Wang et al., 2012a; Yazicioglu et al., 2012) may uncover multiple target sets, as shown by the assessment of the controllability of smaller networks (Luni et al., 2010). Controlling sets, which can assign any prescribed set of centrality values to all other nodes by cooperatively tuning the weights of their out-going edges (Nicosia et al., 2012) may also be promising in the identification of multi-target sets.

- Appropriate reduction of the definition of dominant node sets, i.e. sets of nodes reaching all other nodes of the network, may also be used to determine target sets of multi-target drugs (Milenkovic et al., 2011). Minimal dominant node set determination was recently shown to be equal to finding minimal transversal sets of hypergraphs (i.e. a hitting set of a hypergraph, which has a nonempty intersection with each edge; Kanté et al., 2011), which extends this technique to the powerful hypergraph description, where an edge may connect any groups of nodes and not only two nodes. Definition and determination of appropriately limited dominant edge-sets (Milenkovic et al., 2011) constitute a powerful approach for multi-target identification.

- Analysis of transport between multiple sources and sinks in directed networks (Morris & Barthelemy, 2012), such as in signaling networks or in metabolic networks may reveal preferred source sets (encoding target sets of multi-target drugs) preferentially affecting pre-defined sink sets (encoding the desired effects). Throughflow centrality has been recently defined as an important measure of such network configurations (Borrett, 2012). Methods to find conceptually similar seed sets of information spread in social networks (Shakarian & Paulo, 2012) may also be applied to find multi-target drug sets.

- Recently highly powerful methods were published to design and optimize multitarget ligands for polypharmacology profiles (Ajmani & Kulkarni, 2012; Besnard et al., 2012). Besnard et al. (2012) tested 800 automatically designed multi-target ligands of G-protein coupled receptors, and found that 75% of them had a correctly predicted polypharmacology profile. This area will remain an exciting priority of multi-target drug design efforts.

Some of the above methodologies (such as those based on chemical similarity networks) result in target sets, where lead design is a more feasible process. Target sets, which are highly relevant at the systems-level, but have diverse binding site structures may require the identification of a set of indirect targets selectively influencing the desired target set, but posing a more feasible lead development task. We will describe the network-based identification of such indirect targets in the next section describing allo-network drugs (Nussinov et al., 2011). We note that almost all methods finding target sets of multi-target drugs can be used for drug repositioning summarized in the preceding section. Moreover, all these methods are related to the in silico prediction of drug–drug interactions (detailed in Section 4.3.4) and side-effects (summarized in Section 4.3.5).

### 4.1.6. Allo-network drugs: a novel concept of drug action

Allosteric drugs (binding to allosteric effector sites; Fig. 22) are considered to be better than orthosteric drugs (binding to active centers; Fig. 22) due to 4 reasons. 1.) The larger variability of allosteric binding sites than that of active centers causes less allosteric drug-induced side-effects than that of orthosteric drugs. 2.) Allosteric drugs allow the modulation of therapeutic effects in a tunable fashion. 3.) In most cases the effect of allosteric drugs requires the presence of endogenous ligand making allosteric action efficient exactly at the time when the cell needs it. 4.) Allosteric drugs are non-competitive with the endogenous ligand. Therefore, their dosage can be low (DeDecker, 2000; Rees et al., 2002; Goodey & Benkovic, 2008; Lee & Craik, 2009; Nussinov et al., 2011; Nussinov & Tsai, 2012).

We summarized our current knowledge on allosteric action (Fischer, 1894; Koshland, 1958; Straub & Szabolcsi, 1964; Závodszky et al., 1966; Tsai et al., 1999; Jacobs et al., 2003; Goodey & Benkovic, 2008; Csermely et al., 2010; Zhuravlev & Papoian, 2010; Rader & Brown, 2011; Dixit & Verkhivker, 2012; Szilágyi et al., in press) from the point of view of protein interaction networks in Section 3.2.2. In that section we described the rigidity front propagation model as a possible molecular mechanism of the propagation of allosteric changes (Fig. 14; Csermely et al., 2012).

The concept of allosteric drugs can be broadened to allo-network drugs, whose effects can propagate across several proteins via specific, inter-protein allosteric pathways of amino acids activating or inhibiting the final target (Fig. 22; Nussinov et al., 2011). Earlier data already pointed to an allo-network type drug action. Interprotein propagation of allosteric effects (Bray & Duke, 2004; Fliri et al., 2010) and its possible use in drug design (Schadt et al., 2009) were mentioned sporadically in the literature. Moreover, drug–target network studies revealed that in more than half of the established

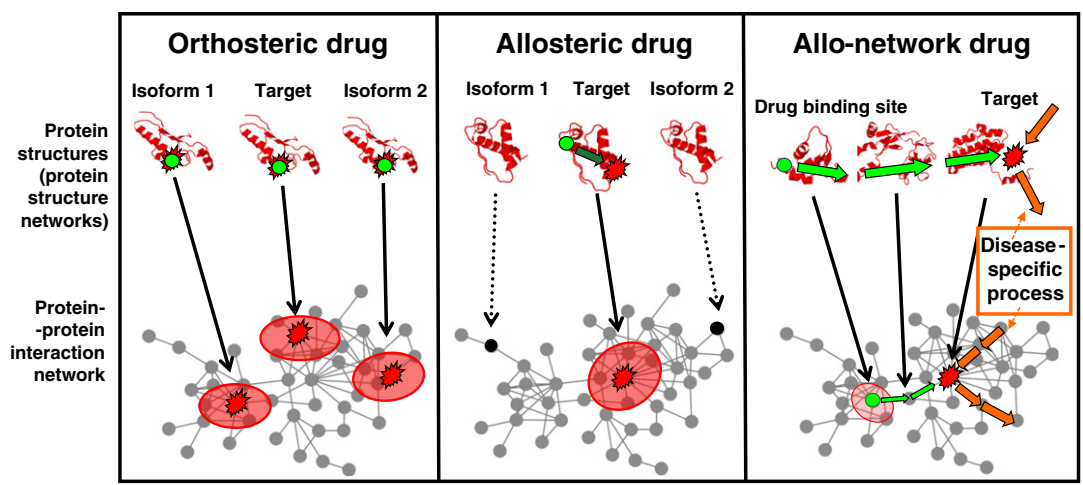

**Fig. 22.** Comparison of orthosteric, allosteric and allo-network drugs. Top parts of the three panels illustrate the protein structures of the primary drug targets showing the drug binding site as a green circle. Bottom parts of the panels illustrate the position of the primary targets in the human interactome. Red ellipses illustrate the 'action radius', i.e. the network perturbation induced by the primary targets. In the top part of the middle panel the allosteric drug binds to an allosteric site and affects the pharmacologically active site of the target protein (marked by a red asterisk) via the intra-protein allosteric signal propagation shown by the dark green arrow. In the top part of the right panel the signal propagation (illustrated by the light green arrows) extends beyond the original drug binding protein, and via specific interactions affects two neighboring proteins in the interactome. The pharmacologically active site is also marked by a red asterisk here. Orange arrows illustrate an intracellular pathway of propagating conformational changes, which is disease-specific in case of successful allo-network drugs. Allo-network drugs allow indirect and specific targeting of key proteins by a primary attack on a 'silent' protein, which is not involved in major cellular pathways. Targeting 'silent', 'by-stander' proteins, which specifically influence pharmacological targets, not only expands the current list of drug targets, but also causes much less side-effects and toxicity.
Adapted with permission from Nussinov et al. (2011).

922 drug–disease pairs drugs do not target the actual disease-associated proteins, but bind to their 3rd or 4th neighbors. However, the distance between drug targets and disease-associated proteins was regarded as a sign of palliative drug action (Yildirim et al., 2007; Barabási et al., 2011), and the expansion of the concept of allosteric drug action to the interactome level has been formulated only recently (Nussinov et al., 2011). Interestingly, targeting neighbors was found to be more influential on the behavior of social networks than direct targeting (Bond et al., 2012).

Allo-network drug action propagates from the original binding site to the interactome neighborhood in an anisotropic manner, where propagation efficiency is highly directed and specific. Binding sites of promising allo-network drug targets are not parts of 'high-intensity' intracellular pathways, but are connected to them. These intracellular pathways are disease-specific in the case of promising allo-network drugs (Fig. 22). Thus allo-network drugs can achieve specific, limited changes at the systems level with fewer side-effects and lower toxicity than conventional drugs. Allosteric effects can be considered at two levels: 1.) small-scale events restricted to the neighbors or interactome module of the originally affected protein; 2.) propagation via large cellular assemblies over large distances (i.e. hundreds or even thousands of Angstroms; Nussinov et al., 2011; Szilágyi et al., in press). Drugs with targets less than 3 steps (or more than 4 steps) from a disease-associated protein were shown to have significantly more side-effects, and failed more often (Wang et al., 2012c); however, rational drug design in recent years proceeded in the opposite direction, identifying drug targets closer to disease-associated proteins than earlier (Yildirim et al., 2007). The above data argue that reversing this trend may be more productive. Allo-network drugs point exactly to this direction.

Databases of allosteric binding sites (Huang et al., 2011a; http://mdl.shsmu.edu.cn/ASD) help the identification possible sites of allo-network drug action. However, allo-network drugs may also bind to sites, which are not used by natural ligands. For the identification of allo-network drug targets and their binding sites, first the interactome has to be extended to atomic level (amino acid level) resolution. For this, docking of 3D protein structures and the consequent connection of their protein structure networks are needed. Thus allo-network drug targeting requires the integration of our knowledge on protein structures, molecular networks, and their dynamics focusing particularly on disease-induced changes. We conclude this section by listing a few possible methods to define allo-network drug target sites.

- A general strategy for the identification of allosteric sites may involve finding large correlated motions between binding sites. This can reveal which residue-residue correlated motions change upon ligand binding, and thus can suggest new allosteric sites (Liu & Nussinov, 2008) even in integrated networks of protein mega-complexes.
- Reverse engineering methods (Tegnér & Bjorkegren, 2007) allow us to discriminate between 'high-intensity' and 'low-intensity' communication pathways both in molecular and atomic level networks, and thus may provide a larger safety margin for allo-network drugs.
- As we summarized in Section 3.2.2, network-based analysis of perturbation propagation is a fruitful method to identify intra-protein allosteric pathways (Pan et al., 2000; Chennubhotla & Bahar, 2006; Ghosh & Vishveshwara, 2007, 2008; Tang et al., 2007; Daily et al., 2008; Goodey & Benkovic, 2008; Sethi et al., 2009; Tehver et al., 2009; Vishveshwara et al., 2009; Park & Kim, 2011; Csermely et al., 2012; Ma et al., 2012a). A successful candidate for the inter-protein allosteric pathways involved in allo-network drug action disturbs network perturbations specific to a disease state of the cell at a site distant from the original drug-binding site. Perturbation analysis (see Section 2.5.2; Antal et al., 2009; Farkas et al., 2011) applied to atomic level resolution of the interactome in combination with disease specific protein expression patterns may help the identification of such allo-network drug targets.
- Central residues play a key role in the transmission of allosteric changes (Section 3.2.2; Chennubhotla & Bahar, 2006; Chennubhotla & Bahar, 2007; Zheng et al., 2007; Chennubhotla et al., 2008; Tehver et al., 2009; Liu & Bahar, 2010; Liu et al., 2010a; Su et al., 2011; Park & Kim, 2011; Dixit & Verkhivker, 2012; Ma et al., 2012a;

Pandini et al., 2012). We may use a number of centrality measures (Kovács et al., 2010), including perturbation-based or game-theoretical assumptions (see Sections 2.5.2 and 2.5.3; Farkas et al., 2011), to find the level of importance of proteins and pathways in interactomes, in signaling networks and important amino acids in their extensions to atomic level resolution (Szalay-Bekő et al., 2012).

- At both the molecular network level and its extension to atomic level resolution we may subtract network hierarchy (Ispolatov & Maslov, 2008; Jothi et al., 2009; Cheng & Hu, 2010; Hartsperger et al., 2010; Rosvall & Bergstrom, 2011; Liu et al., 2012; Mones et al., 2012; Szalay-Bekő et al., 2012) to assess the importance of various nodes (proteins and/or amino acids), or we may find nodes or edges controlling the network by the application of recently published methods (Cornelius et al., 2011; Liu et al., 2011, 2012; Banerjee & Roy, 2012; Cowan et al., 2012; Mones et al., 2012; Nepusz & Vicsek, 2012; Wang et al., 2012a; Pósfai et al., 2013).
- Combination of evolutionary conservation data proved to be an efficient predictor of intra-protein signaling pathways (Tang et al., 2007; Halabi et al., 2009; Joseph et al., 2010; Jeon et al., 2011; Reynolds et al., 2011). Similar approaches may be extended to protein neighborhoods helping to find starting sites for allo-network drug action.
- Disease-associated single-nucleotide polymorphisms (SNPs; Li et al., 2011b) and/or mutations (Wang et al., 2012b) may be part of the propagation pathways of allosteric effects. In-frame mutations are enriched in interaction interfaces (Wang et al., 2012b), and provide a potential dataset to assess the existence of allo-network drug binding sites.

Targeting disease-induced dynamical changes in molecular networks may also be focused on transient interactions specific to disease. Thus allo-network drugs might also provide a novel solution to uncompetitive, 'interfacial' drug action (Pommier & Cherfils, 2005; Keskin et al., 2007). When available, current drugs aim to directly inhibit protein–protein interactions (Gordo & Giralt, 2009). We note that the methods above are suitable to find allo-network drugs, which stabilize/restore/activate a protein, its function or one (or more) of its interactions. The methods we listed here are suitable for finding both primary targets of allo-network drugs in molecular networks and allo-network drug binding sites in the amino acid networks of involved proteins. We will describe additional network-related methods to find binding sites of allo-network drugs proper in Section 4.2.

### 4.1.7. Networks as drug targets

The last two sections on multi-target drugs and allo-network drugs already demonstrated the utility of network-based thinking in the determination of drug-targets. In this closing section on drug target identification we summarize the ideas considering key segments of networks as drug targets.

Considering molecular networks as targets has gained increasing support in recent papers on systems-level drug design (Brehme et al., 2009; Schadt et al., 2009; Zanzoni et al., 2009; Baggs et al., 2010; Pujol et al., 2010; Cho et al., 2012; Erler & Linding, 2012). As we defined in the starting section on drug target identification, from the network point of view it is important to discriminate between two strategies: 1.) the central hit strategy aiming to destroy the network of infectious agents or cancer cells and 2.) the network influence strategy using the systems-level knowledge to find drug target candidates in therapies of polygenic, complex diseases (see Fig. 19 and Section 4.1.1 for further details). Here we list a few major characteristics of both strategies.

Optimal network targeting of the central hit strategy:

- finds hubs and other central nodes or edges of molecular networks or identifies choke points of metabolic networks, i.e. proteins uniquely producing or consuming a certain metabolite;
- finds unique targets of infectious agent- or cancer-specific networks.

Optimal network targeting of the network influence strategy:

- shifts disease-specific changes of cellular functions back to their normal range (Kitano, 2007);
- applies precise targeting of selected network pathways, protein complexes, network segments, nodes or edges avoiding highly influential nodes and edges of molecular networks in healthy cells but converging drug effects at specific pathway sites of diseased cells;
- uses multiple or indirect targeting;
- takes into consideration tissue specificity.

Optimal network targeting of both the central hit and network influence strategies:

- incorporates patient- and disease stage-specific data (such as single-nucleotide polymorphisms, metabolome, phosphoproteome or gut microbiome data) ADMET-related data, side-effect- and drug resistance-related data as detailed in the next section.

We believe that the arsenal of network (re)construction and network analysis methods we listed in this review may offer help and promise for the prediction of novel, systems-level drug targeting possessing the characteristics detailed above.

### 4.2. Hit finding, expansion and ranking

Following target selection discussed in the preceding section, here we will discuss the added-value of network-related methods in the search, confirmation and expansion of hit molecules. Several steps in this process, such as pharmacophore identification, network-based QSAR models, building of a hit-centered chemical library, hit expansion, as well as other network-related methods of chemoinformatics and chemical genomics, have already been discussed in Section 3.1.3. Therefore, the Reader is asked to compare Section 3.1.3 with the current chapter. Here we will first summarize the help provided by network description and analysis in the determination of ligand binding sites, as applicable to network nodes as drug targets. We will continue with network methods to find hot spots, which reside in protein interfaces, and are targets of edgetic drugs. We will conclude the section by a summary of network-related approaches in hit expansion and ranking.

### 4.2.1. In silico hit finding for ligand binding sites of network nodes

Node targeting aims to find a selective, drug-like (low molecular weight, possibly hydrophobic) molecule that binds with high affinity to the target (Lipinski et al., 2001). There are two main network-based approaches for the identification of ligand binding sites. A 'bottom–up approach' uses protein structure networks (see Section 3.2 in detail), while a 'top–down approach' reconstructs binding site features from binding site similarity networks (Section 4.1.3).

For in silico hit prediction, a logical first step is to find pockets (cavities, clefts) on the protein surface. Medium-sized proteins have 10 to 20 cavities. Ligands often bind to the largest surface cavities of this ensemble (Laskowski et al., 1996; Liang et al., 1998b; Nayal & Honig, 2006). Using a protein structural approach, Coleman and Sharp (2010) identified a hierarchical tree of protein pockets using the travel depth algorithm that computes the physical distance a solvent molecule would have to travel from a given protein surface point to the convex hull of the surface. Using the similarity network approach, pocket similarity networks have been constructed, and their small-world character, hubs and hierarchical modules were identified. Pocket groups were found to reflect functional separation (Liu et al., 2008a, 2008b), and may be used for hit identification. However, shape information alone is insufficient to discriminate between diverse binding sites, unless combined with chemical descriptors (http://proline.physics.iisc.ernet.in/pocketmatch; Yeturu & Chandra, 2008; http://proline.physics.iisc.ernet.in/pocketalign; Yeturu & Chandra, 2011). CAVER (http://caver.cz; Chovancova et al., 2012) uses molecular dynamics simulation to predict intra-protein transport pathways.

Protein structure networks (Section 3.1) were relatively seldom used so far to predict ligand binding sites. However, high-centrality segments of protein structure networks were shown to participate in ligand binding (Liu & Hu, 2011). Evolutionary conservation patterns of amino acids in related protein structures identified protein sectors related to catalytic and allosteric ligand binding sites (Halabi et al., 2009; Jeon et al., 2011; Reynolds et al., 2011). Protein structure networks were extended, incorporating ligand atoms, participating ions and water molecules and chemical properties aiming to find network motifs representing a favorable set of protein–ligand interactions used for as a scoring function (Xie & Hwang, 2010; Kuhn et al., 2011). Protein structure network comparison was demonstrated to be useful for the identification of chemical scaffolds of potential drug candidates (Konrat, 2009).

Similarity clusters or network prediction methods of binding site similarity networks (also called as pocket similarity networks, or cavity alignment networks; Zhang & Grigorov, 2006; Kellenberger et al., 2008; Liu et al., 2008a; Park & Kim, 2008; Andersson et al., 2010; Weskamp et al., 2009; Xie et al., 2009a; BioDrugScreen, http://biodrugscreen.org; Li et al., 2010c; Reisen et al., 2010; Ren et al., 2010; Weisel et al., 2010) can be used to predict binding site topology of yet unknown proteins. The complex drug target network, PDTD (http://dddc.ac.cn/pdtd) incorporating 3D active site structures and the web-server TarFishDock enables simultaneous target and target-site prediction of new chemical entities (Gao et al., 2008). The versatile protein–ligand interaction database, CREDO (http://www-cryst.bioc.cam.ac.uk/databases/credo; Schreyer & Blundell, 2009) and the extensive protein–ligand databases, STITCH (http://stitch.embl.de; Kuhn et al., 2012) and BindingDB (http://bindingdb.org; Liu et al., 2007a) offer an important help to search for potential targets and identify their binding sites.

#### 4.2.2. In silico hit finding for edgetic drugs: hot spots

Edgetic drugs (Section 4.1.2) modify protein–protein interactions. Protein–protein interaction binding sites were considered for a long time as "non-druggable", since they are large and flat. However, Clackson and Wells (1995) discovered hot spots of binding surfaces, which are residues providing a contribution to the decrease in binding free energy of larger than 2 kcal/mol. Bogan and Thorn (1998) proposed that hot spots are surrounded by hydrophobic regions excluding water from the hot spot residues. Hot spots are often populated by aromatic residues, and tend to cluster in hot regions, which are tightly packed, relatively rigid hydrophobic regions of the protein–protein interface. Hot spots and hot regions help in finding hits, since 1.) they constitute small focal points of drug binding, which can be predicted within the large and flat binding-interface; 2.) these focal points are relatively rigid, and help in docking. An inhibitor needs to cover 70 to 90 atoms at the protein–protein interaction site, which corresponds to the 'Lipinski-conform' (Lipinski et al., 2001) ~500 Da molecular weight. Several small molecules were found, which are able to compete with the natural binding partner very efficiently (Keskin et al., 2005, 2007; Wells & McClendon, 2007; Ozbabacan et al., 2010). Druggable hot regions have a concave topology combined with a pattern of hydrophobic and polar residues (Kozakov et al., 2011).

Hot spots can be predicted as central nodes of protein structure networks (del Sol & O'Meara, 2005; Liu & Hu, 2011; Grosdidier & Fernández-Recio, 2012). In agreement with this, disease-associated mutations (single-nucleotide polymorphisms) are enriched by 3-fold at the interaction interfaces of proteins associated with the disorder, and often occur at central nodes of the protein structure network (Akula et al., 2011; Li et al., 2011b; Wang et al., 2012b). Using this knowledge, the pyDock protein–protein interaction docking algorithm was improved by protein structure network-based scores (Pons et al., 2011). Intra-protein energy fluctuation pathways were proposed to help in the prediction of hot spot localization (Erman,

2011). Recently the use of associative-memory, water-mediated coarse-grained protein folding model, AWSEM was also demonstrated to predict protein binding surfaces well (Zheng et al., 2012a).

Hit identification of edgetic drugs is helped by the TIMBAL database containing ligands inhibiting protein–protein interactions (http://www-cryst.bioc.cam.ac.uk/databases/timbal; Higueruelo et al., 2009). The machine learning-based technique of the Dr. PIAS server assesses if a protein–protein interaction is druggable (http://drpias.net; Sugaya & Furuya, 2011; Sugaya et al., 2012). Despite the considerable progress of this field in the last decade, we are still at the beginning of using network-related knowledge to identify edgetic drug binding sites. Network-related methods for hot spot and hot region identification are also promising, if applied to aptamers, peptidomimetics or proteomimetics.

#### 4.2.3. Network methods helping hit expansion and ranking

An important step of hit confirmation is the check of the chemical amenability of the hit, i.e. the feasibility up-scaling costs of its synthesis. Core and hub positions or other types of centrality of the hits in the chemical reaction network (Section 3.1.2; Fialkowski et al., 2005; Bishop et al., 2006; Grzybowski et al., 2009) are all predictors of good chemical tractability. Moreover, a simulated annealing-based network optimization uncovered optimal synthetic pathways of selected hits (Kowalik et al., 2012). In the case of multiple hits, hit clustering can be performed by modularization of their chemical similarity networks described in Section 3.1.3. Hubs and clusters of hit-fragments in chemical similarity networks may be used for hit-specific expansion of existing compound libraries (Benz et al., 2008; Tanaka et al., 2009). QSAR-related similarity networks and the other complex similarity networks we listed in Table 5 help the lead development and selection efforts we will detail in the next section.

Hit cluster should usually conform to the Lipinsky-rules of drug-like molecules (Lipinski et al., 2001) restricting the hit-range to small and hydrophobic molecules with a certain hydrogen-bond pattern. Leeson and Springthorpe (2007) warned that systematic deviations from these rules may have a dangerous impact on drug design, increasing late-attritions due to side-effects and/or toxicity. However, natural compounds also contain a set of 'anti-Lipinsky' molecules, which form a separate island in the chemical descriptor space having a higher molecular weight and a larger number of rotatable bonds (Ganesan, 2008). The network-related methods predicting the efficiency, ADME, toxicity, interactions, side-effect and resistance occurrence detailed in the next section may help in decreasing the risk of non-conform hit and lead molecules, and highlight issues of drug safety in an early phase of drug development.

### 4.3. Lead selection and optimization: drug efficacy, drug absorption, distribution, metabolism, excretion, toxicity, drug interactions, side-effects and resistance

Following hit selection and expansion discussed in the preceding section network-related methods may also help the lead selection process. Various aspects of lead selection such as drug toxicity, side-effects and drug–drug interactions are tightly interrelated. The incorporation of personalized data, such as genome-wide association studies/single-nucleotide polymorphisms (GWAS/SNPs), signaling network or metabolome data into the complex network structures which help lead selection may not only predict well the pharmacogenomic properties of the lead, but also help patient profiling in clinical trials, as well as therapeutic guideline determination of the marketed product.

#### 4.3.1. Networks and drug efficacy, personalized medicine

Drug efficacy is the theoretical efficiency of drug action not taking into account the effects in practice, such as patient compliance.

Efficacy is a highly personalized efficiency measure of drug action, which heavily depends on multiple factors including the genetic background (e.g. single-nucleotide polymorphisms and other genetic variants assessed in genome-wide association studies), network robustness and the ADME properties (see next section; Kitano, 2007; Barabási et al., 2011; Yang et al., 2012). Single-nucleotide polymorphisms (SNPs) may alter the interaction properties of at least 20% of the nodes in the human interactome (Davis et al., 2012), and were recently shown to be a reason for the unexpectedly high variability of protein–protein interactions (Hamp & Rost, 2012). A number of studies assessed the effects of SNPs on changing the underlying properties of interactomes and gene–gene association networks (Akula et al., 2011; Cowper-Sal lari et al., 2011; Fang et al., 2011; Hu et al., 2011; Li et al., 2011b, 2012b; Wang et al., 2012b), which may change drug efficacy both directly or indirectly. The integrated Complex Traits Networks (iCTNet, http://flux.cs.queensu.ca/ictnet), including phenotype/single-nucleotide polymorphism (SNP) associations, protein–protein interactions, disease–tissue, tissue–gene and drug–gene relationships, is a rich dataset helping drug efficacy assessments (Wang et al., 2011b).

Incorporation of omics-type data into complex, drug action-related networks will allow the construction of personalized efficacy profiles. Integration of pharmacogenomics, signaling network or metabolome data may improve clinical trial design. However, network-related methodologies for complex drug efficacy profiling have not been developed yet. Similarly, analysis of the semantic networks of medical records by text mining and by network analysis techniques is a future tool to improve the assessment of drug efficiency measures, extending the efficacy with patient compliance and other effects occurring in medical practice (Chen et al., 2009a). Network-related models may help in the development of optimal drug dosage and frequency schedules. As an example of this, the study of Li et al. (2011e) uncovered a 'sweet spot' of drug efficacy dose and schedule regions by the extension of their model to the genetic regulatory network environment of the drug target. Drug dose and schedule considerations are already parts of the ADME characterization, which we will detail in the next section.

### 4.3.2. Networks and absorption, distribution, metabolism and excretion, toxicity: drug absorption, distribution, metabolism and excretion

The integration of early ADME (absorption, distribution, metabolism, excretion) profiling to lead selection is an important element of successful drug design. Prediction of ADME properties using structural networks of lead candidates (Kier & Hall, 2005), molecular fragment networks predicting human albumin binding (Estrada et al., 2006), chemical similarity networks (Brennan et al., 2009), as well as drug–tissue networks (Gonzalez-Diaz et al., 2010b), isotope-labeled metabolomes and drug metabolism networks (Martinez-Romero et al., 2010; Fan et al., 2012), non-linear diffusion models of drug partitioning in lipid network structures, such as the stratum corneum (Schumm et al., 2010), multiple binding to transporters (Kell et al., 2013) and complex networks of major cellular mechanisms participating in ADME determination (Ekins et al., 2006), were all important advances which can help in incorporating better ADME complexity into the lead selection process. Despite these methods, there is room to improve ADME prediction and assessment by network techniques. ADME studies are often combined with toxicity assessments (ADMET), which we will detail in the next section. Toxicity is related to side-effects discussed in Section 4.3.5. Drug combinations may have an especially complex ADME profile due drug–drug interaction effects, which will be described in Section 4.3.4.

### 4.3.3. Networks and drug toxicity

Toxicity plays a different role in drug targets identified using the central hit strategy and the network influence strategy of Section 4.1.1. In the central hit strategy our aim is to kill the cells of the infectious agent or cancer. Therefore, toxicity is a must here—but it has to be selective to the targeted cells. In the network influence strategy targeting other diseases, toxicity becomes generally avoidable. Toxicity is often a network property depending on the extent of network perturbation and robustness (Kitano, 2004a, 2004b, 2007; Apic et al., 2005; Geenen et al., 2012). Network hubs and the essential proteins described in Section 2.3.4 are less frequently targeted by drugs—with the exception of anti-infective and anticancer agents (Jonsson & Bates, 2006; Yildirim et al., 2007). In contrast, those inter-modular bridges, which modulate specific information flows, are preferred drug targets (Hwang et al., 2008). Node centrality in drug-regulated networks correlates with drug toxicity (Kotlyar et al., 2012). All these findings give further support to the utility of network-based toxicity assessments.

Hepatotoxicity is a major reason of drug attritions (Kaplowitz, 2001). The number of network studies addressing this important issue is increasing, and includes cytokine signaling networks related to idiosyncratic drug hepatotoxicity (Cosgrove et al., 2010) and gene–gene interaction networks based on transcriptional profiling (Hayes et al., 2005; Kiyosawa et al., 2010). Importantly, toxicity-related networks should be understood as signed networks containing both toxicity promoting effects and detoxifying effects, such as the glutathione network in liver (Geenen et al., 2012), or hepatic pro-survival (AKT) and pro-death (MAPK) pathways, where specific pathway inhibitors may antagonize drug-induced hepatotoxicity (Cosgrove et al., 2010).

Network-based in silico prediction of human toxicity aims to bridge the gap between animal toxicity studies and clinical trials. Toxicity assessment applications of chemical similarity networks (Section 3.1.3; Kier & Hall, 2005; Brennan et al., 2009), as well as the use of association networks between chemicals and toxicity-related proteins or processes (DITOP, http://bioinf.xmu.edu.cn:8080/databases/DITOP/index.html; Zhang et al., 2007; Audouze et al., 2010; Iskar et al., 2012) open a number of additional possibilities for network-predictions of human toxicity in the future.

### 4.3.4. Networks and drug–drug interactions

Drug interactions may often cause highly unexpected effects. As we already described in Section 4.1.5 on network polypharmacology and multi-target drug design, most of the unexpected drug–drug interactions are not due to direct competition for the same binding site, but are caused by the complex interaction structure of molecular networks. Experimental testing of drug–drug interactions may be used to help infer the underlying molecular network structure, and drug–drug interaction networks (Borisy et al., 2003; Yeh et al., 2006; Lehár et al., 2007; Jia et al., 2009) may be used to predict additional drug–drug interactions using network modularization methods.

A drug–drug interaction network was assembled using drug package insert texts. This network was extended by potential mechanisms, such as drug targets or enzymes involved in drug metabolism, and was included in the KEGG DRUG database (http://genome.jp/kegg/drug; Takarabe et al., 2008; Takarabe et al., 2011; Kanehisa et al., 2012). Recently text mining rules to refine literature-derived drug–drug interaction networks were proposed (Kolchinsky et al., 2013). Drug–drug interaction networks may be perceived as signed networks containing synergistic or antagonistic interactions (Yeh et al., 2006; Jia et al., 2009), and have hubs, i.e. drugs which are involved in most of the observed interactions (Hu & Hayton, 2011). Many of the drug-related databases listed in Table 9 may help to uncover adverse drug–drug interactions. Besides the KEGG DRUG database mentioned above the DTome (http://bioinfo.mc.vanderbilt.edu/DTome; Sun et al., 2012) database also explicitly contains adverse drug interactions. Complex chemical similarity networks and drug–target networks, discussed in Sections 3.1.3 and 4.1.3, respectively, were also used for the prediction of unexpected drug–drug interactions (Zhao & Li, 2010; Yu et al., 2012).

Drugs may affect each other's ADME properties by simple competition, or by more refined network-effects (Jia et al., 2009; Kell et al., 2013), such as the positive synergism of amoxicillin and clavulanate, where clavulanate is an inhibitor of the enzyme responsible for amoxicillin destruction (Matsuura et al., 1980). Drug–herb interactions are important aspects of drug–drug interaction analysis particularly in China, where traditional Chinese medicine is often combined with Western medicine. Here semantic networks and other combined networks of drug and herb effects and targets may offer help in prediction of drug safety (Chen et al., 2009a; Zheng et al., 2012b). Despite the wide variety of approaches listed, network techniques provide many more possibilities in the prediction of drug–drug interaction effects. In practice, all methods listed in Section 4.1.5 on multi-target drugs, such as perturbation, network influence and source/sink analyses, as well as the drug side-effect networks described in the next section may be used for the prediction of drug–drug interactions.

### 4.3.5. Network pharmacovigilance: prediction of drug side-effects

Discovering unexpected side-effects by experimental methods alone, is a daunting task requiring the screen of a large number of potential off-targets. However, side-effects may have their origin in both single and multi-target. Both are systems-level responses, which allow the prediction of drug off-targets by computational methods (Berger & Iyengar, 2011; Zhao & Iyengar, 2012). In this section we introduce several network-related methods of side-effect identification.

Side-effects may come from the involvement of a single drug target in multiple cellular functions or may involve multiple drug targets. In a study on protein–protein interaction networks two third of the side-effect similarities were related to shared targets, while 5.8% of the side-effect similarities were due to drugs targeting proteins close in the human interactome (Brouwers et al., 2011). This result may reflect both the concentration of side-effects in direct drug targets and the efficiency of those allo-network drugs (Section 4.1.6; Nussinov et al., 2011), whose direct target is not the primary binding site, but a neighboring protein in the interactome.

The previous sections uncovered many network-related strategies to avoid side-effects at the level of target selection. We will summarize only a few major considerations here.

- Avoidance of targeting hubs and high centrality nodes of interactomes, signaling networks and metabolomes is a general network strategy of side-effect reduction, especially when using the network influence strategy of Section 4.1.1 against polygenic diseases such as diabetes. Disease specific, limited network perturbation is a key systems-level requirement to avoid drug adverse effects (Guimera et al., 2007b; Hase et al., 2009; Zhu et al., 2009; Yu & Huang, 2012). Network algorithms focusing the downstream components of node-targeting to a certain network segment are important methods to reduce potential side-effects at the level of target identification (Dasika et al., 2006; Ruths et al., 2006; Pawson & Linding, 2008).
- Iterative methods sequentially identified sets of metabolic network edges corresponding to enzymes, whose inhibition can produce the expected inhibition of targets with reduced side-effects in humans and in *E. coli* (Lemke et al., 2004; Sridhar et al., 2007, 2008; Song et al., 2009).
- Unexpected edges between remote targets in ligand binding site similarity networks (also called as pocket similarity networks, or cavity alignment networks) suggest potential side-effects (Zhang & Grigorov, 2006; Liu et al., 2008a; Weskamp et al., 2009).
- Edgetic drugs (Section 4.1.2) are usually more specific and may have generally less side-effects than node-targeting drugs. However, common protein–protein interaction interface motifs are important indicators of potential side-effects of edgetic drugs (Engin et al., 2012).
- Future analysis may uncover nodes and edges having a major influence on the occurrence of the disease-specific critical network-transitions mentioned in Section 2.5.2. These influential nodes will most probably represent the 'Achilles-heel' of network in the disease state, and their targeting will induce fewer side-effects than the average.

Side-effect prediction is tightly related to drug–target prediction (Section 4.1) involving the comparison of novel target(s) with those of existing drugs. The selective optimization of side-effects (Wermuth, 2006) is a known lead development technique. Consequently both drug–target interaction networks (Section 4.1.3; Xie et al., 2009b; Yang et al., 2010; Azuaje et al., 2011; Xie & Bourne, 2011; Yang et al., 2011; Takarabe et al., 2012; Yu et al., 2012) and drug–disease networks (Hu & Agarwal, 2009) may be used for the prediction of side-effects. Analysis of drug–disease networks may be extended using pathway analysis (Ye et al., 2012). Complex chemical similarity networks (Section 3.1.3) also use a combination of network-related data including e.g. interactomes for the prediction of off-target effects (Hase et al., 2009; Yamanishi et al., 2010; Zhao & Li, 2010). The web-servers SePreSA (http://SePreSA.Bio-X.cn; Yang et al., 2009a) and DRAR-CPI (http://cpi.bio-x.cn/drar; Luo et al., 2011) were constructed to show possible adverse drug reactions based on drug–target interactions. Practically all methods listed in Section 4.1.5 on multi-target drugs may be used to predict side-effects. As an example, the Monte Carlo simulated annealing network perturbation method of Yang et al. (2008) correctly predicted the well-known side-effects of non-steroidal anti-inflammatory drugs and the cardiovascular side-effects of the recalled drug, Vioxx. Moreover, side-effect determination may be extended to any complex similarity networks we listed in Table 5 (such as that containing disease-specific genome-wide gene expression data; Huang et al., 2010a; the Cytoscape plug-in DvD program, http://www.ebi.ac.uk/saezrodriguez/DVD; Pacini et al., 2013) and to those future network representations, which will include signaling network or metabolome data. These datasets may be used to construct personalized or patient cohort-specific side-effect profiles enabling a better focusing of therapeutic indications and contraindications.

In recent years, many types of side-effect networks, drug target/adverse drug reaction networks or drug target/adverse target networks were constructed.

- Campillos et al. (2008) and later Yamanishi et al. (2010) and Takarabe et al. (2012) combined structural similarity and side-effect similarity to construct a side-effect similarity network of drugs, and used this network to identify novel drug targets for drug repositioning (Section 4.1.4).
- Correlation analysis of drug protein-binding profiles and side-effect profiles revealed the enrichment of drug targets participating in the same biological pathways (Mizutani et al., 2012).
- Text mining of drug package insert text was used for the construction of side-effect networks showing a gross similarity of preclinical and clinical compound profiles (Fliri et al., 2005; Oprea et al., 2011). Text mining of scientific papers may result in an extended drug–target network revealing potential side-effects (Garten et al., 2010).
- A drug–target/adverse drug reaction network was constructed from chemical similarity-based prediction of off-targets and related side-effects of 656 drugs (Lounkine et al., 2012).
- A network of 162 drugs causing at least one serious adverse drug reaction and their 845 targets showed similar target profiles for similar serious adverse drug reactions. The MHC I (Cw*4) protein was identified and confirmed as the possible target of the sulfonamide-induced toxic epidermal necrolysis adverse effect (Yang et al., 2009b).
- Yang et al. (2009c) used the CitationRank network centrality algorithm and a dataset of gene/serious adverse drug reaction associations (collected by text mining from PubMed records) to identify the association strength of genes with 6 major serious adverse drug reactions (http://gengle.bio-x.cn/SADR).

Side-effect similarity networks were used for efficient refinement of primary side-effect identification based on similarities in drug structures (Atias & Sharan, 2011). Network prediction methods detailed in Section 2.2.2 and network modularization methods may help to decipher novel side-effects from side-effect networks in the future.

The side-effect database, SIDER (http://sideeffects.embl.de; Kuhn et al., 2010) considerably enhanced side-effect network studies. The SIDER-derived side-effect network was extended by biological processes related to Gene Ontology terms and text mining of PubMed data (Lee et al., 2011). Combination of SIDER data with those on disease-associated genes showed that drugs having a target less than 3 or more than 4 steps away from a disease-associated protein in human signaling networks had significantly more side-effects, and failed more often (Wang et al., 2012c).

Sources of unexpected side-effects can sometimes be focused on a certain tissue or cellular process. Analysis of tissue-specific network dynamics, such as that of the kidney metabolic network revealing hypertensive side-effects (Chang et al., 2010), might be a promising method to predict tissue-specific side-effects. Csoka and Szyf (2009) raised the possibility of epigenetic side-effects, where a drug modifies the chromatin structure, and thus indirectly influences a number of other genes. Similarly, microRNA related side-effects may be developed by the interaction of a drug with the complex microRNA signaling network (Section 3.4), where a change in the transcription of a microRNA may influence a set of rather unrelated proteins and related functions.

### 4.3.6. Resistance and persistence

In recent years antibiotic resistance became a major threat of human health (Bush et al., 2011). Antibiotic persistence is a form of antibiotic resistance, which is related to a dormant, drug-insensitive subpopulation of bacteria (Rotem et al., 2010). Resistance development against chemotherapeutic agents is a key challenge of anti-cancer therapies (Section 3.4.3; Kitano, 2004a; Logue & Morrison, 2012). Resistance development is involved in the application of the central hit strategy defined in Section 4.1.1 aiming to destroy pathogen- or cancer-related networks.

Ligands may be optimized against resistance by targeting conserved amino acids and main-chain atoms with strong interactions instead of weaker interactions pointing towards mutatable residues (Hopkins et al., 2006). Tuske et al. (2004) defined the substrate-envelope for HIV reverse transcriptase as the space occupied by various conformations of naturally occurring ligands and their targets. Lamivudine and zidovudine induced resistance by protruding beyond this substrate-envelope, while tenofovir, which did not have handles projecting beyond the substrate envelope was more resistant against resistance development (Tuske et al., 2004). Protein structure network studies may help in designing more resistant-prone lead molecules.

Development of drug resistance is often a phenomenon involving network-robustness, when the affected cell activates alternative or counter-acting pathways to minimize the consequences of drug action (Kitano, 2007). Oberhardt et al. (2010) offer a comprehensive analysis of metabolic network adaptation of *Pseudomonas aeruginosa* to a host organism during a 44-months period. Co-targeting of an additional crucial point of drug-affected network pathways is an efficient tool to fight against resistance. Drug combinations and multitarget drugs develop less resistance (Zimmermann et al., 2007; Pujol et al., 2010; Rosado et al., 2011; Savino et al., 2012). Analysis of pathogen interactomes involving random walks or known drug resistance-related proteins plus gene expression changes revealed pathways often involved in resistance development helping cotarget determination (Raman & Chandra, 2008; Chen et al., 2012b). Resistance-related proteins defined a subset of pathogen interactome, called resistome. Drug-induced gene expression changes and betweenness centralities of their interactions were used as weights of resistome edges. Resistome hubs may serve as important co-targets (Padiadpu et al., 2010). Differential assessment of molecular networks of normal and resistant pathogens allows more efficient drug resistant target and/or co-target identification (Kim et al., 2010). As we described in Section 3.4.3, the combination of anti-tumor drugs and stress response targeting increases therapeutic efficiency (Rocha et al., 2011; Tentner et al., 2012). The Hebbian learning rule, i.e. the property of neuronal networks to increase edge-weights along frequently used pathways (Hebb, 1949) may be extended to molecular networks, and studied as a possible source of systems-level resistance development.

Importantly, the most efficient synergistic drug combinations typically preferred in clinical settings may develop a faster resistance, which might argue for other, e.g. antagonistic drug combinations (Chait et al., 2007; Hegreness et al., 2008). Synthetic rescues (when the inhibition of a target compensates for the inhibition of another; Section 3.6.3) are good candidates for anti-resistant antagonistic co-target action (Motter, 2010). Network simulation of resistance transmission in bacterial populations also underlined the need for potent antimicrobials and high-enough doses to kill the susceptible population segment as soon as possible (Gehring et al., 2010). Network-related methods to fight drug-resistance may help both anti-infective and anti-cancer strategies as we will describe in the next two sections.

## 5. Four examples of network description and analysis in drug design

In this section we will illustrate the usefulness of network-related methods in drug design with examples of four major threats of human health: infectious diseases, cancer, diabetes (extended to metabolic diseases) and neurodegenerative diseases. The first two disease groups (infectious diseases and cancer) are examples of the central hit strategy aiming to destroy the network of infectious agents or cancer cells (see Section 4.1.1 for a definition of the two strategies). The last two therapeutic areas (diabetes and neurodegenerative diseases) are examples of the network influence strategy aiming to re-wire molecular networks of diseased cells to restore normal function (see Section 4.1.7 for a summary of these two strategies). The sections on the use of network science to combat these diseases will not give a comprehensive summary, but will only highlight a few key solutions and their results.

### 5.1. Anti-viral drugs, antibiotics, fungicides and antihelminthics

Drugs against infectious agents are central hit strategy drugs in the classification that we introduced in Section 4.1.1. Efficient central hit strategy drugs interfere with viral replication or kill the infectious cells with high efficiency (instead of temporal growth inhibition, which may induce resistance), and avoid any toxic effects in humans. We list a number of selected illustrative examples of network approaches for drug target identification against pathogenic agents in Table 10.

Integrated host–pathogen networks proved to be very efficient in complex targeting strategies in the case of viral/host interactomes (Uetz et al., 2006; Calderwood et al., 2007; de Chassey et al., 2008; Chautard et al., 2009; Navratil et al., 2009, 2011; Brown et al., 2011; Prussia et al., 2011; Xu et al., 2011b; Lai et al., 2012; Schleker et al., 2012; Simonis et al., 2012). Complex databases of viral/host interactions were also assembled (Zhang et al., 2005). Anti-viral target proteins often emerge as bridges between host/pathogen and human network modules, as well as hubs or otherwise central proteins of the virus-targeted human interactome (Uetz et al., 2006; Calderwood et al., 2007; de Chassey et al., 2008; Navratil et al., 2011; Lai et al., 2012). Targets of viral proteins were shown to be major perturbators of human networks (de Chassey et al., 2008; Navratil et al., 2011). A machine learning technique with a learning set including viral/host interactome-derived topological and functional information identified several formerly validated viral targets of the influenza A virus, and predicted novel drug target candidates (Lai et al., 2012). The combination of viral/host interactome data with siRNA, transcriptome, microRNA, toxicity and other data may significantly extend the prediction efficiency of antiviral targets (Brown et al., 2011).

**Table 10**
Illustrative examples of the use of network methods in anti-viral drugs, antibiotics, fungicides and antihelminthics.

| Drug design area | Network method | References |
|---|---|---|
| *Protein–protein interaction networks* | | |
| Drug target identification | Identification of (pathogen-specific) hubs as potentially essential proteins (http://hub.iis.sinica.edu.tw/Hubba) | Lin et al., 2008; Kushwaha & Shakya, 2009 |
| | Identification of clique-forming, high-centrality, or otherwise topologically essential proteins | Real et al., 2004; Estrada, 2006; Florez et al., 2010; Milenkovic et al., 2011; Raman et al., 2012; Zhang et al., 2012 |
| *Metabolic networks* | | |
| Drug target identification | Comparative load point (high-centrality) and choke point (unique reaction) analysis of pathogenic and non-pathogenic bacteria (with the identification of conserved critical amino acids forming similar cavities: UniDrugTarget server, http://117.211.115.67/UDT/main.html) | Perumal et al., 2009; Chanumolu et al., 2012 |
| | Selection of essential metabolites | Kim et al., 2011 |
| | Selection of super-essential reactions | Barve et al., 2012 |
| Drug target identification, drug repositioning | Strain-specific anti-infective therapies by comparative metabolic network analysis | Shen et al., 2010 |
| *Drug–target, drug–drug similarity and complex dataset networks* | | |
| Target identification, drug repositioning | Drug target network of *Mycobacterium tuberculosis* | Kinnings et al., 2010 |
| Prediction of drug activity against different pathogens | Multi-tasking QSAR drug–drug similarity network analysis | Gonzalez-Diaz & Prado-Prado, 2008; Prado-Prado et al., 2008; Prado-Prado et al., 2009; Prado-Prado et al., 2010; Prado-Prado et al., 2011 |
| Drug target identification | Interactome, signaling network and gene regulation network of *Mycobacterium tuberculosis* | Vashisht et al., 2012 |

Analysis of integrated bacterial/fungal/parasite and human metabolic networks also became a widely used tool to predict potential drug target efficiency (Bordbar et al., 2010; Huthmacher et al., 2010; Fatumo et al., 2011; Riera-Fernandez et al., 2012). Chavali et al. (2012) and Kim et al. (2012) offered comprehensive collections of datasets and analyses of antimicrobial drug target identification using metabolic networks. Combinations of the metabolic network and the interactome of *Mycobacterium tuberculosis* were used to identify the most influential network target singletons, pairs, triplets and quadruplets (Raman et al., 2008, 2009; Kushwaha & Shakya, 2010). Multiple targets are useful to prevent the development of resistance (see preceding section; Raman & Chandra, 2008; Chen et al., 2012b). However, recent studies showed that synergistic drug combinations, which are preferred in clinical settings due to their high efficiency, may develop a faster resistance. Therefore, antagonistic drug combinations should also be tried (Yeh et al., 2006; Chait et al., 2007; Hegreness et al., 2008).

Complex chemical similarity networks including chemical-genetic interactions (i.e. hypersensitivity data of mutant strains for chemical compounds; for additional examples see Table 5) help in the identification of drug targets in anti-infective therapies (Parsons et al., 2006; Hansen et al., 2009). The (random) upstream attack strategy proposed by Liu et al. (2012) may uncover even more influential targets than currently known in directed networks such as metabolic or signaling networks. Mészáros et al. (2011) warned that the selective targeting of bacterial proteins may involve complex domain architecture. Complex similarity networks (Vilar et al., 2009) may allow patient- and disease stage-specific target search in the anti-cancer therapies detailed in the next section. As another approach linking the two central hit strategy-type drug design areas, anti-infective and anti-cancer therapies, the assessment of interactome and transcriptome perturbations by DNA tumor virus proteins highlighted the Notch- and apoptosis-related pathways that also go awry in cancer (Rozenblatt-Rosen et al., 2012).

## 5.2. Anti-cancer drugs

Cancer is a systems-level disease (Hornberg et al., 2006), where the rapidly proliferating system is characterized by an increase of network entropy (West et al., 2012), i.e. an increase of network flexibility and plasticity. This increase in network flexibility may characterize the initial stages of tumor development (such as adenomas) better than later stages of malignant transformation (such as carcinomas; Dezsö Módos, Tamás Korcsmáros and Péter Csermely, in preparation). Key aims of anti-cancer pharmacology include the identification of targets and the efficient combination of drugs to overcome the robustness of cancer-specific cellular networks with the least toxicity and resistance development possible (Kitano, 2004a, 2004b, 2007; Werner, 2011; Cheng et al., 2012c; Rivera et al., 2012). Similarly to the anti-pathogenic drugs described in the preceding section, anti-cancer drugs mostly belong to central hit strategy drugs (see Section 4.1.1). However, central targets of cancer cells are more often central components of healthy cells, than those of infectious agents discussed in the preceding section. Therefore, as we will discuss in Section 5.2.5, a few anti-cancer drugs (presumably those, which target the more rigid networks of advanced cancer types) start to resemble the network influence strategy-type drugs.

In the following sections we will show the help of interactomes, metabolic and signaling networks to find cancer-specific drug targets and drug combinations. To illustrate the special importance of network-level thinking in anti-cancer drug design, we start the section with the description of autophagy, which is a very promising area to develop novel anti-cancer drugs—but only if treated in a systems-level context using network description and analysis.

### 5.2.1. Autophagy and cancer—an
### example for the need of systems-level view

Autophagy (cellular self-degradation) has a highly ambiguous role in cancer. On the one hand, autophagy has tumor suppressing functions a.) by limiting chromosomal instability; b.) by restricting oxidative stress, which is also an oncogenic stimulus; and c.) by promoting oncogene-induced senescence. On the other hand, autophagy is used by tumor cells to escape hypoxic, metabolic, detachment-induced and therapeutic stresses as well as to develop metastasis and dormant tumor cells (Apel et al., 2009; Morselli et al., 2009; White & DiPaola, 2009; Kenific et al., 2010; Chen & Klionsky, 2011). Thus autophagy should be modulated in a cell-specific manner. In cancer cells over-activation of autophagy can induce cell death, while autophagy inhibitors sensitize cancer cells to chemotherapy. In normal cells, autophagy stimulators may be useful for cancer prevention by enhancing damage mitigation and senescence, while autophagy inhibitors can induce tumorigenesis (White & DiPaola, 2009; Ravikumar et al., 2010; Chen & Karantza, 2011). Network analysis of the regulation of autophagy may point out such context-specific intervention points. The recent work of Serra-Musach et al. (2012) showed that in contrast to most cancer-related proteins, proteins involved in autophagy are more 'failure-prone', i.e. can be saturated by perturbations faster. This gives an additional rationale to employ autophagy-related proteins as drug targets corresponding to the 'Achilles-heel' of cancer cells. Network approaches described in all the following sections may be promising for the identification of autophagy-related drug target candidates.

### 5.2.2. Protein–protein interaction network targets of anti-cancer drugs

Cancer-specificity in the anti-cancer drug targets is a primary requirement to avoid toxicity. Target-specificity may increase by selecting cancer-related mutation events or proteins having altered gene expression. In addition, all these data can be combined at the network-level (Pawson & Linding, 2008).

Large-scale sequencing identified thousands of genetic changes in tumors, which were collected in databases, such as COSMIC (http://www.sanger.ac.uk/genetics/CGP/cosmic; Forbes et al., 2011) or the Network of Cancer Genes (http://bio.ieo.eu/ncg/index.html; D'Antonio et al., 2012). Section 1.3.3 and Tables 2 and 3 listed a number of network-related methods to identify cancer-associated proteins. From the large number of tumor-associated genes, only a few play a key role in tumor pathogenesis (called driver mutations). Driver mutations can be characterized by their pathway association. In many tumors p53, Ras and PI3K are the major signaling pathways containing driver mutations (Li et al., 2009b; Pe'er & Hacohen, 2011). Genes with co-occurring mutations in the COSMIC database prefer direct signaling interactions. Genes having a less coherent neighborhood in the network of co-occurring mutations tend to have a higher mutation frequency (Cui, 2010). Driver mutations are in cancer-modules and are neighbors of signature genes, whose expression can be used as a prognostic marker of metastasis and survival in breast tumors (Li et al., 2010d). Recently pathway and network reconstitution methods were suggested using patient survival-related mutation data (Vandin et al., 2012).

In human interactomes, proteins with cancer-specific mutations are hubs. They form a rich-club, acting as bridges between modules of different functions, and behave as bottlenecks providing exclusive connections between network segments or are otherwise central nodes (Jonsson & Bates, 2006; Chuang et al., 2007; Sun & Zhao, 2010; Rosado et al., 2011; Xia et al., 2011). Preferential connectedness of cancer-related proteins may contribute to their increased robustness to transmit a large volume of perturbations without being damaged (Serra-Musach et al., 2012). In agreement with the above observations, targets of anticancer drugs have a significantly larger number of neighbors than targets of drugs against other diseases (Hase et al., 2009). Inter-modular interactome hubs were found to associate with oncogenesis better than intra-modular hubs (Taylor et al., 2009). Integration of the interactome, protein domain composition, evolutionary conservation and gene ontology data in a machine learning technique predicted target genes, whose knockdown greatly reduced colon cancer cell viability (Li et al., 2009b). The interactomes of cancer associated cells, such as cancer associated fibroblasts may also highlight important prognostic markers of the disease (Bozóky et al., in press).

Differential gene expression analysis became one of the key approaches to identify genes important in diagnosis and prediction of cancer progression. The Oncomine resource includes more than

18,000 gene expression profiles (http://oncomine.org; Rhodes et al., 2007a). Oncomine data were extended by drug treatment signatures and target/reference gene sets providing a network of Molecular Concepts Map (http://private.molecularconcepts.org; Rhodes et al., 2007b). Differentially expressed proteins in human cancers were cataloged in the dbDEPC database (http://lifecenter.sgst.cn/dbdepc/index.do; He et al., 2012). Gene expression profiles may be used for reverse engineering of cancer specific regulatory networks (Basso et al., 2005; Ergun et al., 2007). Gene expression subnetworks showed increased similarity with the progression of chronic lymphocytic leukemia, suggesting that degenerate pathways converge into common pathways that are associated with disease progression (Chuang et al., 2012).

Interactome nodes may be marked according to their up- or down-regulation in cancer, and may identify clusters of proteins involved in cancer progression, such as in metastasis-formation (Rhodes & Chinnaiyan, 2005; Jonsson et al., 2006; Hernandez et al., 2007). Network analysis measures (e.g. degree, betweenness centrality, shortest path) of integrated interactome and expression data ranked cancer related proteins for target prediction, and showed their central network position (Wachi et al., 2005; Platzer et al., 2007; Chu & Chen, 2008; Mani et al., 2008).

However, altered expression of mRNAs is generally not enough to predict target efficiency (Yeh et al., 2012). mRNAs are often regulated by microRNAs, thus the inclusion of microRNA pattern analysis improves prediction as we will show in the next section. Moreover, the analysis of proteomic changes is also necessary in most cases (Gulmann et al., 2006; Pawson & Linding, 2008). Changes in protein levels may act synergistically (Maslov & Ispolatov, 2007). Starting from this idea, random walk-based interactome analysis identified sub-networks, which were around 'seeds' changing their protein levels in colorectal cancer, and screened these subnetworks using the level of the synergistic dysregulation of the associated mRNAs in colorectal cancer (Nibbe et al., 2010). Inclusion of additional data in the interactome and gene expression datasets, such as protein domain interactions, gene ontology annotations, cancer-related mutations, or cancer prognosis information refined predictions further (Franke et al., 2006; Pujana et al., 2007; Chang et al., 2009; Lee et al., 2009; Wu et al., 2010; Xiong et al., 2010; Yeh et al., 2012).

### 5.2.3. Metabolic network targets of anti-cancer drugs

The metabolism of cancer cells is adapted to meet their proliferative needs in predominantly anaerobic conditions (Warburg, 1956). Network modeling uses a number of cancer-specific pathways of energy metabolism (Resendis-Antonio et al., 2010; Vazquez et al., 2011; Khazaei et al., 2012; Kung et al., 2012). Metabolic networks of several cancer-types such as that of colorectal cancer were constructed recently (Martinez-Romero et al., 2010). Li et al. (2010e) used the k-nearest neighbor model to predict the metabolic reactions of the NCI-60 set (a set of 60 human tumor cell lines derived from various tissues of origin) influenced by approved anti-cancer drugs, and extended their method to suggest possible enzyme targets for anti-cancer drugs. Through the analysis of cancer-specific human metabolic networks Folger et al. (2011) predicted 52 cytostatic drug targets, of which 40% were targeted by known anti-cancer drugs, and the rest were new target-candidates. However, it should be kept in mind that key enzymes of cancer-specific metabolism, such as the PKM2 isoenzyme of pyruvate kinase playing a predominant role in the Warburg-effect (Warburg, 1956; Steták et al., 2007; Christofk et al., 2008; Kung et al., 2012), were also shown to play a direct role in cancer-specific signaling (Gao et al., 2012). Both metabolic and signaling networks are directed networks, where a hierarchy can be established, and where targeting of upstream, more influential nodes may be a fruitful strategy (Liu et al., 2012). We will review the use of signaling networks in anti-cancer therapies in the next section.

### 5.2.4. Signaling network targets of anti-cancer drugs

Signaling-related anti-cancer therapies increasingly outnumber metabolism-related chemotherapy options. From the network point of view this trend is due to the more developed signaling in humans than in pathogens, and to the increased selectivity of signaling interactions as compared to metabolism-related targeting.

Mass spectrometry can be effectively used for the analysis of post-translational modifications during the progression of cancer. Post-translational modifications, e.g. phosphorylation may change due to changes in the cellular environment and regulation under physiological conditions, but also due to a mutation at the phosphorylation site, or at a protein binding interface regulating kinase or phosphatase activity (Pawson & Linding, 2008). The bioinformatics resources NetworKIN (http://networkin.info; Linding et al., 2007) and NetPhorest (http://netphorest.info; Miller et al., 2008) can help in the analysis cancer-related signaling changes.

Rewiring of cancer-related changes of signaling networks is a primary aim in signal transduction-related anti-cancer therapies (Papatsoris et al., 2007). Cancer-specific changes in gene expression may activate or inactivate noncanonical edges in signal transduction networks (Klinke, 2010). Higher complexity of cancer-specific signaling network was shown to correlate with shorter survival (Breitkreutz et al., 2012). Proteins with cancer-related mutations are often hubs of human signaling network and are enriched in positive signaling regulatory loops (Awan et al., 2007; Cui et al., 2007; Cloutier & Wang, 2011; Li et al., 2012a). Alteration in cross-talking, multi-pathway, inter-modular proteins of signaling networks was proposed to be a key process in tumorigenesis (Hornberg et al., 2006; Taylor et al., 2009; Korcsmáros et al., 2010; Rivera et al., 2012).

The mammalian target of rapamycin (mTOR) is an important example of multi-pathway effects. mTOR has a key role in cell growth and regulation of cellular metabolism. In most tumors, mTOR is mutated, causing a hyper-active phenotype (Zoncu et al., 2011). Though mTOR activity was expected to be a promising therapeutic target, drugs showed poor results in clinical trials. mTOR could not meet node-targeting expectations because of its multi-pathway position, participating in at least two major signaling complexes, mTORC1 and mTORC2 (Huang et al., 2004; Caron et al., 2010; Catania et al., 2011; Pe'er & Hacohen, 2011; Fingar & Inoki, 2012).

Edgetic drugs specifically targeting mTOR interactions may selectively influence cancer-specific mTOR functions (Section 4.1.2; Ruffner et al., 2007). Another example of edgetic anti-cancer therapy options is that of nutlins, which block the interaction between p53 and its negative modulator MDM2 activating the tumor suppressor effect of p53 (Vassilev et al., 2004). Cancer-related proteins have smaller, more planar, more charged and less hydrophobic binding interfaces than other proteins, which may indicate low affinity and high specificity of cancer-related interactions (Kar et al., 2009). These structural features make lead compound development of cancer-related edgetic drugs a challenging task.

MicroRNAs are increasingly recognized as highly promising, non-protein intervention points of the signaling network (see also in Section 3.4). Loss- or gain-of-function mutations of microRNAs have been identified in nearly all solid and hematologic types of cancer (Calin & Croce, 2006; Spizzo et al., 2009). In addition, microRNAs were recently found as a form of intercellular communication (Chen et al., 2012c). Thus, alteration of microRNA content may have an effect on the microenvironment of tumor cells. Drug-induced changes in the expression of specific microRNAs can induce drug sensitivity leading to an increased inhibition of cell growth, of invasion and of metastasis formation (Sarkar et al., 2010). However, microRNAs have a dual role in cancer, acting both as oncogenes targeting mRNAs coding tumor-suppression proteins, or tumor suppressors targeting mRNAs coding oncoproteins (Iguchi et al., 2010; Gambari et al., 2011). This suggests the use of systems-level, network approaches to select microRNA targets.

Combination of cancer-specific mRNA and microRNA expression data may be used to infer cancer-specific regulatory networks (Bonnet et al., 2010). MicroRNAs involved in prostate cancer progression preferentially target interactome hubs (Budd et al., 2012). MicroRNA networks obtained from 3312 neoplastic and 1107 nonmalignant human samples showed the dysregulation of hub microRNAs. Cancer-specific microRNA networks had more disjoined subnetworks than those of normal tissues (Volinia et al., 2010). The fast growing complexity of signaling networks still awaits a comprehensive treatment in anticancer therapies.

### 5.2.5. Influential nodes and edges in network dynamics as promising drug targets

As we mentioned in the introduction of Section 5, the central hit strategy of anti-cancer drug design would often hit a protein, which is also central in the networks of healthy cells. Therefore, here the more indirect targeting of the network influence strategy may also prove useful. Many targets of anti-cancer therapies are not directly cancer-related (Hornberg et al., 2006; Cheng et al., 2012c). Context can influence network behavior in at least four different ways: a.) the genetic background (e.g., single-nucleotide polymorphisms and other mutations); b) gene expression changes (caused by e.g. transcription factor, epigenetic or microRNA changes); c.) neighboring cells; and finally d.) exogenous signals (e.g. nutrients or drugs) all providing increment to the patient-specific, context-dependent responses to anti-cancer therapy (Klinke, 2010; Sharma et al., 2010b; Pe'er & Hacohen, 2011). Differential gene expression and phosphorylation studies were already shown to be useful to distinguish among different stages of cancer development in the preceding sections. The next challenging step is to examine the cancer-induced dynamic changes on a network-level.

The examination of differential networks of cancer stages, or networks of drug treated and un-treated cells, is one of the first steps in possible solutions (Ideker & Krogan, 2012). Network level integration of cancer-related changes (such as mutations, gene expression changes, post-translational modifications, etc.) may capture key differences in network wiring (Pe'er & Hacohen, 2011).

Network dynamics may be assessed by the dissipation of perturbations, which can be used for the prioritization of drug target candidates. The early work of White and Mikulecky (1981) used a small network to assess the dynamics of methotrexate action. Stites et al. (2007) studied changes of Ras signaling in cancer using a differential equation model applied to a limited signaling network-set. They concluded that a hypothetical drug preferably binding to GDP-Ras would only induce a cancer-specific decrease in Ras signaling. Shiraishi et al. (2010) identified 6585 pairs of bistable toggle switch motifs in regulatory networks forming a network of 442 proteins. Among the 24 conditions examined, mRNA expression level changes reversed the ON/OFF status of a significantly high number of bistable toggle switches in various types of cancer, such as in lung cancer or in hepatocellular carcinoma. Serra-Musach et al. (2012) found that cancer-related proteins have an increased robustness to transmit a large volume of perturbations without being damaged. Extensions of such investigations to network-wide perturbations (modulated by neighboring cells and exogenous signals) will be an important research area for finding influential nodes/edges serving as drug target candidates.

### 5.2.6. Drug combinations against cancer

As we have shown in the preceding sections cancer is a systems-level disease, where magic-bullet type drugs may fail. Partially redundant signaling pathways are hallmarks of cancer robustness. Thus an inhibitor of a particular hallmark may not be enough to block the related function. Moreover, when inhibitors of a specific cancer hallmark are used separately, they may even strengthen another hallmark, like certain types of angiogenesis inhibitors increased the rate of metastasis. In most failures of anti-cancer therapies, unwanted off-target effects and undiscovered feedbacks prevented the desired pharmacological goal. Combination therapies and multi-target drugs may both overcome system robustness and provide less side-effects (see Section 4.1.5; Gupta et al., 2007; Berger & Iyengar, 2009; Wilson et al., 2009; Azmi et al., 2010; Glaser, 2009; Hanahan & Weinberg, 2011).

Cancer-specific subsets of the human interactome can provide a guide for the development of multi-target therapies. Mutually exclusive gene alterations which share the same biological process may define cancer type-related interactome modules (Ciriello et al., 2012b). Other types of cancer-related network modules were identified as sub-interatomes, as in colorectal cancer. These were centered on proteins, which markedly change their levels, and showed a synergistic dysregulation at their mRNA levels (Nibbe et al., 2010). Simultaneous targeting of these modules may be an efficient therapeutic strategy.

Multiple-targets can be identified using cancer-specific metabolic network models. Combinations of synthetic lethal drug targets were predicted in cancer-specific metabolic networks (Moreno-Sanchez et al., 2010; Folger et al., 2011). Ensemble modeling, which exploited a perturbation of known targets in a subset of 58 central metabolic reactions, was used to predict target sets of key enzymes of central energy metabolism (Khazaei et al., 2012).

Potential drug target sets were identified by an algorithm, which calculates the downstream components of a prostate cancer-specific signaling network affected by the inhibition of the target set (Dasika et al., 2006). In the particular example of EGF receptor inhibition, subsequent applications of drug combinations were shown to have a dramatically improved effect. This unmasked an apoptotic pathway, and via complex signaling network effects dramatically sensitized breast cancer cells to subsequent DNA-damage (Lee et al., 2012d). These findings substantiate Kitano's earlier emphasis on the importance of cancer chronotherapy (Kitano, 2004a, 2004b, 2007).

Tumors contain a highly heterogeneous cell population. Drug combinations may act via an intracellular network of a single cell; but also via inhibiting subsets of the heterogeneous population of malignant cells. Cell populations and their drug responses can be perceived as a bipartite graph. Applying minimal hitting set analysis allowed the search for effective drug combinations at the intercellular network level (Vazquez, 2009).

As we showed in this section, analysis of network topology and, especially, network dynamics can predict novel anti-cancer drug targets. Incorporation of personalized data, such as mutations, singalome or metabolome profiles to the molecular networks listed in this section may enhance patient- and disease stage-specific drug targeting in anti-cancer therapies.

## 5.3. Diabetes (metabolic syndrome including obesity, atherosclerosis and cardiovascular disease)

Diabetes is the first of our two examples showing the applications of the network influence strategy defined in Section 4.1.1, where therapeutic interventions need to push the cell back from the attractor of the diseased state to that of the healthy state. Diabetes is a multigenic disease tightly related to central obesity, atherosclerosis and cardiovascular disease, a connection also revealed by network representations (Ghazalpour et al., 2004; Lusis & Weiss, 2010; Stegmaier et al., 2010). Here we summarize network-related methods to predict novel drug target candidates in diabetes and related metabolic diseases.

Type 2 diabetes is the most common form of diabetes that is characterized by insulin resistance and relative insulin deficiency. T2D-db is a database of molecular factors involved in type 2 diabetes (http://t2ddb.ibab.ac.in; Agrawal et al., 2008) providing useful information for the construction of various diabetes-related networks. Combination of interactome and diabetes-related gene expression data identified the possible molecular basis of several endothelial, cardiovascular and kidney-related complications of diabetes, and revealed novel links between diabetes, obesity, oxidative stress and inflammatory abnormalities (Sengupta et al., 2009; Mori et al., 2010). Similar studies suggested a network of protein–protein interactions bridging insulin signaling and the peroxisome proliferator-activated receptor-(PPAR)-related nuclear hormone receptor family (Liu et al., 2007b). Refinement of interactome data containing domain–domain interactions combined with the earlier observation that disease-related genes have a smaller than average clustering coefficient (Feldman et al., 2008) led to the prediction of type 2 diabetes-related genes (Sharma et al., 2010a). Inter-modular interactome nodes between type 2 diabetes-, obesity- and heart disease-related proteins may play a key role in the dysregulation of these complex syndromes (Nguyen & Jordan, 2010; Nguyen et al., 2011).

Type 1 diabetes is primarily related to the dysregulation of insulin secretion of pancreatic ß-cells, where ß-cell dedifferentiation was recently shown to play an important role (Talchai et al., 2012). The ß-cell endoplasmic reticulum stress signaling network is an important regulator of this process (Fonseca et al., 2007; Mandl et al., 2009). Integration of interactome and genetic interaction data revealed novel protein network modules and candidate genes for type 1 diabetes (Bergholdt et al., 2007).

Reconstruction of changes of the human metabolic network of skeletal muscle in type 2 diabetes enabled the identification of potential new metabolic biomarkers. Analysis of gene promoters of proteins associated with the biomarker metabolites led to the construction of a diabetes-related transcription factor regulatory network (Zelezniak et al., 2010). Recently an integrated, manually curated and validated metabolic network of human adipocytes, hepatocytes and myocytes was assembled. Several metabolic states, such as the alanine-cycle, the Cori-cycle and an absorptive state, as well as their changes between obese and diabetic obese individuals were characterized (Bordbar et al., 2011). Such studies will highlight key enzymes of metabolic network, where a drug-induced activity and/or regulation change may significantly contribute to the rewiring of the metabolic network to its normal state.

Insulin signaling is in the center of the etiology of metabolic diseases. Several studies highlighted diabetes-responsible segments of the human signaling network enriching and re-focusing the traditionally known insulin signaling pathway. The mammalian target of rapamycin (mTOR) protein is one of the focal points of the insulin signaling network. From the two mTOR-related signaling complexes mentioned in the preceding section, Complex 1 (mTORC1) is a key player in nutrient-related signaling involving the hypothalamus, peripheral organs, adipose tissue differentiation and ß-cell dependent insulin secretion (Catania et al., 2011; Fingar & Inoki, 2012). An siRNA knockout screen of 300 genes involved in the lipolysis of 3T3-L1 adipocytes led to the identification of a core, insulin resistance-related sub-network of the insulin signaling pathway highlighting a number of novel genes related to insulin-resistance, such as the sphingosine-1-phosphate receptor-2 (Tu et al., 2009). Reconstruction of the subnetwork of human interactome related to insulin signaling and the determination of its hubs and bottleneck proteins (Durmus Tekir et al., 2010) is an ongoing work, which will uncover many important novel targets of therapeutic interventions in the future. As an additional extension of insulin signaling, recent studies started to uncover the changes and most influential members of the microRNA regulatory network in diabetes (Huang et al., 2010c; Zampetaki et al., 2010). Phosphoproteome-studies help to extend the insulin signaling network further, and to uncover its time-dependent changes (Schmelze et al., 2006).

Tissue-specific gene expression data identified metabolic disease-specific regulatory network modules, and revealed the involvement of both macrophages and the inflammatome in the pathogenesis of metabolic diseases (Schadt et al., 2009; Lusis & Weiss, 2010; Wang et al., 2012f). These studies show the inter-pathway and inter-organ complexity reached in the network understanding of metabolic disease. In Section 4.1.7 we summarized the network influence strategy; that is, to rewire the cellular networks from their diseased state to healthy state as a tool to help in successful drug design. This includes avoiding network segments which are essential in healthy cells, and focusing on targeting pathway sites specific to diseased cells, and the use of multiple or indirect targeting. For this, metabolic disease network studies need to apply network dynamics methods such as we listed in Section 2.5. Systematic, network-based identification of edgetic, multi-target and allo-network drugs (see Section 4.1) could also be beneficial. Refined network methods should also incorporate patient- and disease stage-specific data. These are intimately related to the network consequences of aging, which will be described in the next section.

### 5.4. Promotion of healthy aging and neurodegenerative diseases

Aging is one of the most complex processes of living organisms. Aging was described as a network phenomenon (Kirkwood & Kowald, 1997; Sőti & Csermely, 2007; Simkó et al., 2009; Chautard et al., 2010). In the first half of this section we will summarize the few initial network studies on age-related multifactorial changes. Besides cancer and the metabolic syndrome described in the preceding sections, neurodegeneration is one of the major aging-associated diseases. In the concluding part of the section we will describe network-related studies on the prediction of potential drug targets to prevent and slow down various forms of neurodegeneration, such as Alzheimer's, Parkinson's, Huntington's and prion-related diseases.

#### 5.4.1. Aging as a network process

Aging organisms show similar early warning signals of critical phase transitions (i.e.: slower recovery from perturbations, increased self-similarity of behavior and increased variance of fluctuation-patterns) as described for a wide variety of complex systems (Section 2.5.2; Scheffer et al., 2009; Sornette & Osorio, 2011; Dai et al., 2012). Aging can be perceived as an early warning signal of a critical phase transition, where the phase transition itself is death (Farkas et al., 2011). However, this sobering message also has a positive implication: phase transitions of complex systems can be slowed down, postponed, or prevented by nodes having an independent and unpredictable behavior (Csermely, 2008). The identification of these nodes may lead to the discovery of novel molecular agents promoting healthy aging.

The complexity of the aging process is illustrated well by the duality of possible aging-related trends in network changes. Aging-related disorganization causes an increase of non-specific edges, and an aged organism has fewer resources predicting the loss of network edges during aging. Thus, small-worldness may often be lost during the aging process, and the hub-structure may get reorganized. Aging networks are likely to become more rigid, and may have less overlapping modules (Sőti & Csermely, 2007; Csermely, 2009; Kiss et al., 2009; Simkó et al., 2009; Gáspár & Csermely, 2012). The longest documented lifespan is currently 122 years achieved by a French woman (Allard et al., 1998). It is currently an open question, whether lifespan has any upper limits. It will be interesting if future aging-related studies of network topology and behavior will predict any upper limit of human lifespan.

Aging-associated genes form an almost fully connected sub-interactome (also called longevity networks; Budovsky et al., 2007), and occupy both hub (Promislow, 2004; Ferrarini et al., 2005; Budovsky et al., 2007; Bell et al., 2009) and inter-modular positions (Xue et al., 2007). Aging-associated genes are concentrated in 4 modules of the yeast interactome (Barea & Bonatto, 2009). Similarly, age-related gene expression changes preferentially affect only a few modules of the human brain and *Drosophila* interactomes (Xue et al., 2007). The sub-interactome of aging-genes can be extended by their neighbors and the related network edges. The extended network provides a target-set to identify novel aging-related genes (Bell et al., 2009). The sub-interactomes of aging-associated genes and major age-related disease genes highly overlap with each other.

Aging-genes bridge other genes related to various diseases (Wang et al., 2009; Wolfson et al., 2009).

Longevity networks are enriched by key signaling proteins (Reja et al., 2009; Simkó et al., 2009; Wolfson et al., 2009; Borklu Yucel & Ulgen, 2011). The complexity of age-related processes is exemplified well by the extensive cross-talks of age-related signaling pathways (de Magalhaes et al., 2012). As an example, the growth hormone-related pathways, the oxidative stress-induced pathway and the dietary restriction pathway all affect the FOXO (Daf-16) transcription factor (Greer & Brunet, 2008). The yeast gene regulatory network was reconstituted by reverse engineering methods using age-associated transcriptional changes. The regulatory network revealed novel aging-associated regulatory components (Lorenz et al., 2009). MicroRNAs play an important role in aging-related signaling events (Chen et al., 2010b). Network analysis will help the identification of critical nodes of age-related signaling. These nodes may serve as potential targets of drugs promoting healthy aging.

During the aging process, the nuclear pore complexes become more permeable (D'Angelo et al., 2009). It is likely that age-induced increase of permeability is a general phenomenon involving other cellular compartments (Simkó et al., 2009) and increasing in the number of non-specific edges of the inter-organelle network.

Though drug development efforts are rapidly increasing in the field, currently there are only a few drugs which directly target the aging process (Simkó et al., 2009; de Magalhaes et al., 2012). To date, it is an open question, if central hit strategy-type or network influence strategy-type drug targeting (aiming to target key network nodes, or aiming to influence aging-related changes, respectively) will be the most efficient route for finding appropriate drugs for the promotion of healthy aging. Most probably the network influence strategy will be the 'winner', and the anti-aging drugs of the future will be multi-target drugs, providing an indirect influence on key processes of aging networks. For this additional studies on aging-related dynamics of molecular networks are needed.

#### 5.4.2. Network strategies against neurodegenerative diseases

As Lipton (2004) remarked, according to some predictions by 2050 the entire economy of the industrialized world could be consumed by the costs of caring for the sick and elderly. Neurodegenerative diseases, such as Alzheimer's, Parkinson's, Huntington's and prion diseases constitute one of the major aging-related disease-class besides cancer and metabolic diseases. Although several symptomatic drugs are available, a disease-modifying agent is still elusive making novel approaches especially valuable (Dunkel et al., 2012; Mei et al., 2012; Funke et al., in press).

We listed the major network-related methods to uncover novel neurodegenerative disease-associated genes, potential drug targets, or for drug repositioning in Table 11. Two major network methodologies emerge, which are widely used in connection with neurodegenerative diseases. One of them constructs, or extends disease-related protein–protein interaction networks and predicts novel disease-associated proteins. This appears a straightforward technique, since a neurodegenerative disease causes a major reconfiguration of cellular protein complexes. The other major method uses network analysis of differentially expressed genes in disease-affected patients or model organisms. This method identifies novel regulatory and signaling components involved in disease progression. Both methods may identify disease-affected pathways, which may be used to construct "heat-maps" identifying novel drug target candidates (Dunkel et al., 2012; Mei et al., 2012).

When summarizing neurodegenerative disease-related network efforts, it was surprising that, besides a few initial attempts in Alzheimer's disease, how little attention was devoted to chemical similarity networks, metabolic networks, signaling networks and drug–target networks in this field. Dysregulated, over-acting signaling pathways have a major contribution to all neurodegenerative diseases, and their network analysis would deserve more attention. A good anti-neurodegenerative drug is typically a network influence strategy-type drug reconfiguring the distorted pathways of disease-associated networks (Lipton, 2004; Dunkel et al., 2012). Learning more on changes in network dynamics during neurodegenerative disease progression would be a major advance of drug design efforts in this crucially important field.

**Table 11**
Illustrative examples of network strategies against neurodegenerative diseases.

| Type of network | Drug design benefit | References |
| --- | --- | --- |
| *Alzheimer's disease* | | |
| Protein–protein interaction network (extended with drug interactions) | Prediction of novel disease-related genes and novel disease-associated drugs from existing ones by interactome proximity | Krauthammer et al., 2004; Li et al., 2009a; Yang & Jiang, 2010; Hallock & Thomas, 2012; Raj et al., 2012 |
| Differentially co-expressed gene networks of normal and Alzheimer's disease affected patients | Identification of co-expressed gene modules and disease-related transcription factors | Ray et al., 2008; Satoh et al., 2009; Liang et al., 2012 |
| Network of differentially expressed MicroRNAs of Alzheimer's affected patients | Prediction of novel disease-associated signaling pathways and regulators | Satoh, 2012 |
| Drug binding site similarity networks | Prediction of novel drug targets | Yang et al., 2010 |
| Drug target networks of anti-Alzheimer's herbal medicines | Prediction of novel drug targets | Sun et al., in press |
| | | |
| *Parkinson's disease* | | |
| Differentially co-expressed gene networks of normal and Parkinson's disease affected patients | Identification of central disease-associated genes | Moran & Graeber, 2008 |
| | | |
| *Poly-glutamine (polyQ) expansion diseases (Huntington's disease, ataxias)* | | |
| Protein–protein interaction network of poly-glutamine proteins (and their known interactome neighbors) | Identification of novel modifiers of disease progression | Goehler et al., 2004; Lim et al., 2006; Kaltenbach et al., 2007; Kahle et al., 2011 |
| | | |
| *Prion disease* | | |
| Differentially co-expressed gene networks of normal and prion disease affected mice | Identification of disease-associated pathways and modules | Hwang et al., 2009; Kim et al., 2011b |

# 6. Conclusions and perspectives

The value of every drug design technology must be assessed by asking: "*How much does the new technology help to solve one of the two central problems: the identification and validation of a disease-specific target or the identification of a molecule that can modify this target in a way that makes therapeutic sense?*" (Brown & Superti-Furga, 2003; Drews, 2003). Network description and analysis may offer novel leads in both questions. In this concluding section we will highlight the promises and perspectives of network-aided drug development.

## 6.1. Promises and optimization of network-aided drug development

One of the major promises of network description and analysis are their help to overcome the "one-effect/one-cause/one-target" magic bullet-type drug development paradigm (Ehrlich, 1908). Magic bullets do work—sometimes. When designing "central hit strategy-type" drugs (Sections 4.1.1 and 4.1.7), which target key nodes of the network to eliminate pathogens or malignant cells, eradicating single hit may be beneficial. However, pathogen resistance or unexpected toxicity of anti-cancer drugs (and resistance against them) may dog the outcome. In the development of "network influence strategy-type drugs", where an efficient reconfiguration of rigid networks needs to be achieved, network dynamics has to be reset from its disease-affected state back to normal (Sections 4.1.1 and 4.1.7). Under these circumstances, the traditional approach of rational drug discovery selecting a single and central target often fails. The paucity of disease-modifying anti-neurodegenerative drugs described in the preceding section is an example for the need for novel approaches in network influence strategy-type drug design.

James Black described well a wide-spread behavior saying that 'the most fruitful basis for the discovery of a new drug is to start with an old drug' (Chong & Sullivan, 2007). We started our review with the statement that 'business as usual' is no longer an option in drug industry (Begley & Ellis, 2012). Currently, there is a broad consensus that this state, where the vast majority of new drugs are related to existing ones, needs to be changed (Section 1.1; Cokol et al., 2005; Yildirim et al., 2007; Iyer et al., 2011a).

Nonetheless, the question is how to find 'surprisingly novel drugs'. The failure of some efforts using the reductionist approach of rational drug design shifted the thinking to the other extreme that 'we need unbiased research methods to cover complexity'. Indeed, unbiased methods (including network analysis or machine learning) may successfully predict novel drug targets. However, clearly, artificial intelligence may miss 'true' surprises (Section 2.2.2). This leads to our first major conclusion, which we summarized in Fig. 23: network description and analysis should be combined with human creativity and background knowledge.

Network analysis helps in comprehending the vast amounts of systems-level data, which accumulated over the last decade. However, network analysis alone is clearly insufficient and has to be complemented with the intuitions coming from background knowledge (Valente, 2010). In this process creativity (marked as the 'surprise factor' in Fig. 23), which strives for novelty, and identifies it in networks as the 'prediction of the unpredictable' (Section 2.2.2) cannot be overlooked. Combined, the current boom in network dynamics-related methods can help in discovering the key actors in the cellular community, which are the hidden masterminds of cellular changes in health and in disease (Fig. 23).

Our second major conclusion is that a protocol of network-aided drug development would be aided by alternating exploration and optimization phases of drug design (Fig. 23). In the exploration phase background knowledge may be temporarily suppressed. In contrast, in the optimization phase we need to suppress the playfulness and ambiguity tolerance of the exploration phase, and rank the options by rigorous application of background knowledge including all the well-orchestrated rules of the drug development process (Csermely, in press; Gyurkó et al., in press). Importantly, the sequence of exploration and optimization phases may be applied repeatedly, providing a more detailed 'zoom-in' of the optimal (drug) target than a single round of exploration/optimization (Fig. 23). The utility of repeated exploration/optimization rounds was shown in a number of examples ranging from thermal cycles of simulated annealing optimization

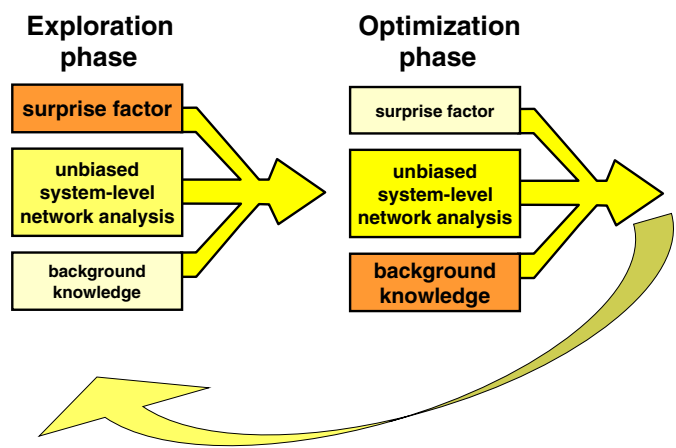

**Fig. 23.** Optimized protocol of network-aided drug development. The figure illustrates the two major phases of discovery having three segments marked as boxes on the left side of the triple arrows. The "surprise factor" box denotes originality (as the highest level of human creativity), a strong drive to discover the unexpected, including playfulness and ambiguity tolerance. The "unbiased systems-level network analysis" box marks the network methods described in this review. The "background knowledge" box includes all our contextual, background knowledge on diseases, drugs and their actions, as well as the validation procedures guiding our judgment on the quality of the drug discovery process. In the exploration phase the surprise factor is dominant. At this phase background knowledge may be temporarily suppressed. On the contrary, at the optimization phase we need to suppress the surprise factor, and rank our previous options by the rigorous application of our background knowledge. The arrow at the bottom of the figure marks the heretofore not rigorously applied method that the sequence of exploration and optimization phases may be applied repeatedly, which gives a much more precise 'zoom-in' to the optimal (drug) target than a single round of exploration/optimization.

(Möbius et al., 1997) to the human learning process (Bassett et al., 2011).

Drug design-related networks became increasingly complex during the past decade. Albert Einstein's saying that "the supreme goal of all theory is to make the irreducible basic elements as simple and as few as possible without having to surrender the adequate representation of a single datum of experience" (Einstein, 1934) (also called 'Einstein's razor', extending the Occam's razor theorem advocating only the simplest solution) encourages finding the optimal network representation, which is simple enough, but not too simple. Finding the optimal complexity of network representation in the drug discovery process is an important task. Eventually, the recurring application of exploration and optimization phases shown in Fig. 23 suggests that network data coverage should be extended in consecutive phases separated by recurrent network simplifications based on background knowledge.

Thomas Singer wrote a few years ago "Extrapolation of preclinical data into clinical reality is a translational science and remains an ultimate challenge in drug development." (Singer, 2007). Addressing this challenge our third and last major conclusion stresses the importance of network prediction of these human data, which are not available experimentally. This includes overcoming three major hurdles (Fig. 2; Brown & Superti-Furga, 2003; Austin, 2006; Bunnage, 2011; Ledford, 2012): 1.) insufficient drug efficacy; 2.) unexpected major adverse effects; 3.) unexpected forms of human toxicity. Network analysis may help ameliorate the efficacy by taking into account patient–, disease stage–, age–specificities (Section 4.3.1); it may help obtain a better prediction of side-effects (Section 4.3.5) and predictive human toxicology (Section 4.3.3; Henney & Superti-Furga, 2008).

Network science is a novel area of biology; and this is particularly the case with respect to drug design. We often lack rigorous comparisons of existing methods, which could have allowed a more critical approach to some of them. It is an ongoing effort of the current years to develop benchmarks, gold-standards and rigorous assessment tools in network science.

### 6.2. Systems-level hallmarks of drug quality and trends of network-aided drug development helping to achieve them

In this closing section we identify the systems-level hallmarks of drug quality, and list the major trends of network-aided drug development helping to achieve them. From the network point of view we propose two strategies in finding drug targets: 1.) the central hit strategy aiming to destroy the network of infectious agents or cancer cells and 2.) the network influence strategy aiming to shift the network dynamics of polygenic, complex diseases back to normal (Fig. 19; Sections 4.1.1 and 4.1.7). Both strategies converge to the same level of network complexity in hit finding, hit expansion, lead selection and optimization phases. We note that these two strategies appear as general strategies to design the most efficient attack of flexible, plastic systems (using the central hit strategy) or rigid systems (using the network influence strategy) valid from molecular structures, through molecular and cellular networks to social and engineered networks or ecosystems.

Table 12 lists the systems-level hallmarks of drug target identification and validation, hit finding and development, as well as lead selection and optimization. We believe that the systematic application of these systems-level hallmarks will not only help the identification of novel drug targets, but will also streamline the drug design process to be more selective, less attrition-prone and more profitable.

Table 12 also includes the most important network-related drug design trends helping the accomplishment of various systems-level hallmarks. We highlight the development of edgetic drugs (Section 4.1.2), multi-target drugs (Section 4.1.5) and allo-network drugs (Section 4.1.6) among the richness of network strategies to find novel drug targets. We believe that there are a large number of unexplored strategies, which are the hidden masterminds of cellular regulation. Analysis of network dynamics can help to find them. Incorporation of disease-stage, age-, gender- and human population-specific genetic, metabolome, phosphoproteome and gut microbiome data; the development of human ADME and toxicity network models; and the use of side-effect networks to judge drug safety, may greatly increase the efficiency of the drug development process.

Network-related methods—if applied systematically (and carefully)—will uncover a number of novel drug targets, and will increase the efficiency of the drug development process. Analysis of the structure and dynamics of molecular networks, extended by the network dynamics of constituting proteins and in particular their binding sites, provides a novel paradigm of drug discovery.

### Conflict of interest statement

The authors declare that there are no conflicts of interest.

**Table 12**
Systems-level hallmarks of drug quality and trends of network-related drug design helping to achieve them.

| Systems-level hallmark of drug quality | Network-related drug design trend |
|---|---|
| Drug target identification<br>  • Using the central hit strategy: drug hits central (or otherwise essential) network nodes, whose efficient inhibition selectively destroys infectious agent or cancer cell<br>  • Using the network influence strategy: drug hits disease-specific network segments (nodes, edges or their sets), whose manipulation shifts disease-affected functions back to normal<br>Drug target validation<br>  • Network dynamics-based, disease-specific early and robust human biomarkers are used for drug target validation, drug added-value assessment over current standard care, and translation for later monitoring in clinical trials | • Disease stage-related differential interactome, signaling network, metabolic network data (*including protein abundance, human/comparative genetic data and microRNA profiles, optionally combined with protein, RNA and chromatin structure information, as well as with subcellular localization*)<br>• Network comparison and reverse engineering<br>• Disease-specific models of network dynamics (*including deconvolution, perturbation, hierarchy, source/sink/seeder analysis and network influence*)<br>• Drug target, patient and therapy-related networks helping multi-target design and drug repositioning<br>• Use of weighted, directed, signed, colored and conditional edges or hypergraph structures<br>• Network prediction methods (sensitized for finding the unexpected)<br>• In the central hit strategy: network centrality measures; host/parasite, host/cancer network combinations at the local ecosystem level<br>• In the network influence strategy: network controllability, influence, dynamic network centrality; compensatory deletions; edgetic, multi-target and allo-network drugs; chrono-therapies (temporal shifts in administration of drug combinations) besides application of the trends listed above |
| Hit finding and development<br>  • Hit finding and ranking are helped by network chemoinformatics<br>  • Hit expansion and library design are helped by chemical reaction networks | • Complex chemical similarity (QSAR) networks including chemical similarity networks, multi-QSAR networks, pocket similarity networks, and chemical descriptors of ligand binding sites, integrated bio-entity networks<br>• Chemical reaction networks<br>• Network analysis of protein structures and correlated segments of protein dynamics<br>• Analysis of the substrate envelope to avoid drug resistance development<br>• Hot spot, and hot region identification |
| Lead selection and optimization<br>  • Optimization of drug efficacy, selection of robust efficacy end-points and patient populations are guided by network pharmacogenomics, as well as by disease-stage, age-, gender- and population-specific metabolome, phosphoproteome and gut microbiome data<br>  • ADME and toxicity data are 'humanized', side-effect, drug–drug interaction and drug resistance evaluation are helped, as well as indications and contraindications are defined by extensive network data | Besides application of the trends listed above<br>• Network extension by disease-stage, age-, gender- and human population-specific genetic, metabolome, phosphoproteome and gut microbiome data<br>• Analysis of semantic networks from medical records<br>• Human ADME and toxicity network models<br>• Network methods of multi-target drug design to uncover adverse drug–drug interactions<br>• Assessment of side-effect networks<br>• Antagonistic drug combinations to avoid drug resistance development |

## Acknowledgments

Authors thank Aditya Barve and Andreas Wagner (University of Zürich, Switzerland) for sharing the human homology of enzymes encoding superessential metabolic reactions, Haiyuan Yu, Xiujuan Wang (Department of Biological Statistics and Computational Biology, Weill Institute for Cell and Molecular Biology, Cornell University, Ithaca NY, USA) and Balázs Papp (Szeged Biological Centre, Hungarian Academy of Sciences, Szeged, Hungary) for the critical reading of Sections 1.3.3 and 3.6, respectively. Authors thank Zoltán P. Spiró (École Polytechnique Federale de Lausanne, Switzerland) for help in drawing Fig. 11, and the anonymous referee, members of the LINK-Group (www.linkgroup.hu), as well as more than twenty additional colleagues for reading the original version of the paper and for valuable suggestions. Work in the authors' laboratory was supported by research grants from the Hungarian National Science Foundation (OTKA K83314), by the EU (TÁMOP-4.2.2/B-10/1-2010-0013), by a Bolyai Fellowship of the Hungarian Academy of Sciences (TK) and by a residence at the Rockefeller Foundation Bellagio Center (PC). This project has been funded, in part, with federal funds from the NCI, NIH, under contract HHSN261200800001E. This research was supported, in part, by the Intramural Research Program of the NIH, National Cancer Institute, Center for Cancer Research. The content of this publication does not necessarily reflect the views or policies of the Department of Health and Human Services, nor does mention of trade names, commercial products, or organizations imply endorsement by the U.S. Government.